\documentclass[12pt,a4paper]{article}
\usepackage{graphicx,color}
\usepackage{amsmath}
\usepackage{amssymb}
\pdfoutput=1
\usepackage{multirow}
\usepackage{url}
\usepackage{subfigure}
\usepackage{authblk}
\usepackage{fullpage}
\usepackage{lscape}
\begin{document}
\vspace*{-0.5in}

\vspace{0.5in}
\begin{center}
\begin{LARGE}
{\bf Proposal:\\
A Search for Sterile Neutrino at J-PARC Materials and Life Science Experimental Facility\\}
\vspace{5mm}
%\today\\
September 2, 2013\\
\end{LARGE}
\vspace{15mm}
{\large
M.~Harada, S.~Hasegawa, Y.~Kasugai, S.~Meigo,  K.~Sakai, \\
S.~Sakamoto, K.~Suzuya \\
{\it JAEA, Tokai, Japan}\\
\vspace{5mm}
E.~Iwai, T.~Maruyama, K.~Nishikawa, R.~Ohta\\
{\it KEK, Tsukuba, JAPAN}\\
\vspace{5mm}
M.~Niiyama\\
{\it Department of Physics, Kyoto University, JAPAN}\\
\vspace{5mm}
S.~Ajimura, T.~Hiraiwa, T.~Nakano, M.~Nomachi, T.~Shima\\
{\it RCNP, Osaka University, JAPAN}\\
\vspace{5mm}
T.~J.~C.~Bezerra, E.~Chauveau, T.~Enomoto, H.~Furuta, H.~Sakai, \\
F.~Suekane\\
{\it Research Center for Neutrino Science, Tohoku University, JAPAN}\\
\vspace{5mm}
M.~Yeh\\
{\it Brookhaven National Laboratory, Upton, NY 11973-5000, USA}\\
\vspace{5mm}
G.~T.~Garvey, W.~C.~Louis, G.~B.~Mills, R.~Van~de~Water\\
{\it Los Alamos National Laboratory, Los Alamos, NM 87545, USA}\\
}
\vspace{5mm}
\end{center}
\footnote{\large{Spokes person : Takasumi Maruyama (KEK) \\
takasumi.maruyama@kek.jp} }
\renewcommand{\baselinestretch}{2}
\large
\normalsize

\setlength{\baselineskip}{5mm}
\setlength{\intextsep}{5mm}

\renewcommand{\arraystretch}{0.5}

\newpage

\begin{center}
%%%%%%%%%%%%%%%%%%%%%%%%%%%%%%%%%%%%%%%%%%%%%%%%%
{\bf Executive Summary}
\end{center}
%%%%%%%%%%%%%%%%%%%%%%%%%%%%%%%%%%%%%%%%%%%%%%%%%%

In the last fifteen years, several experiments have reported neutrino phenomena,
which may indicate the existence of more than three kinds of neutrinos. 
The experimental indications include an excess of $\bar{\nu}_e$  events in
a predominantly $\mu^+$ decay, a deficiency of $\nu_e$ events from a $\beta$
source, and a $\bar{\nu}_e$ deficiency in nuclear reactors. 
There is also an indication of an excess of electromagnetic shower events
in a predominantly $\nu_{\mu}$ and $\bar{\nu}_{\mu}$ neutrino beams.
If these phenomena are confirmed and are shown to be due to neutrino
oscillations, then the corresponding $\Delta m^2$ is about an $eV^2$, which is
orders of magnitude larger than those in solar and atmospheric
neutrinos oscillations. 
This requires the existence of a new mass state(s) near an eV in addition to the
three standard mass states. 
This, in turn, requires an additional neutrino state(s) in nature both in
mass and flavor (in addition to $\nu_e$, $\nu_{\mu}$, and 
$\nu_{\tau}$). 
Considering the Z-boson width,  the new flavor state does not couple to
the Z-boson. Therefore, this is a new kind of lepton, namely a sterile 
neutrino, which does not interact
electromagnetically or weakly.

We propose a definite search for the existence of neutrino oscillations
with $\Delta m^2$ near $1~eV^2$ at the J-PARC Materials and Life Science
Experimental Facility (MLF).  With the 3 GeV Rapid Cycling Synchrotron 
(RCS) and spallation neutron target, an intense neutrino beam from muon
decay at rest ($\mu DAR$) is available. Neutrinos come predominantly from
$\mu^+$ decay : $\mu^+\rightarrow e^+ +\bar{\nu}_{\mu} +\nu_e$.
The oscillation to be searched for is $\bar{\nu}_{\mu} \rightarrow \bar{\
\nu}_e$  which is detected by the inverse $\beta$ decay interaction 
$\bar{\nu}_e+p \rightarrow e^++ n$, followed by a $\gamma$ from
neutron capture. 

The unique features of the proposed experiment, compared with the prior
experiment at LSND and experiments using conventional horn focused beams, are;

(1) The pulsed beam with about 600~ns spill width from J-PARC RCS
and muon long lifetime allow us to select neutrinos from $\mu DAR$. This
can be easily achieved by gating out for about 1$\mu$s from the start
of the proton beam spill. This eliminates neutrinos from pion and kaon
decay-in-flight.

(2) Due to nuclear absorption of $\pi^-$ and $\mu^-$, neutrinos from $\mu
^-$ decay are suppressed to about the $10^{-3}$ level. 
The resulting
neutrino beam is predominantly $\nu_{e}$ and $\bar{\nu}_{\mu}$ from $\mu^{+}$
with contamination from
other neutrino species at the level of $10^{-3}$.

(3) Neutrino cross sections are well known. $\bar{\nu}_e$ interacts by
inverse $\beta$ decay. The cross section is known to a few percent
accuracy.

(4) The neutrino energy can be calculated from positron
energy by adding $\sim$ 1.8~ MeV. 

(5) The $\bar{\nu}_{\mu}$ and $\nu_{e}$ fluxes have different and well defined
spectra. This allows us to separate $\bar{\nu}_e$ due to 
$\bar{\nu}_{\mu} \rightarrow \bar{\nu}_e$ oscillations from those due to 
$\mu^-$ decay contamination.\\

We propose to proceed with the oscillation search in steps. 
The region of $\Delta m^2$ to be examined is large, i.e. a positive 
signal can be found anywhere between $sub-eV^2$ to several tens of 
$eV^2$. 
We would like to start by examining the large $\Delta m^2$ region, which 
can be done with short baseline. At close distance to the MLF 
target, a high neutrino flux is available and allows us to use relatively small detector. 

If no definitive positive signal is found by this experiment, 
a future option exists to cover small $\Delta m^2$ region. 
This needs a relatively long baseline and requires a large detector to 
compensate for the reduced neutrino flux.

\newpage

\tableofcontents
\vspace*{0.5in}
\setcounter{figure}{0}
\setcounter{table}{0}
\indent
\clearpage
\pdfoutput=1
%%%%%%%%%%%%%%%%%%%%%%%%%%%%%%%%%%%%%%%%%%%%%%%%%%
\section{Physics Goals}
%%%%%%%%%%%%%%%%%%%%%%%%%%%%%%%%%%%%%%%%%%%%%%%%%%
~~

In 1998 the Super-Kamiokande collaboration announced the observation of 
neutrino oscillations with atmospheric neutrinos~\cite{cite:SK,cite:atm}. 
Oscillations have been observed also in accelerator~\cite{cite:acc}, solar~\cite{cite:solar} and reactor~\cite{cite:reactor} neutrinos since then. These 
observations are evidence of the fact that each neutrino flavor state is a super-position of mass states and that 
neutrino mass states have different masses, i.e. at least two non-zero values.
For three flavors of neutrinos, three mass states exist.  
The relation between flavor and mass states can be described by a
3x3 Unitary matrix~\cite{cite:PMNS} and can be parametarized by three mixing angles, one Dirac and two Majorana phases. 

From solar, atmospheric, accelerator and reactor experiments, the difference of square of masses have two distinct values, namely
\begin{eqnarray}
m_2^2-m_1^2=(7.54\pm0.21) \times 10^{-5}eV^2, \mid m_3^2-m_2^2\mid =(2.42\pm0.12) \times 10^{-3} eV^2 \nonumber
\end{eqnarray}
There is no way to construct a mass squared difference to be very different from those numbers within three neutrino scheme.

In the last fifteen years, several experiments have reported neutrino 
phenomena, which are consistent with the existence of neutrino oscillations
with $\Delta m^2 \geq eV^2$. If they are confirmed to be neutrino oscillations, 
it requires more than three mass states, and therefore, more than three kinds 
of neutrinos are required in nature. On the other hand, collider experiments 
have measured the number of neutrinos to be three by the measurement of the
invisible width of the Z-boson, $Z\rightarrow \nu+\bar{\nu}$\cite{cite:LEP}. 
The fourth state, if it exists,  is a new kind of lepton, which does not have 
electromagnetic or weak intearction, namely a sterile neutrino. 

The standard three neutrio scheme is a minimal phenomenological extension of 
the Standard Model of particle physics, requiring a lepton mixing matrix that 
is analogous to the quark sector and non-zero neutrino masses.
Despite its success,  the present description of neutrino states does not 
address fundamental questions such as how many fermions exist in nature, 
why the neutrino sector has small masses and large mixing angles compared to 
the quark sector etc. There needs to be a critical test regarding
whether the three generation neutrino scheme completes the discription of the
lepton sector. In the quark sector, the three generation scheme has been 
tested extensively in term of the test of the unitarity triangle of the 
Kobayashi-Maskawa matrix. 

One of the critical tests of the three neutrino scheme is to examine the 
existence of a 4th mass state beyond three generations of neutrinos.
Sterile neutrinos are naturally present in many theories beyond the standard model~\cite{cite:theory}, in particular in several manifestations of the seesaw mechanism. 
the number of fermion species is important not only in particle physics but also in cosmology. 
If the existence of the sterile neutrino is confirmed, the following implications show up;\\

(1) The 3x3 active neutrino mixing matrix (PMNS matrix) is not unitary, and 
neutrino mixing involves at least 6 mixing angles and 6 phases, 3 of which are 
Majorana phases.

(2) The electron neutrino is a superposition of four neutrino mass eigen-states.  The direct mass measurements and the rate of neutrino-less double $\beta$ decay  must take into account the effect of the 4th neutrino state. 

(3) Depending on the model of the early Universe, the existence of more than 3 
light neutrinos influence the expansion rate of the early Universe.

(4) Depending on the nature of sterile neutrinos, possiblities for the 
 existence of more species and thier masses and mixing are future subjects 
 of investigation.

In this Proposal, we describe a short baseline neutrino experiment at the J-PARC 3 GeV MLF facility to search for sterile neutrinos. 

\pdfoutput=1
%%%%%%%%%%%%%%%%%%%%%%%%%%%%%%%%%%%%%%%%%%%%%%%%%%
\section{Present status and the principle of measurement}
%%%%%%%%%%%%%%%%%%%%%%%%%%%%%%%%%%%%%%%%%%%%%%%%%
%%%%%%%%%%%%%%
\subsection{Experimental status}
%%%%%%%%%%%%%%
~~
Experimental evidence for sterile neutrinos would come from disappearance or appearance of active flavor with different $\Delta m^2$, which cannot be explained by $\Delta m^2_{12}$ or $\Delta m^2_{23}$. 
Table\ref{tab:LSNDetc} is the summary of observed anomalies and their significance.
\begin{table}[h]
\begin{center}
	\begin{tabular}{|l|c|c|c|}
	\hline
	Experiment     & neutrino source      & Signal                                & $\sigma$ \\ \hline \hline
	LSND               & $\pi$ decay at rest & $\bar{\nu}_{\mu}\rightarrow\bar{\nu}_e$  & $3.8\sigma$     \\ \hline
	MiniBooNE       & $\pi$ decay in flight & $\nu_{\mu}\rightarrow\nu_e$         & $3.4\sigma$         \\ \hline
	MiniBooNE       &  $\pi$ decay in flight & $\bar{\nu}_{\mu}\rightarrow\bar{\nu}_e$ & $2.8\sigma$ \\ \hline
	Gallium/SAGE  & e capture         & $\nu_e\rightarrow\nu_x$   & $2.7\sigma$  \\ \hline
	Reactor           &  $\beta$ decay           & $\bar{\nu}_e\rightarrow\bar{\nu}_x$ & $3.0\sigma$   \\ 
	\hline
	\end{tabular}
	\caption{Possible large $\Delta m^2$ anomalies}
        \label{tab:LSNDetc}
\end{center}
\end{table}

The first indication was reported by the LSND experiment. LSND reported an excess of $87.9\pm22.4\pm6.0$ $\bar{\nu_e}$ events ($3.8\sigma$) in 1998\cite{LSND}. MiniBooNE results are presented recently. The MiniBooNE experiment observed excesses of $\nu_e, \bar{\nu}_e$ candidates in the 200-1250 MeV energy range in neutrino mode ($3.4\sigma$) and in anti-neutrino mode ($2.8\sigma$). The combined excess is $240.3\pm34.5\pm52.6$ events, which corresponds to $3.8\sigma$~\cite{cite:MiniBooNE}. It is not clear whether the excesses are due to oscillations. If they are due to oscillations, both LSND and MiniBooNE indicate a flavor conversion of $\bar{\nu}_{\mu}$ to $\bar{\nu}_e$ at a probability of about 0.003 with a
 $\Delta m^2$ of $\sim 1 eV^2$.

The second indication is a deficiency observed in the calibrations of low energy radio-chemical solar neutrino experiments. The results indicated a deficiency in neutrino event rates.  Mono-energetic neutrino sources ($^{51}Cr$ and $^{37}Ar$) were used in these experiments. Their results were presented in terms of the ratio of the observed and the predicted rate. The predictions are based on theoretical calculations of neutrino cross sections by Bahcall and by Haxton. The quoted numbers are $R_{obs}/R_{pred}=0.86 \pm 0.05 (\sigma_{Bahcall}), 0.76\pm 0.085 (\sigma_{Haxton})$~\cite{GaAnomaly}.

The so-called reactor anomaly indicates a 6\% deficit of detected $\bar{\nu}_e$ from nuclear reactors at baselines less than 100 m. The ratio of observed and expected rate is $0.927\pm 0.023$. This is entirely based on the re-analysis of existing data. The deficit is caused by three independent effects which all tend to increase the expected neutrino event rate. There have been two re-evaluations of reactor anti-neutrino fluxes and both indicate an increase of flux by about
3\%. The neutron lifetime decreased from 887-899s to 885.7s and thus the inverse $\beta$-decay cross section increased by a corresponding amount. The contribution from long-lived isotopes to the neutrino spectrum was previously neglected and enhances the neutrino flux at low energies~\cite{cite:ReactorAnomaly}.

All these hints have a statistical significance around $3 - 3.8\sigma$ and may be caused by one or more sterile neutrinos with a mass of roughly 1 eV.  If they are due to neutrino oscillation with new mass state $m_4(\sim eV)$, the disappearance and the appearance of active neutrinos are related by 
($m_4\gg m_{1,2,3}$ and $U_{s4} \sim 1 \gg U_{e \mu \tau,4}$. )
\begin{eqnarray}
P(\nu_e, \nu_{\mu}\rightarrow\nu_s)&=&-4\sum_{i > j}Re(U_{si}U^*_{\mu,e i} U^*_{sj}U_{\mu,e j})\sin^2\Delta_{ij}                \nonumber \\
 &-& 2\sum_{i > j}Im (U_{si}U^*_{\mu,e i}U^*_{sj}U_{\mu,e j})\sin2\Delta_{ij}
\nonumber \\
P(\nu_{\mu}\rightarrow\nu_e)&=&-4\sum_{i > j}Re(U_{ei}U^*_{\mu i} U^*_{ej}U_{\mu j})\sin^2\Delta_{ij}  \nonumber\\
 &-& 2\sum_{i > j}Im (U_{ei}U^*_{\mu i}U^*_{ej}U_{\mu j})\sin2\Delta_{ij} \nonumber \\
\Delta _{ij}&=&(m^2_j-m^2_i)L/4E_{\nu}  \nonumber 
\end{eqnarray}
For a short baseline experiments $(L(m)/E(MeV)\sim 1)$ and if only one sterile neutrino involved in mixing,\\
\begin{eqnarray}
P(\nu_{e,\mu}\rightarrow\nu_s)&\sim&-4\sum_{j}Re(U_{s4}U^*_{\mu,e 4} U^*_{sj}U_{\mu,e j})\sin^2(m^2_4L/4E_{\nu})  \nonumber\\
 &-& 2\sum_{j}Im (U_{s4}U^*_{\mu,e 4}U^*_{sj}U_{\mu,e j})\sin2(m^2_4L/4E_{\nu})
 \nonumber\\
 &=&4\mid{U_{s4}}\mid^2\mid{U_{\mu,e4}}\mid^2\sin^2(m^2_4L/4E_{\nu})   
\nonumber\\
P(\nu_{\mu}\rightarrow\nu_e)&\sim&-4\sum_{i}Re(U_{e4}U^*_{\mu 4} U^*_{ei}U_{\mu i})\sin^2(m^2_4L/4E_{\nu})  \nonumber\\
 &-& 2\sum_{j}Im (U_{e4}U^*_{\mu 4}U^*_{ej}U_{\mu j})\sin2(m^2_4L/4E_{\nu})
 \nonumber\\
 &=&4\mid{U_{e4}}\mid^2\mid{U_{\mu4}}\mid^2 \sin^2(m^2_4L/4E_{\nu})   
\nonumber\\
\end{eqnarray}
 Thus
$P(\nu_{\mu}\rightarrow\nu_s) \cdot P(\nu_e\rightarrow\nu_s)\sim P(\nu_{\mu}\rightarrow\nu_e)$.

In order for the LSND and MiniBooNE data to be  consistent with the sterile neutrino hypothesis,  $\nu_{\mu}$ disappearance at $\Delta m^2\sim eV^2$ should exist in addition to the possible $\nu_e$ deficiencies, which has been observed in $\beta$ source and reactor measurements. So far only several \% level upper limit exists for $\nu_{\mu}$. Thus some tensions exist in this respect~\cite{cite:tension}.\\

\begin{figure}[h]
 \centering
 \includegraphics[width=1.1 \textwidth]{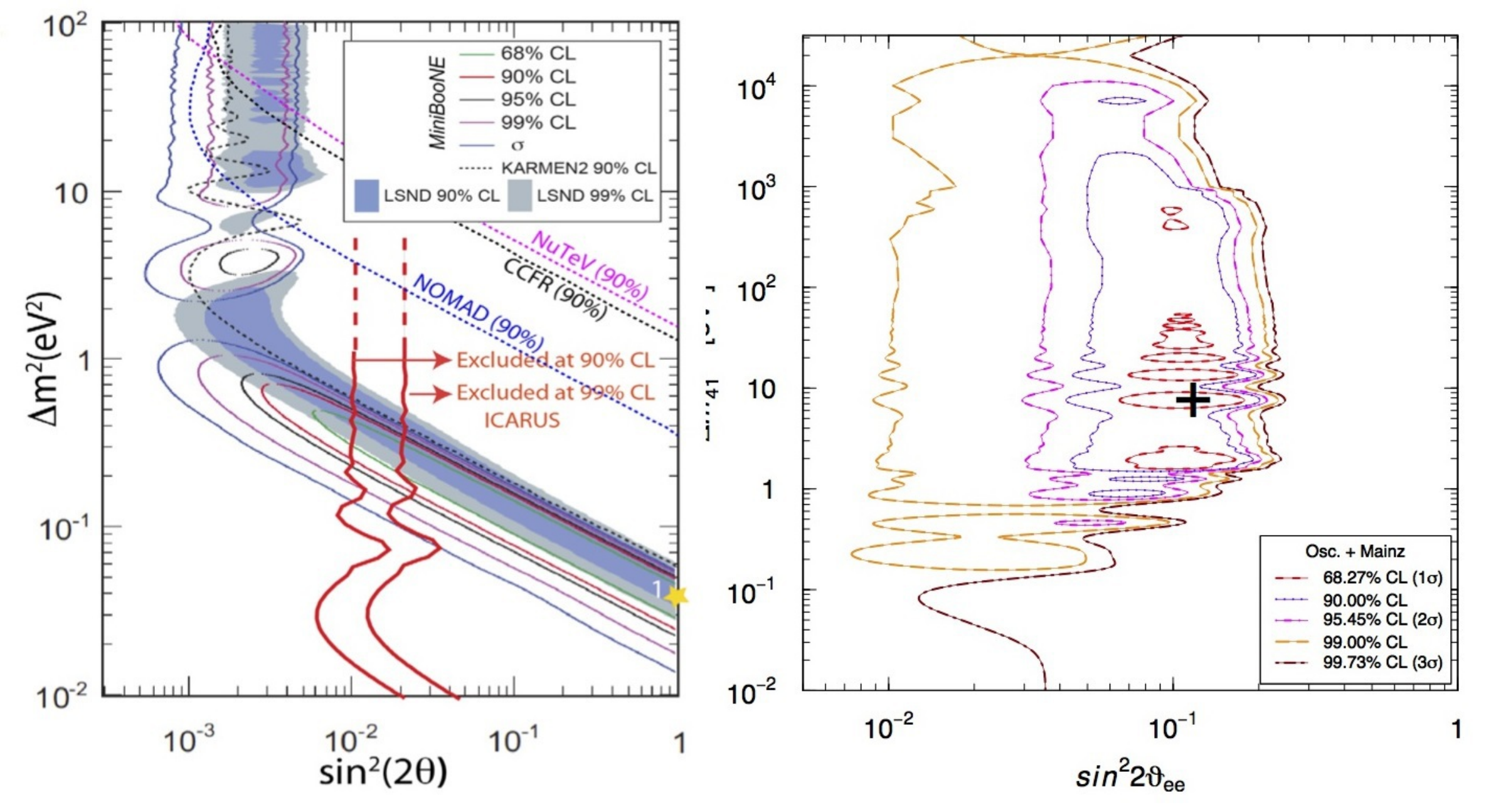}
 \caption{\setlength{\baselineskip}{4mm} 
Left figure : Allowed region for $\bar{\nu_{\mu}}\rightarrow\bar{\nu_e}$ appearance channel as a result of combining LSND, MiniBooNE and ICARUS~\cite{cite:contmue}.
Right figure : Allowed region for disappearance channel with Reactor and $\beta$ source anomalies, taken into account KATRIN and neutrino-less double $\beta$ decay limits~\cite{cite:contee}.
}
 \label{fig:LSNDALL}
 \end{figure}

The indicated allowed regions are shown in Figure \ref{fig:LSNDALL} for the appearance channel (left figure) ($\bar{\nu_{\mu}}\rightarrow\bar{\nu_e}$)  and for the disappearance channel (right figure) ($\nu_e\rightarrow\nu_s$).

%%%%%%%%%%%%
\subsection{The principle of measurement\\
-Advantages of pion/muon decay at rest neutrino source}
%%%%%%%%%%%%
~~

The measurement will be based on the following features of the $\mu$ Decay-At-Rest(DAR; $\pi^{+} \rightarrow \nu_{\mu} + \mu^{+}$ ; $\mu^{+} \rightarrow e^{+} + \nu_{e} + \bar{\nu}_{\mu}$ ) beam at J-PARC MLF;

\hspace*{0.3mm}

(1) Low duty factor of the pulsed proton beam.

(2) No decay in flight components by timing cut.

(3) Selection of the neutrino flavor by detecting Inverse Beta Decay (IBD; $\bar{\nu}_{e} + p \rightarrow e^{+} + n$) signal in liquid scintillator.

(4) Well known different spectrum shapes for $\bar{\nu}_{\mu}$ and $\bar{\nu}_e$ from $\mu^-$ decay contamination.

(5) The IBD cross section is well measured in neutron $\beta$ decay~\cite{cite:IBD}.\\

(6) Ease of $E_{\nu}$ reconstruction. :  $E_{\nu} \sim E_e$ + 1.8 MeV \footnote{\setlength{\baselineskip}{4mm} 
Note that recoil neutrons carry kinetic energy up to 5 MeV where $E_{\nu} = $50
MeV; however, the incident neutrino energy can be determined by the energy and
angle of the positron.}.

(7) The flux of $\bar{\nu}_{\mu}$ can be monitored by the rate of $\nu_e$ 
interactions.

%%%%%%%%%%%%%%%%%
\subsubsection{Neutrino spectrum from decay at rest of $\mu^+$}
%%%%%%%%%%%%%%%%
~~
In this proposal, we concentrate on the neutrino beam from muon decay at rest ($\mu DAR$). To use neutrinos from pion and kaon decay at rest, 
extensive studies and experience in dealing with the backgrounds during the proton beam bunch will be needed.

\begin{figure}
 \centering
 \includegraphics[width=1.\textwidth]{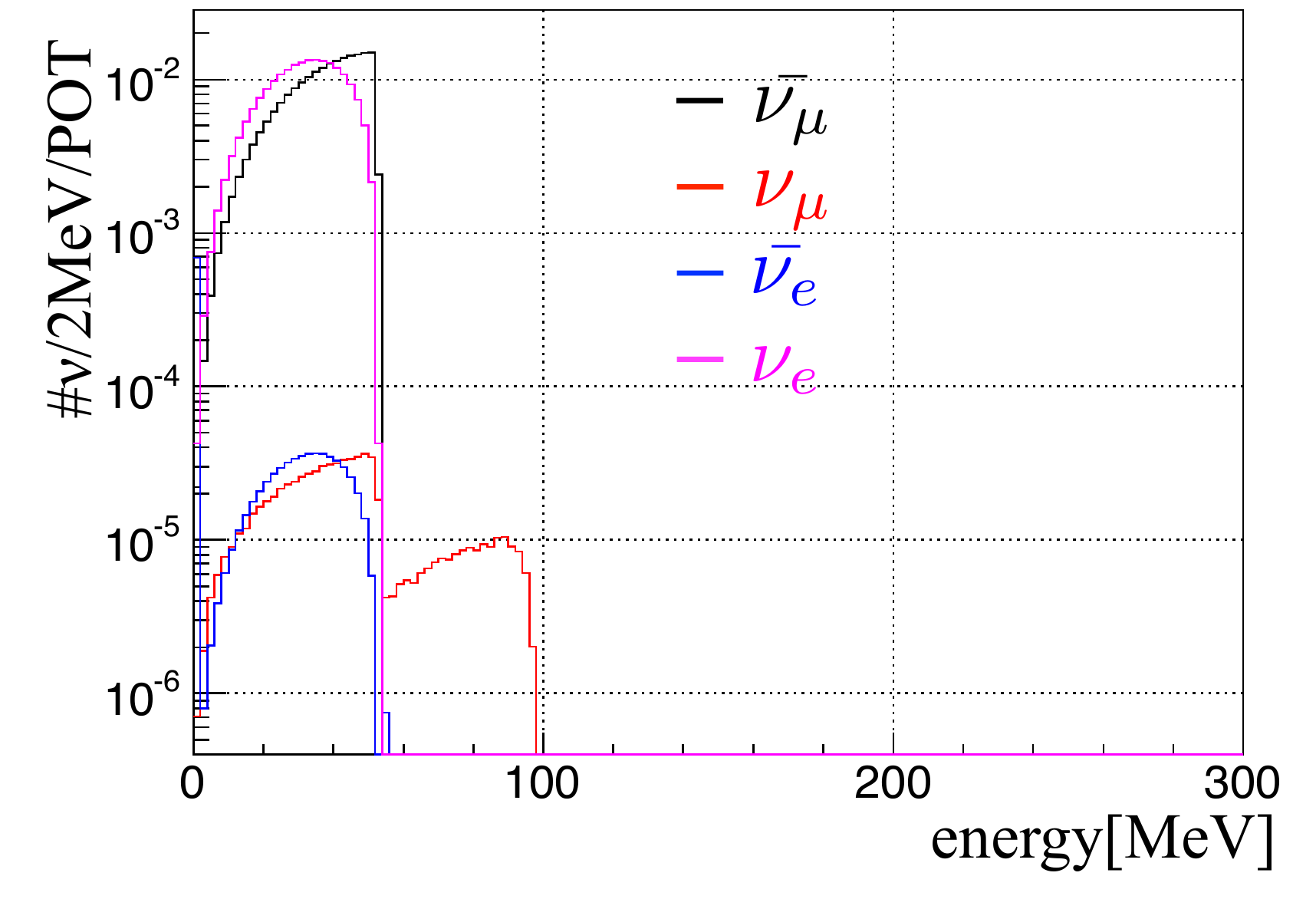}
 \caption{\setlength{\baselineskip}{4mm}
Estimated neutrino flux after 1 $\mu$s from the start of proton beam. The $\mu^{+} DAR$ components are selected and main background come from $\mu^-$ decays. 
}
 \label{fig:muDARflux}
 \end{figure}

The $\mu DAR$ component can be selected by gating out the first 1 $\mu$s from the start of the  proton beam. The resulting neutrino fluxes for each type of neutrino species are shown in Figure ~\ref{fig:muDARflux}.
Note that the resulting $\bar{\nu}_{\mu}$ and $\nu_e$ fluxes
have different spectrum with endpoint energy of 52.8 MeV. A possible survived $\mu^-$ decay will be at the level of $10^{-3}$ and produce $\nu_{\mu}$ and $\bar{\nu_e}$ with same spectrum as those of $\bar{\nu}_{\mu}$ and $\nu_e$, respectively.

%%%%%%%%%%%%%%%%%%%%%%%
\subsubsection{Interactions of neutrino from $\mu^+$ decay at rest}
%%%%%%%%%%%%%%%%%%%%%%%
There are four kinds of neutrino interactions in scintillator detector ($CH_2$).\\

(1)Inverse beta decay (IBD)\\
The signal we are looking for is $\bar{\nu}_{\mu} \rightarrow \bar{\nu}_e$ and $\bar{\nu_e}+p \rightarrow e^++n$ with neutron capture $\gamma$ as a delayed coincidence.  \\
\begin{eqnarray}
\sigma_{IBD}&=&\frac{G_F^2E_{\nu}^2}{\pi} (g_V^2+3g_A^2) \sqrt{1-\frac{2Q}{E_{\nu}}+ \frac{Q^2-m_e^2}{E_{\nu}^2}} \theta(E_{\nu}-Q)  \nonumber \\
&\sim& 9.3\times10^{-48}E_{\nu}^2(MeV) m^2 \nonumber
\end{eqnarray}

(2)Charged current interaction of $\nu_e$\\
$\nu_e + ^{12}C \rightarrow e+N^{*}$. For $N_{gs}$ final state, $N_{gs}-\beta$ 
decay gives a delayed coincidence as a clear signature with 16 MeV endpoint energy.\\

(3) Neutral current interaction with nucleus\\
All active neutrinos interact by neutral current interaction with nucleus; 
$\nu_{e,\mu}+X \rightarrow \nu_{e,\mu}+X')$.  A dominant process in 
scintillator ($CH_2$) detector to produce an electro magnetic particle is\\
$\nu_{e,\mu}+^{12}C\rightarrow \nu_{e,\mu}+C(15.11)$ producing 15 MeV $\gamma$.\\

(4) Atomic electron target reaction\\
$\nu_{e,\mu}+e\rightarrow \nu_{e,\mu} + e$. These are negligible contribution.\\

\begin{figure}
 \centering
 \includegraphics[width=1.\textwidth]{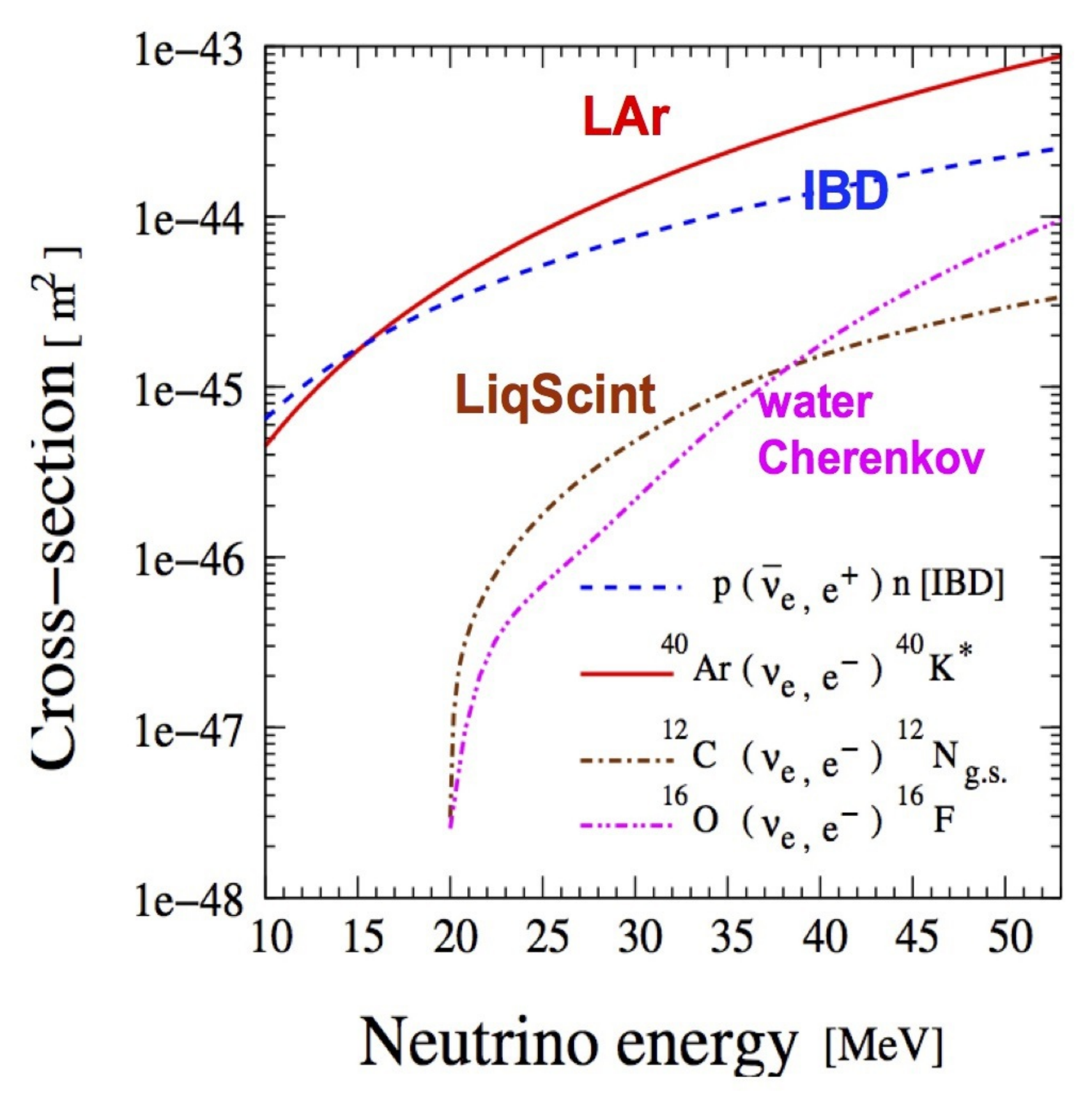}
 \caption{\setlength{\baselineskip}{4mm}
Cross sections for IBD and $^{12}$C($\nu_e$,$e^-$)$^{12}$N$_{g.s.}$ are shown as a function of $E_{\nu}$~\cite{cite:XSECgraph}. 
% Croos sections for $\nu_e$ and $\bar{\nu_e}$.
}
 \label{fig:munuXSEC}
 \end{figure}

Figure \ref{fig:munuXSEC} shows the cross sections as functions of $E_{\nu}$ for each interaction.
 
%%%%%%%%%%%%%%%%%%%%%%%
\subsubsection{Signatures of the oscillation}
%%%%%%%%%%%%%%%%%%%
~~

A sensitive search for $\bar{\nu}_e$ appearance ($\bar{\nu}_{\mu} \rightarrow \bar{\nu}_e$ from $\mu^+ DAR$) can be performed by searching for the two-fold signature of $\bar{\nu}_e + p\rightarrow e^{+} + n$  scattering with a positron with 52.8 MeV endpoint energy followed by gammas due to Gd neutron capture.

The main background coming from $\mu^-$ decays as shown in 
Figure~\ref{fig:muDARflux}, is highly suppressed by $\pi^-$  and $\mu^-$ capture in heavy metals like Hg. However, $\mu^-$s, which stopped in a light metal such as Be, does usually decay before absorption. This background can be estimated
from the $E_{\nu}$ distribution, which is well defined and different from oscillated events. 

\begin{figure}
 \centering
 \includegraphics[width=1.0\textwidth]{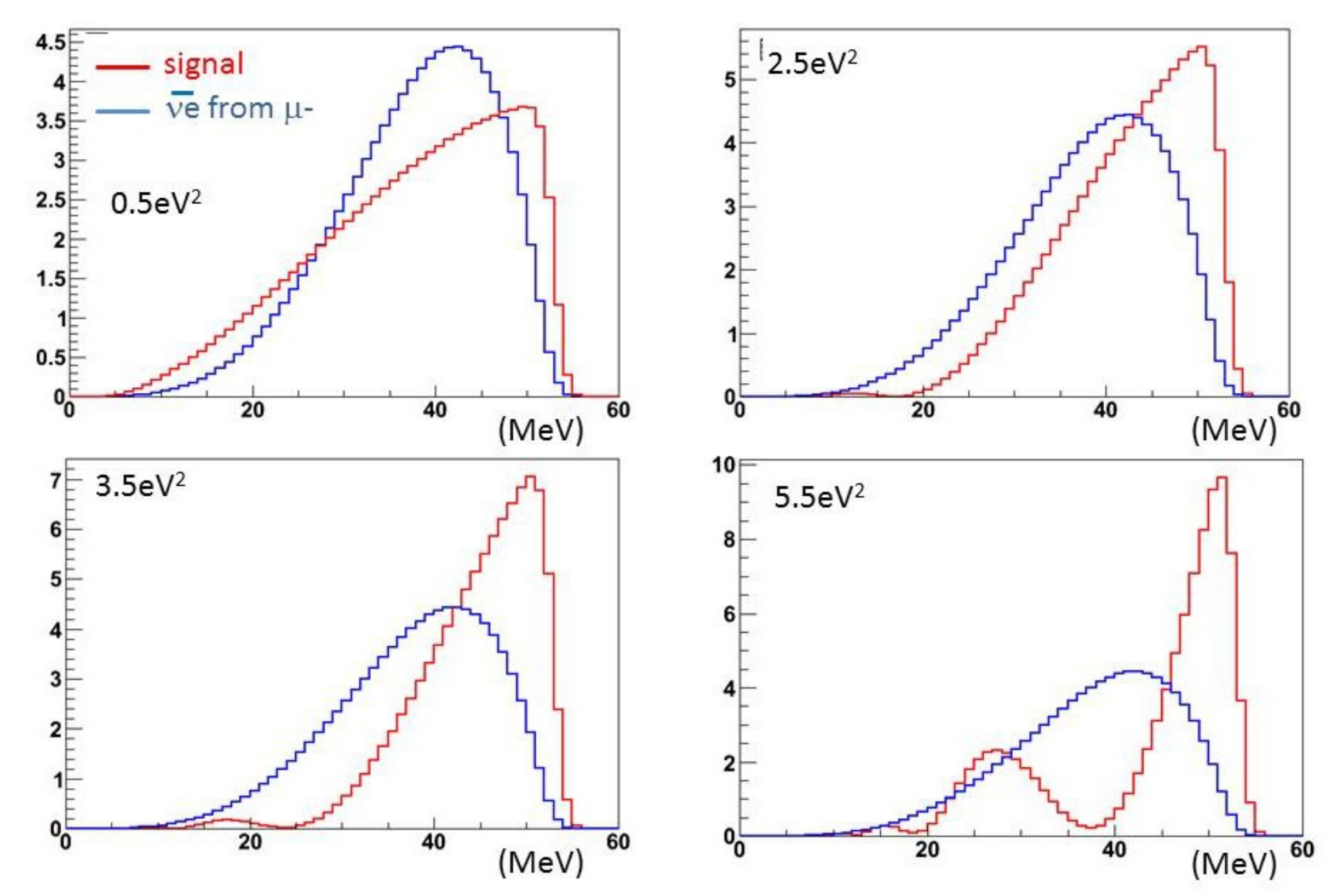}
 \caption{\setlength{\baselineskip}{4mm}
Examples of oscillation signals at typical $\Delta m^2$ for a baseline of 17 m. The red graphs are shapes of $\bar{\nu}_e$ appearance signal and the blue graphs are shapes of signal from $\mu^-$ decays.
Energy is smeared by 15$\%$/$\sqrt{E}$, which is the assumption of 
the energy resolution of the detector. All plots are normalized by 
area. 
 }
 \label{fig:Examposc}
 \end{figure}

Since the oscillation probability is given by
\begin{eqnarray}
P=\sin^2 2\theta \sin^2(\frac{1.27 \Delta m^2(eV^{2}) L(m)}{E_{\nu} (MeV)} )\nonumber
\end{eqnarray}
, there are two distinct signatures of oscillation signal.
One is the energy spectrum of the oscillated signal, which is a convolution of
the energy spectrum of the original neutrino (in this case, $\bar{\nu}_{\mu}$ ) and the oscillation probability.  The other signature is the distribution of
events as a function of distance from the source. The background $\bar{\nu}_e$ from $\mu ^{-}$ decay has a different spectrum from that of $\bar{\nu}_{\mu}$ oscillations.
Figure \ref{fig:Examposc} shows $E_{\bar{\nu}}$ distributions of oscillation signals at some typical $\Delta m^2$s for a baseline of 17 m.

%%%%%%%%%%%%%%%%%%
\subsubsection{Signal identification}
%%%%%%%%%%%%%%%%%
\label{sec_signatures}
~~
The signal for $\bar{\nu_e}$ appearance from 
$\bar{\nu}_{\mu} \rightarrow \bar{\nu}_e$ 
is a primary positron signal followed by delayed signal. 

The primary signal is $\bar{\nu_e} + p \rightarrow e+n$ (Inverse Beta Decay (IBD)) 
and the delayed signal is the neutron capture gamma.
For the normalization of $\mu^+$ decay, $\nu_e + C \rightarrow e+N_{gs}$ events will be measured. The primary signal is an electron and the delayed signal is a positron from $N_{gs}$ $\beta$ decay.

The time gate for the primary signal should be from $1 \mu$s to $10 \mu$s, 
corresponding to the muon lifetime and avoiding pion decay from both 
decay at rest and decay in flight. 
Table~\ref{tab:signal} is a summary of primary and delayed signal.
\begin{table}[h]
\begin{center}
	\begin{tabular}{|l|c|c|c|c|}
	\hline
	~          & primary   &   primary  & delayed & delayed \\ 
       ~          &  timing         &  energy       & timing      &  energy \\ \hline \hline
$\bar{\nu}_{\mu} \rightarrow \bar{\nu}_e$ & 1-10 $\mu$s & 0-53 MeV & 10-100$\mu$s & 8 MeV \\ \hline
$\nu_e C\rightarrow e N_{gs} $, $N_{gs}\rightarrow C e^+ \nu_e$&  1-10 $\mu$s & 0-37 MeV & 100$\mu$s-10 ms & 0-16 MeV \\ \hline \hline
	\end{tabular}
	\caption{The expected signal timing relative to proton beam bunch and signal energy}
	\label{tab:signal}
\end{center}
\end{table}

%%%%%%%%%%%%%%%%%%%%%%%%%%%%%%%%%%%%%%%%%%%%%%%%%%%%%%%%%%%%
\subsection{Strategy of this experiment}
%%%%%%%%%%%%%%%%%%%%%%%%%%%%%%%%%%%%%%%%%%%%%%%%%%%%%%%%%%%%
\indent

We select a strategy to put a detector with 50 tons of liquid
scintillator using a 
short baseline ($\sim$ 17 m; the third floor of the MLF, maintenance area of the target. See section~\ref{sec_detectorloc} for more details.) for the
following reasons;

\begin{figure}
\centering
\includegraphics[width=0.9\textwidth]{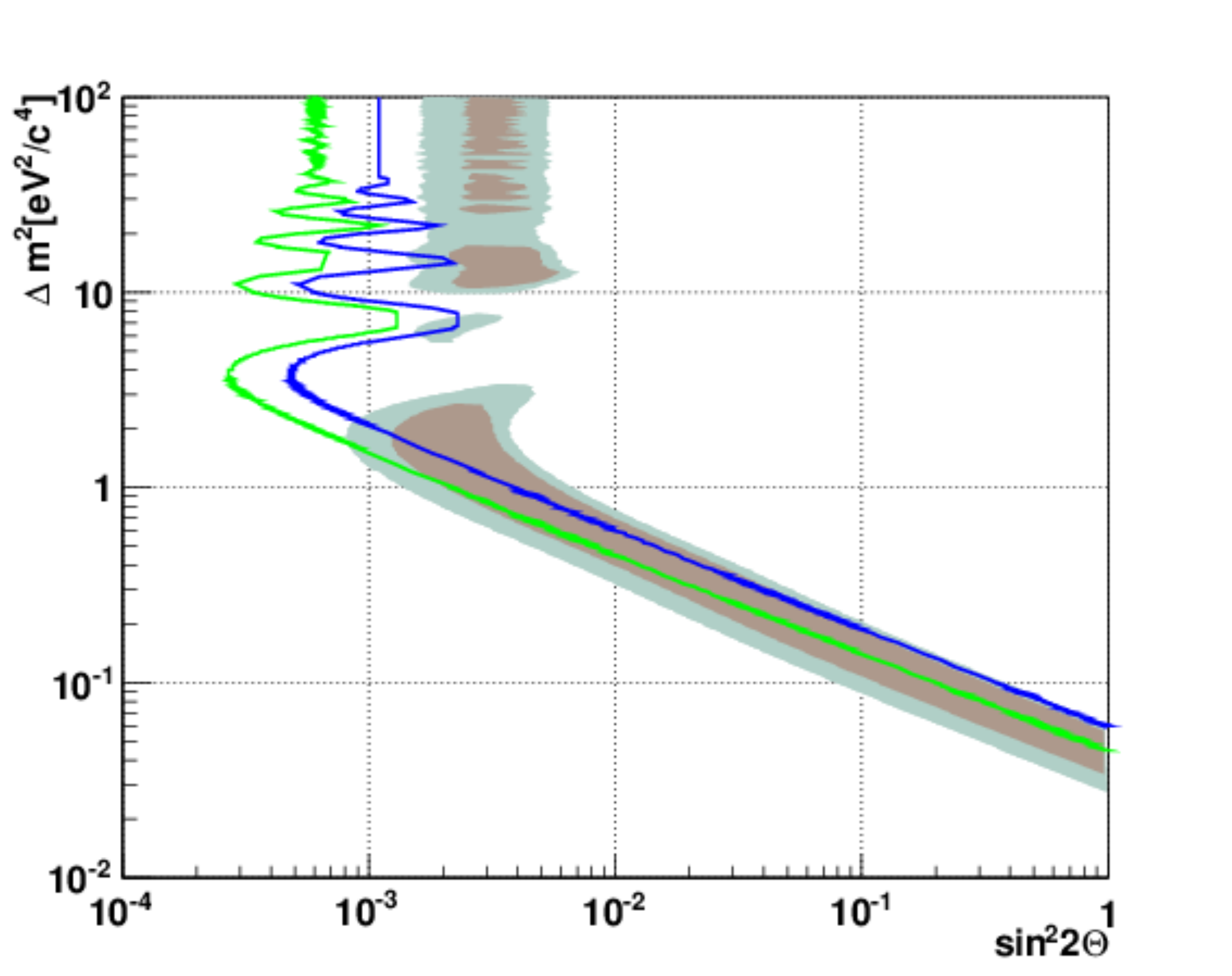}
\caption{\setlength{\baselineskip}{4mm}
The sensitivity of the MLF experiment assuming 4 years operation
(4000 hours / year) assuming the $\sim$50$\%$ 
detection efficiency and a 17 m baseline. The blue line shows the 5 $\sigma$ C.L., while green one corresponds to 3 $\sigma$.
}
\label{fig:sensitivity_2sec}
\end{figure}

\begin{enumerate}
\item The distortion of the neutrino energy spectrum is clear in 
     the $\Delta m^{2} >$ 2.0 eV$^{2}$ region if neutrino oscillations exist
     as shown in Figure~\ref{fig:Examposc}.
\item This short baseline provides a large number of 
     events due to 1/L$^{2}$ law from the target, i.e., the neutrino 
     flux is reduced as a function of 1/L$^{2}$.     
\item As a result, the sensitivity to search for sterile neutrinos
     above a few eV$^{2}$ region is comparable to or better 
     (see Fig.~\ref{fig:sensitivity_2sec})
     than the experiment using a larger detector at a longer distance.
\end{enumerate} 

If no definitive positive signal is found by this configuration,
a future option exists to cover the small $\Delta m^{2}$ region.
This needs a relatively long baseline and requires a large detector to 
compensate for reduced neutrino flux (see appendix).

\pdfoutput=1
%%%%%%%%%%%%%%%%%%%%%%%%%%%%%%%%%%%%%%%%%%%%%%%%%
\section{The J-PARC MLF as a DAR Neutrino Source}
%%%%%%%%%%%%%%%%%%%%%%%%%%%%%%%%%%%%%%%%%%%%%%%%%
\label{sec_beam}
~
Neutrino beam from stopped pion decay from the 
J-PARC Rapid Cycling Synchrotron (RCS) is one of the best suited facilities 
for searches of neutrino oscillations in the mass range $\Delta m^2\sim eV^2$, 
because of the following reasons: \\
(1) available beam power\\
(2) mercury target absorbs $\mu^-$ and suppresses free decay,\\
(3) short duty factor of the pulsed beam enables us to eliminate decay-in-flight components and to separate $\mu DAR$ from other background sources, and the resulting  $\nu_e, \bar{\nu_e}$ have well defined spectrum and well defined cross section.

%%%%%%%%%%%%%%%%%%%%%%%%%%%
\subsection{The RCS beam and the target}
%%%%%%%%%%%%%%%%%%%%%%%%%%%
~~
The proton intensity is expected to reach 0.33 mA (1 MW) after major upgrades 
of the accelerator. The protons are produced with a repetition rate of 25 Hz, where each spill contains two 80 ns wide pulses of protons spaced 540 ns apart. 1 MW beam provides 3$\times$10$^{22}$ protons-on-target (POT) during 4000 hours / year operation (i.e.;
3.6$\times$10$^{8}$ spills are provided during one year).
The short pulsed beam (540~ns, 25~Hz) provides the ability to distinguish between neutrinos from pion decay and those from muon decay.

\begin{figure}
 \centering
 \includegraphics[bb=0 0 885 661, width=1.1 \textwidth]{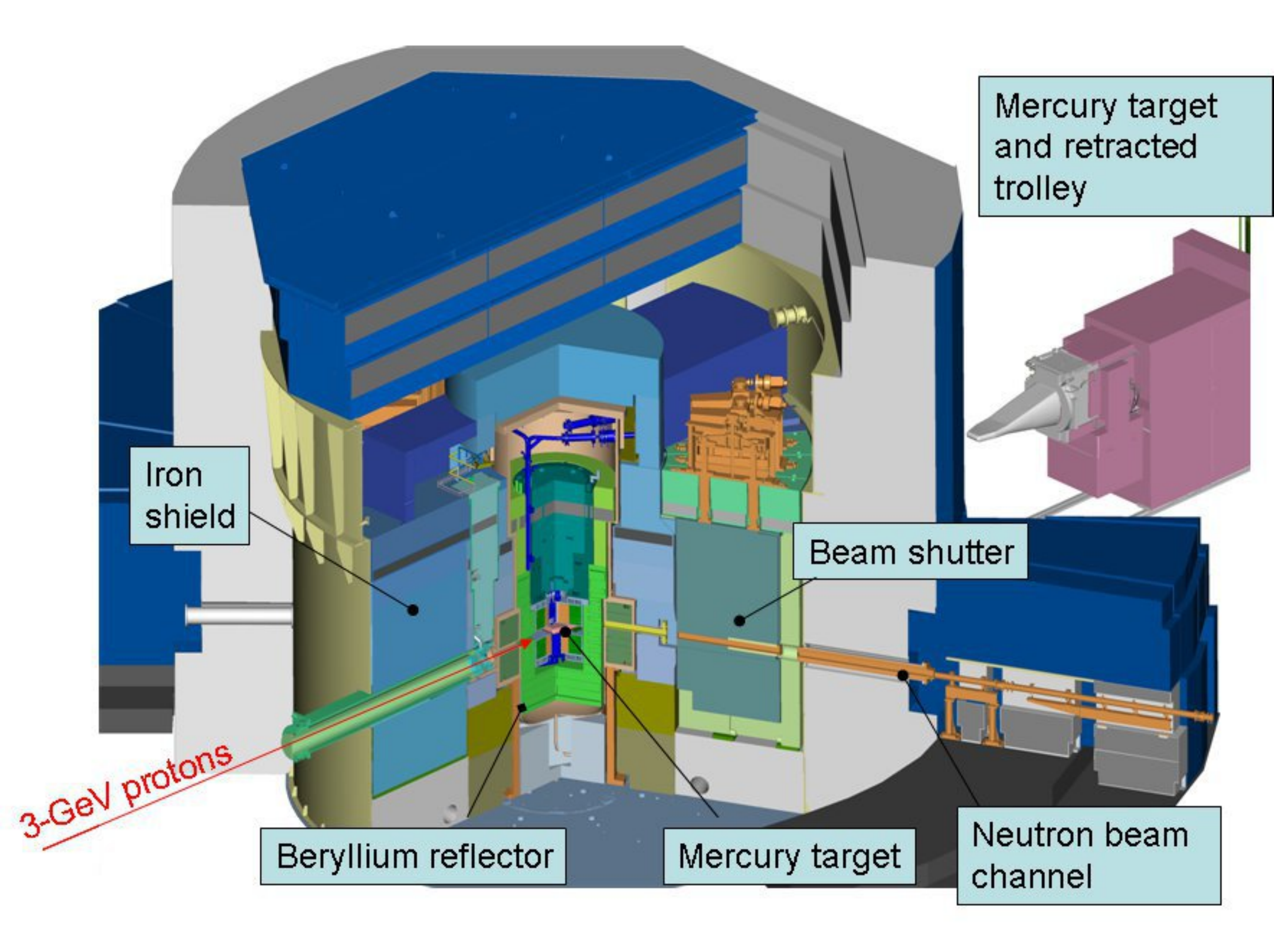}
\caption{\setlength{\baselineskip}{4mm}
A schematic view of the MLF facility in J-PARC.
}
 \label{fig:MLF}
\end{figure}

Figure \ref{fig:MLF} shows a plan view of Materials and Life Science and Experimental Facility (MLF) in J-PARC. After penetrating a muon production target made of carbon graphite of 2 cm thickness, protons are introduced to a mercury target. A schematic drawing of the J-PARC spallation neutrino source is shown in Figure \ref{fig:MLF}. 3 GeV protons interact in the mercury spallation target, producing pions and kaons that decay into 
$\nu_e$ and $\nu_{\mu}$ and their anti-neutrinos after heavy shielding. Surrounding the target are cooling pipes, beryllium reflectors, and steel shielding.

\begin{figure}
 \centering
 \includegraphics[bb=0 0 590 296, width=1.1 \textwidth]{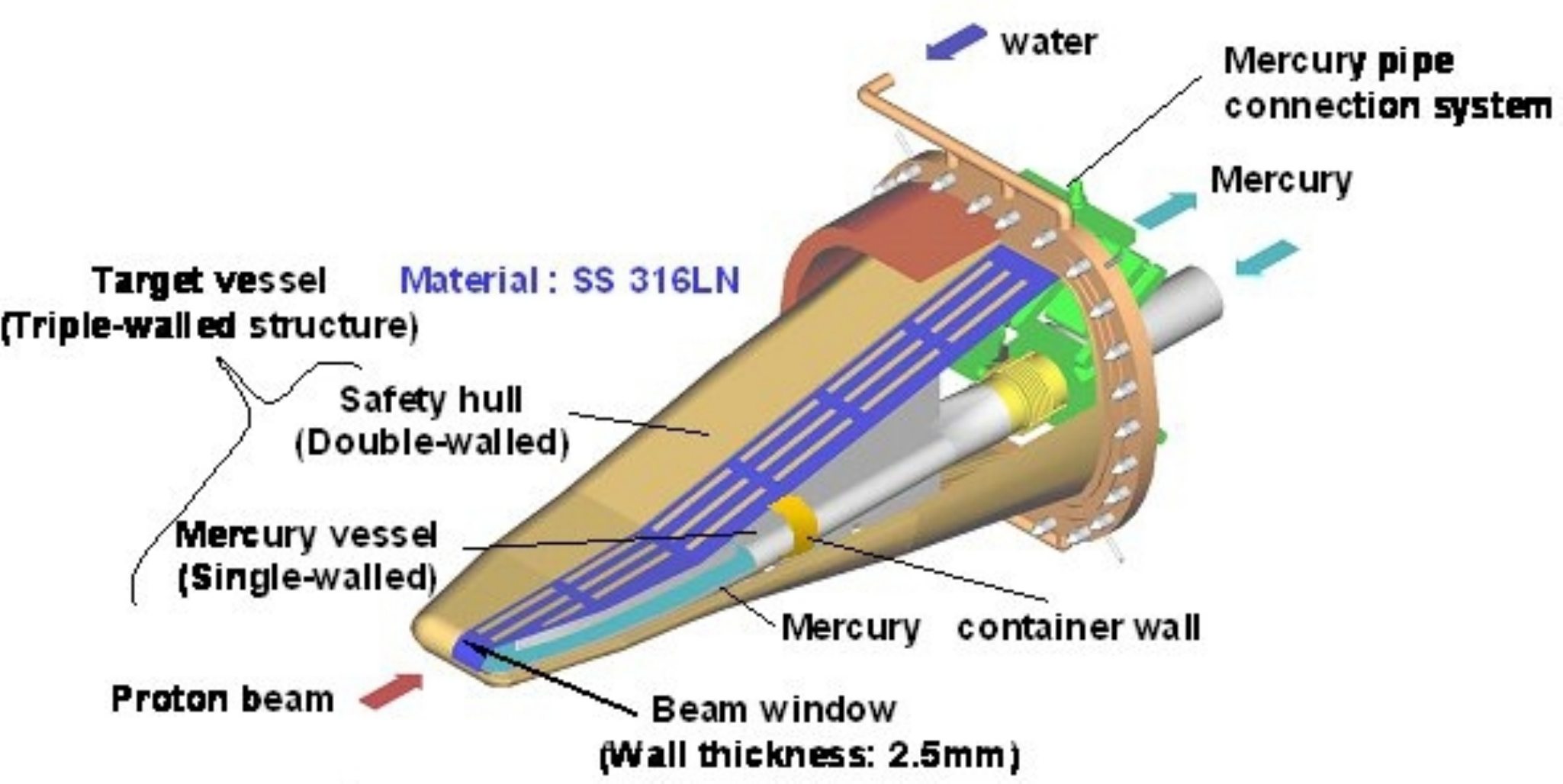}
\caption{\setlength{\baselineskip}{4mm}
A schematic drawing of the mercury target in the J-PARC.
}
 \label{fig:MLFTarget}
\end{figure}

A beam of protons enters from the left and strikes the target. The beam has a wide spot size such as 3.3 cm by 1.3 cm in root mean square (rms) for reduction of the local heat load in the target. The target, shown in Figure \ref{fig:MLFTarget}, has dimensions of 54 cm in width by 19 cm in height by 210 cm in length. Mercury is contained within a multiple wall structure made of  stainless steel. To remove heat, the mercury of the target is constantly circulated at a rate of 154 kg/sec. The cryogenic liquid hydrogen moderators are located at the top and bottom of the target. Target and moderators are surrounded by a beryllium reflector and an iron shielding  which extends at least to a radius of 5 m around the target. There are 23 neutron channels looking at the moderators, rather than at the target. Shutters are provided on each channel.

%%%%%%%%%%%%%%%%%%%%%%%%%%%
\subsection{Neutrino Beam}
\begin{figure}
 \centering
 \includegraphics[width=0.9 \textwidth]{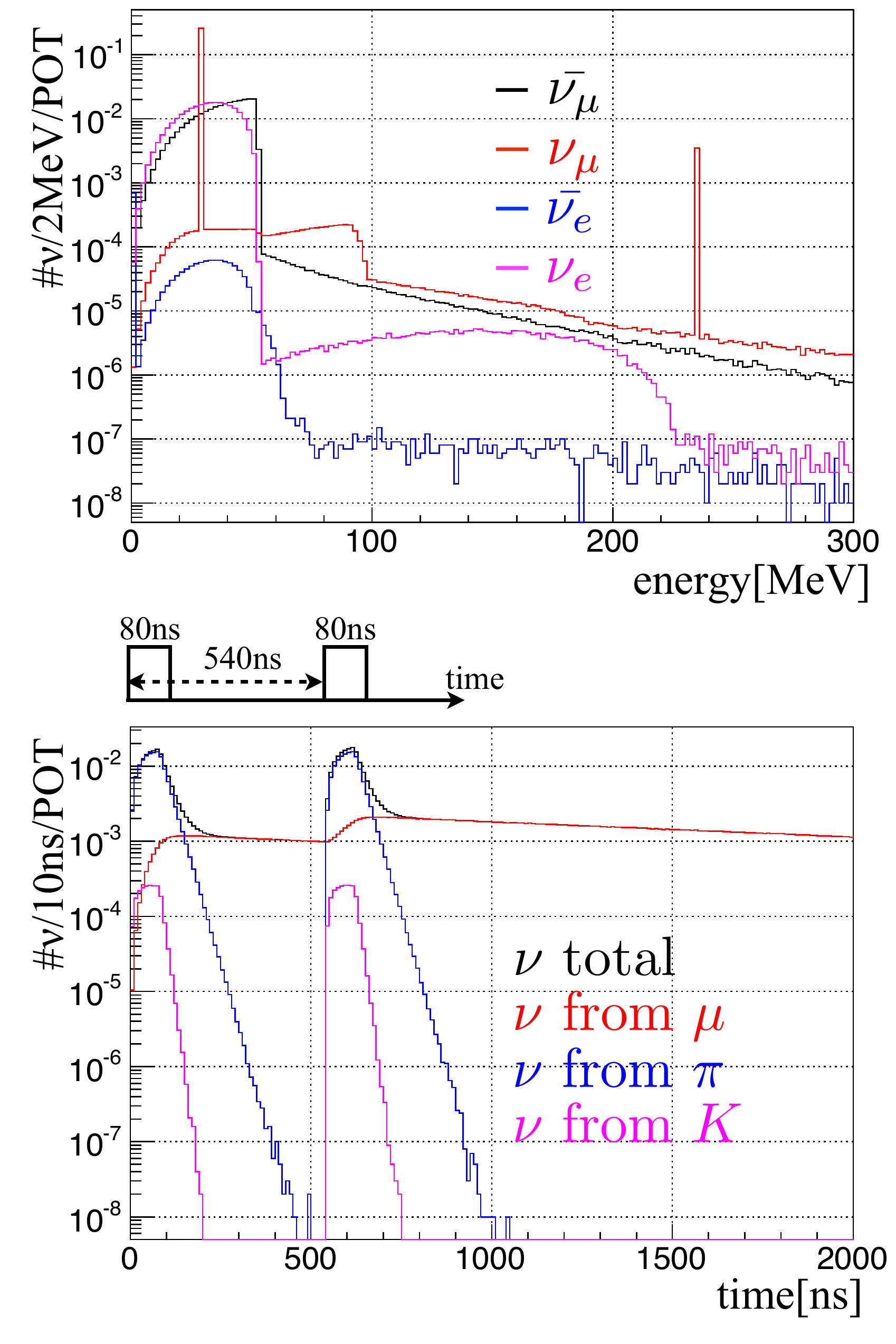}
\caption{\setlength{\baselineskip}{4mm}
The energy spectra of neutrinos from pion and kaon decays which are based on Geant4~\cite{GEANT4}
calculations (top). This tends to be at the low end of neutrino yeild 
estimates of various particle production models.
Time distribution of neutrinos 
from pion, muon and kaon decays is shown in the bottom plot. Neutrino beams from muon decay at 
rest only survive after 1 $\mu$s from the start of proton beam.
}
 \label{fig:NeutrinoAll}
\end{figure}

%%%%%%%%%%%%%%%%%%%%%%%%%%%
~~
There are two time structures of the neutrino beam.
\\
One  is 'On-bunch' (neutrinos produced during proton bunch and pion or kaon 
lifetime).
\begin{itemize}
\item $\pi^+ \rightarrow \mu^+~\nu_{\mu}$ decay at rest with monochromatic neutrino energy of 30 MeV
\item $\mu^- +A \rightarrow \nu_{\mu} +A$ with end point at 105 MeV
\item $K^+ \rightarrow \mu^+~\nu_{\mu}$ decay at rest with monochromatic energy of 236 MeV
\item $K^+ \rightarrow \mu^+~\pi^0 ~\nu_{\mu}$ decay at rest with a end point energy of 215 MeV
\item $K^+ \rightarrow e^+~\pi^0 ~\nu_e$ decay at rest with end point energy at 228 MeV
\item Small components from $\pi$ and K decay in flight
\end{itemize}
The other is 'Off-bunch' (during muon lifetime) component, which is produced by muon decay at rest.
\begin{itemize}
\item $\mu^+ \rightarrow e^+~\nu_e~\bar{\nu_{\mu}}$
\item If $\mu^-$ stop in a light material, $\mu^-$ also decay partially by $\mu^- \rightarrow e^+~\bar{\nu_e}~\nu_{\mu}$
\end{itemize}
These can be selected by gating out the first 1 $\mu$s from the start of the  proton beam. 
Figure \ref{fig:NeutrinoAll} shows the expected neutrino energy spectrum 
from MLF 
target (top) and time distributions from various sources (bottom).
In this proposal, we concentrate on neutrino beam from muon decay at rest. With extensive studies and experience in dealing with the backgrounds during the proton beam, neutrinos from pion and kaon decay at rest may become usable. 

The $ \mu$ decay at rest neutrino beam was simulated by the following steps. Table~\ref{tab:pionprod} and \ref{tab:pionprod2} are summary tables for
the production of neutrinos from $\mu$ decays.
 \begin{enumerate}
\item Particle production by 3 GeV proton \\
The interaction of the 3 GeV proton beam with the mercury target and beam line components has been simulated with FLUKA~\cite{FLUKA} and QGSP-BERT (in Geant4~\cite{GEANT4}) hadron interaction simulation packages.
 \item $\pi ^{\pm}$ interactions and decay \\
After the production, both $\pi^+$ and $\pi^-$ lose their energy mainly by ionization. In addition, they disappear by charge exchange reaction $\pi^{\pm} (n,p) \rightarrow \pi^0 (p, n)$, $\pi^0\rightarrow \gamma \gamma$.
The survived $\pi^+$s stop and decay with 26 ns lifetime.
On the other hand, the survived $\pi^-$s are absorbed by forming a $\pi$-mesic atom and get absorbed promptly. The decay-in-flight takes place with very suppressed rate of about $\sim 8\times 10^{-3}$ of produced $\pi^{\pm}$s.
\item $\mu ^{\pm}$ absorption and decay \\
All $\mu^+$ decay by $\mu^+\rightarrow e^+\nu_e \bar{\nu_{\mu}}$. Because of the muon lifetime and energy loss process, the decay-in-flight is negligible.
$\mu^-$ is captured by nucleus by forming a mu-mesic atom and eventually produce $\nu_{\mu}$ with an end point energy of 100 MeV. The absorption rate depends on the nucleus and becomes faster for heavier nuclei.
The total nuclear capture rates for negative muons have been measured in terms of effective muon lifetime~\cite{mudecay}.
\end{enumerate}
The resulting neutrino fluxes for each type of neutrino species are shown in Figure \ref{fig:muDARflux}, which has been already shown previously.
Table~\ref{tab:pionprod} and \ref{tab:pionprod2} are examples of expected production rates of $\pi ^{\pm}$ by 3 GeV  protons on mercury target and 
resulting $\mu^+$ and $\mu^-$ decay neutrino per proton, based on a pion production model.  

\begin{table}[htb]
\begin{center}
	\begin{tabular}{|c|c|c|}
\hline
 ~  &  $\pi^+\rightarrow \mu^+ \rightarrow \bar{\nu_\mu}$ &   $\pi^-\rightarrow \mu^- \rightarrow \bar{\nu_e}$   \\ \hline \hline
$\pi$/p  &   $6.49\times 10^{-1} $ &  $4.02\times 10^{-1}$   \\ \hline
 $\mu$/p   &  $3.44\times 10^{-1} $ &  $3.20\times 10^{-3}$ \\ \hline
$\nu$/p  & $3.44\times 10^{-1} $ & $7.66\times 10^{-4}$  \\ \hline
 $ \nu$ after $1\mu$s & $2.52\times 10^{-1}$ & $4.43\times 10^{-4}$
\\ \hline \hline
	\end{tabular}
	\caption{\setlength{\baselineskip}{4mm}
	Summary of an estimate of $\mu DAR$ neutrino production by 3 GeV proton by FLUKA hadron simulation package.}
	\label{tab:pionprod}
\end{center}
\end{table}

\begin{table}[htb]
\begin{center}
	\begin{tabular}{|c|c|c|}
\hline
 ~  &  $\pi^+\rightarrow \mu^+ \rightarrow \bar{\nu_\mu}$ &   $\pi^-\rightarrow \mu^- \rightarrow \bar{\nu_e}$   \\ \hline \hline
$\pi$/p  &   $5.41\times 10^{-1} $ &  $4.90\times 10^{-1}$   \\ \hline
 $\mu$/p   &  $2.68\times 10^{-1} $ &  $3.90\times 10^{-3}$ \\ \hline
$\nu$/p  & $2.68\times 10^{-1} $ & $9.34\times 10^{-4}$  \\ \hline
 $ \nu$ after $1\mu$s & $1.97\times 10^{-1}$ & $5.41\times 10^{-4}$
\\ \hline \hline
	\end{tabular}
	\caption{\setlength{\baselineskip}{4mm}
	Summary of an estimate of $\mu DAR$ neutrino production by 3 GeV protons by QGSP-BERT hadron simulation package.
	}
	\label{tab:pionprod2}
\end{center}
\end{table}

Needless to say, there are many sources of ambiguities in pion production, for example production by secondary particles in thick target, target geometrical modeling, and pion production from mercury. We use these calculations as estimates
and the actual $\mu^-$ backgrounds should be determined from the data based on their known spectrum and known cross section (see section 7.1).

For this proposal, numbers from Table~\ref{tab:pionprod} are used 
to estimate the central values, and those in Table~\ref{tab:pionprod2} are used for the cross checks.

%%%%%%%%%%%%%%%%%%%%%%
\subsection{Estimated neutrino event rates}
%%%%%%%%%%%%%%%%%%%%%%
\label{ENER}
~~
The estimated neutrino event rates (50~tons of total fiducial mass, 4~years measurement) is shown in Table~\ref{tab:RATE}. The proton intensity is assumed to be 0.33 mA, delivering $3\times 10^{22}$ protons on target (POT) per 4000 hour operation in one year. The stopping $\nu$/p ratio is estimated from FLUKA simulations to be 0.344. The $\bar{\nu}_{\mu}$ flux from the $\pi^+\rightarrow \nu_{\mu}+\mu^{+}; \mu^{+} \rightarrow e^{+} + \nu_{e} + \bar{\nu}_{\mu}$ chain at 17 m is then equal to 1.9$\times$10$^{14} \nu$ year/cm$^2$. The event reconstruction efficiencies are assumed to be 100~\% for all processes here.

\begin{table}[h]
\begin{center}
\begin{tabular}{|c|c|c|c|}
\hline
Reactions           & XS ($cm^2$) averaged by energy    & Events & Comment    \\ \hline \hline
&&& $\Delta m^2=3.0eV^{2}$\\ 
$\bar{\nu_e} p\rightarrow e^+ n$ (signal)&$9.5\times 10^{-41}$ & 1690 &$sin^2(2\theta)=3.0\times10^{-3}$\\
&&&(Best $\Delta m^{2}$ for MLF exp.)\\\hline
&&& $\Delta m^2=1.2eV^{2}$\\ 
&& 703 &$sin^2(2\theta)=3.0\times10^{-3}$\\
&&&(Best fit values of LSND)\\\hline
$\bar{\nu_e} p\rightarrow e^+ n$(from $\mu^{-}$) & $7.2\times 10^{-41}$ & 804 & FLUKA (shown in Table~\ref{tab:pionprod}) \\ \hline
$\nu_e +$$^{12}C\rightarrow e^{-}+$$^{12}N_{gs}$  & $8.9 \times 10^{-42}$ & 22934  & \\ \hline
$\nu_e +$$^{12}C\rightarrow e^{-}+$$^{12}N^{*}$  & $4.3 \times 10^{-42}$ & 11008  & \\ \hline
%$\nu_{e,\mu}$$^{12}C\rightarrow \nu_{e,\mu}$$^{12}C^*$ & $2.8 \times 10^{-45}E_{\nu}$ &  400 & \\ 
%(15.11)&&&\\\hline
	\end{tabular}
	\caption{\setlength{\baselineskip}{4mm}Number of neutrino events with 50~tons of total fiducial mass times 4~years operation. }
	\label{tab:RATE}
\end{center}
\end{table}

\pdfoutput=1
%%%%%%%%%%%%%%%%%%%%%%%%%%%%%%%%%%%%%%%%%%%%%
\section{Backgrounds Studies}
%%%%%%%%%%%%%%%%%%%%%%%%%%%%%%%%%%%%%%%%%%%%%

\subsection{Overview and brief summary of this section}
\indent

As described in Section \ref{sec_signatures}, we use inverse $\beta$-decay (IBD; $\bar{\nu_e} + p \to e^+ + n$) to detect $\bar{\nu_\nu} \to \bar{\nu_e}$ oscillations.
The neutron from IBD is thermalized in detector material and captured by a 
nucleus, and afterwards the nucleus emits a gamma(s).
The signature of the $\bar{\nu_e}$ signal is thus the coincidence of an $e^+$ as a `prompt signal' and gammas by the neutron capture\footnote{The neutron capture reaction by Gd releases a sum of 8-MeV energy in a cascade of 3-4 gammas.} as a `delayed signal'.

Beam associated and natural sources of backgrounds (gammas, neutrons etc.) can mimic the prompt and delayed signals.
To estimate amount of background at the detector location (the 3rd floor of MLF), we carried out background measurements in one of the MLF beam lines, BL13.
In addition to gammas and neutrons as backgrounds for delayed signal,
Michel electrons from muon decays were observed as a background for prompt signal.
Followings are the results of measurements:

\begin{itemize}
\item Michel electrons made by fast neutrons ($n + p \to  X + \pi^{\pm}$, 
then $\pi^{\pm} \to \mu^{\pm} \to e^{\pm}$) are observed as an "IBD prompt" 
background. Probability of the fake prompt signal made by Michel electrons is $5.6 \times$10$^{-4}$ /spill/ton/300 kW/9 $\mu$s (1$<{\rm t[\mu s]}<$10).  
\item Gammas ($6<E[{\rm MeV}]<12$) made from neutron captured around the 1 ton detector, which is the background of "IBD delayed" signal, are observed. Total amount of the background is 0.9 /spill/ton/300 kW/100 $\mu$s.
\item The amount of the neutrons below 1 MeV are also estimated by the detector.
Total amount of the background is 14 /spill/ton/300 kW/100 $\mu$s.
\end{itemize}

The expected amount of background at the 3rd floor was then extrapolated by using the MC predictions at both BL13 and the 3rd floor.
The background rates at the 3rd floor at the surface of the supposed detector described in Section \ref{sec_detector} with 1 MW beam power, are estimated as follows;

\begin{itemize}
\item Michel electrons at the 3rd floor is $2 \times 10^{-7}$ /spill/detector/MW/9 $\mu$s.
\item  Total amount of gammas ($6<E[{\rm MeV}]<12$) is 14 /spill/detector/MW/100 $\mu$s. 
\item  Total amount of neutrons ($E<1$ MeV) is 40 /spill/detector/MW/100 $\mu$s. 
\end{itemize}

By taking the probability of the backgrounds coming to the fiducial volume in the detector into account (see Section \ref{ESBALL}),
these values are low enough to observe the oscillated $\bar{\nu_e}$ signals by IBD at the 3rd floor.

%%%%%%%%%%%%%%%%%%%%%%%%%%%%%%%%%
\subsection{Measurements at BL13}
\begin{figure}[hbtp]
	\begin{center}
		\includegraphics[scale=0.5]{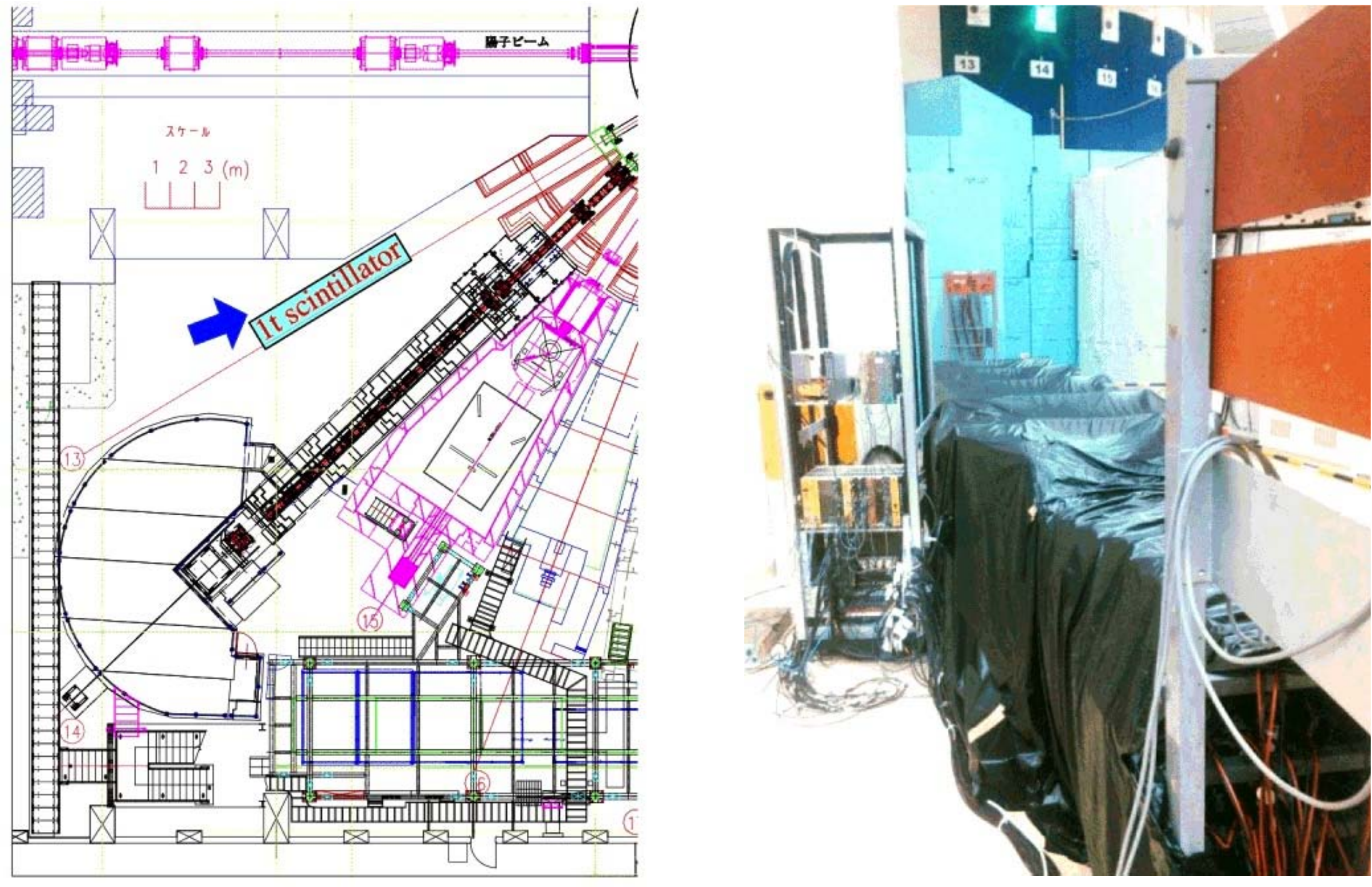}
		\caption{\setlength{\baselineskip}{4mm}Top view of the BL13 and the location of the scintillator detector (left),
		and a photograph of the scintillator detector at BL13 (right).
		This photograph was taken at the blue arrow shown in the left figure.}
		\label{1t_view}
	\end{center}
\end{figure}

%%%%%%%%%%%%%%%%%%%%%%%%%%%%%%%%%
~
Figure \ref{1t_view} shows the top view of BL13 beam line.

The amount of background was measured with a one-ton scintillator detector and 370g NaI
during 300kW beam operation and beam-off.

		We placed a plastic scintillator detector at BL13 in MLF.
		The target mass of the detector was 1 ton.
		The details of the detector are described in Appendix \ref{appendix_1t}.
		Figure \ref{1t_tvse} shows the correlation between energy and timing of the observed activities.
		The two bunch structure of the RCS proton beam was clearly seen as the groups of the activities.
		
		We also placed a NaI detector at BL13.
		The details of the NaI detector are described in Appendix \ref{appendix_NaI}.
		
		\begin{figure}[hbtp]
			\begin{center}
				\includegraphics[scale=0.6]{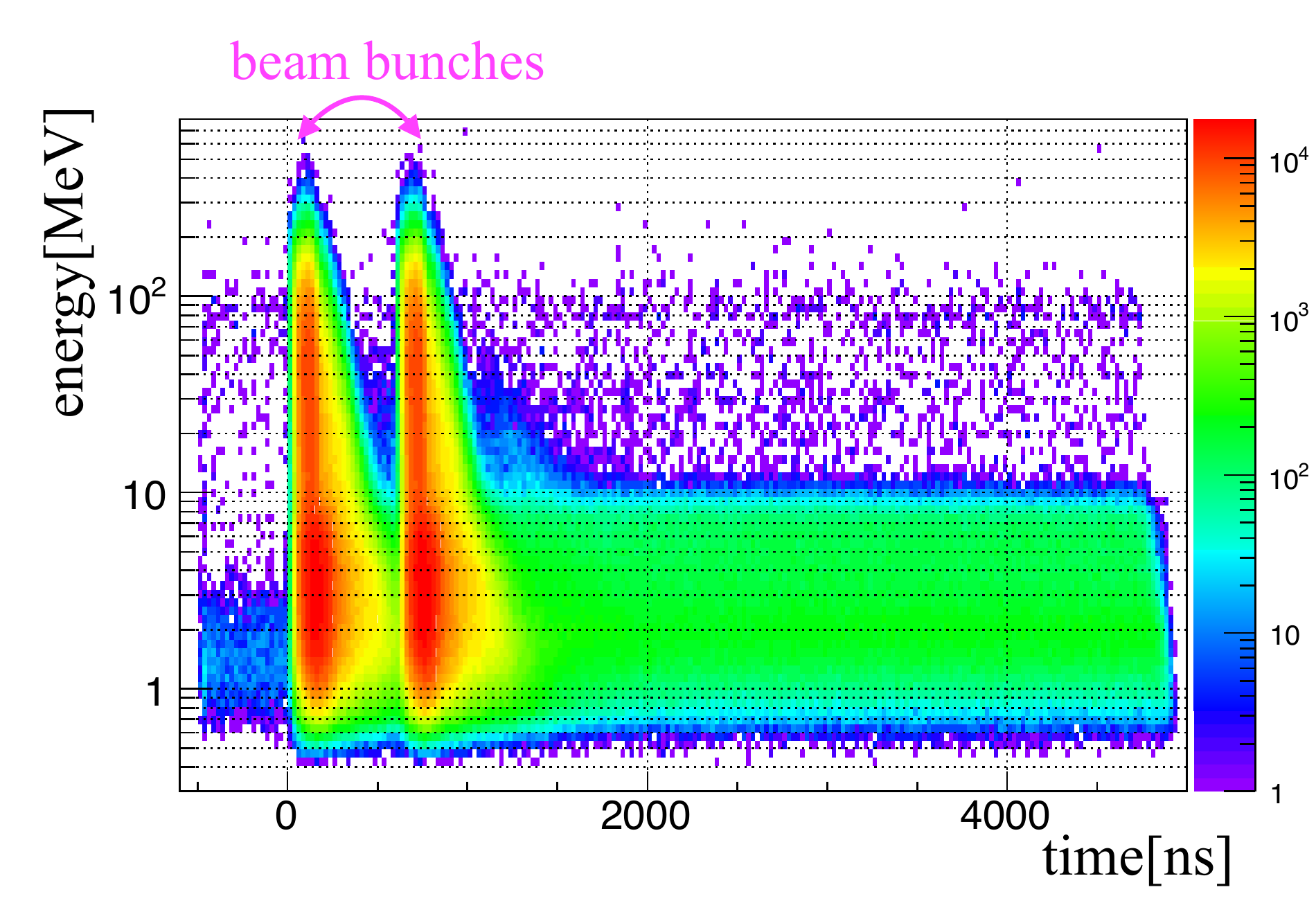}
				\caption{\setlength{\baselineskip}{4mm}
				2D plot for energy (vertical) and timing of the activities (horizontal) observed with the 1 ton scintillator detector placed at BL13.
				The clear two bunch structure of the RCS proton beam is seen.}
				\label{1t_tvse}
			\end{center}
		\end{figure}
%%%%%%%%%%
\subsubsection{Backgrounds for the prompt signal at BL13}
\begin{figure}[hbtp]
				\begin{center}
					\includegraphics[scale=0.6]{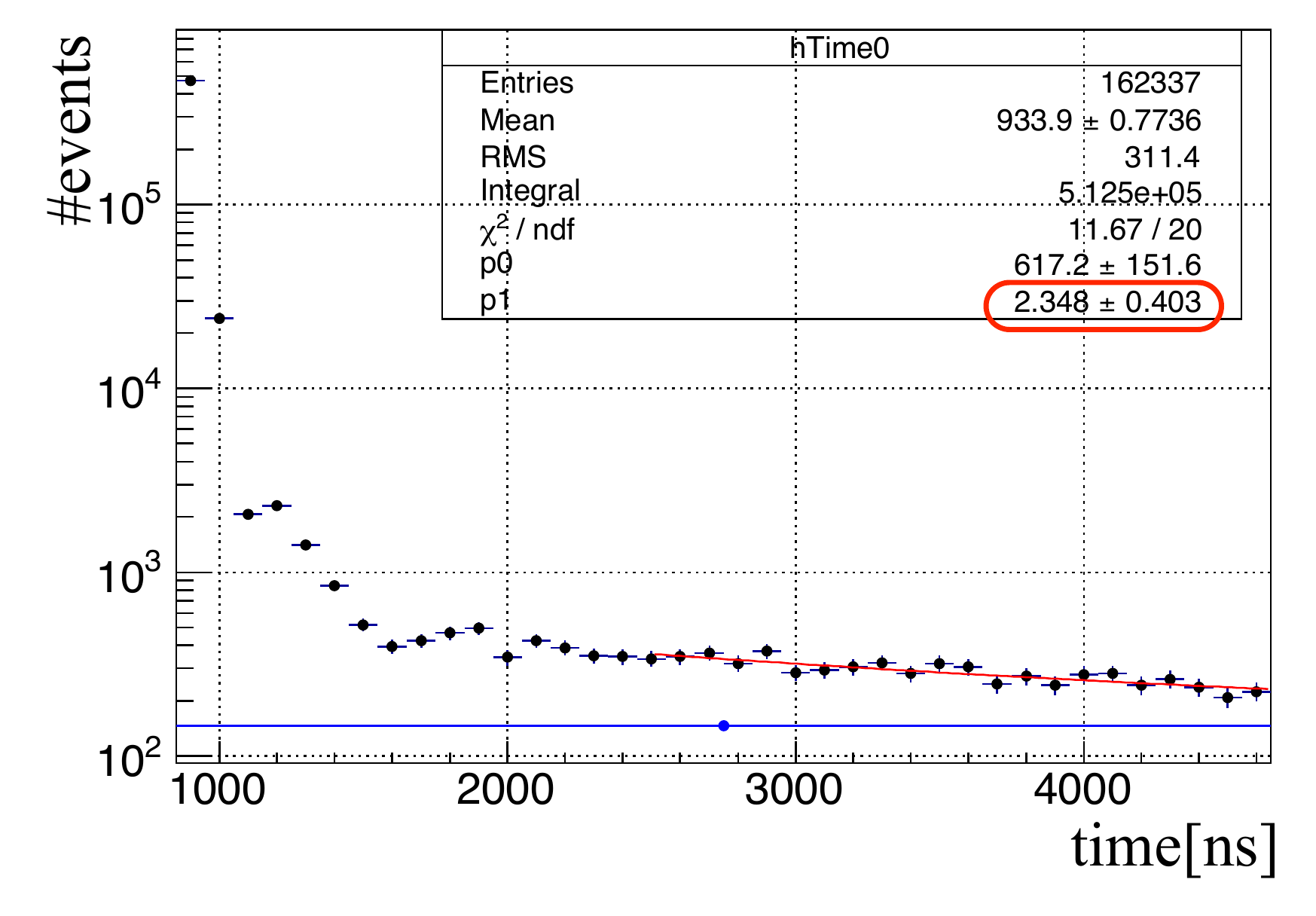}
					\caption{\setlength{\baselineskip}{4mm}
					Timing distribution from the 1st bunches for the activities: $20 < E [{\rm MeV}] < 60$.
					The distribution was fitted with the Equation \ref{eq_lifetime} (red line).
					The constant (time-independent) term was estimated with Beam-off data (blue line with a dot).
					The lifetime of the activities is $\tau = 2.3 \pm 0.4\ {\rm \mu s}$ and is consistent with the muon lifetime.}
					\label{prompt_lifetime}
				\end{center}
			\end{figure}

%%%%%%%%%%
			Michel electrons were observed as a background for prompt signal.
			Figure \ref{prompt_lifetime} shows the timing distribution from the first bunches for the activities: $20 < E [{\rm MeV}] < 60$.
			We fitted the distribution with the function:
			\begin{equation}
				\label{eq_lifetime}
				f(t) = A \exp{\left( -\frac{t}{\tau} \right)} + B, 
			\end{equation}
			where $A\ (B)$ is the amount of the time dependent (independent) term and $\tau$ is the lifetime of the activities.
			When we fixed the time independent term, $B$, with the estimated value from Beam-off data,
			we obtained $\tau = 2.3 \pm 0.4\ {\rm \mu s}$ and it is consistent with the muon lifetime.
			
			Figure \ref{prompt_energy} shows the energy distributions of the activities just after beam bunches and during Beam-off with and without the cosmic-muon veto\footnote{\setlength{\baselineskip}{4mm}As shown in Figure~\ref{1t_setup}, the 1 ton detector consists of 24 pieces of plastic scintillators.
			Pairs of scintillators on the top, bottom, and both horizontal sides, surrounding scintillators in the middle, were used as cosmic-muon veto counters.
			Activities associated with the scintillators in the second row from top and bottom of the detector were also vetoed.}.
			The distributions above 60 MeV are consistent with each other between Beam-on and Beam-off,
			and activities below 60 MeV remain even with the cosmic-muon veto.
			Figure \ref{prompt_lifetime_wVeto} shows the timing distribution from the 1st bunches for the activities: $20 < E [{\rm MeV}] < 60$ with the cosmic-muon veto.
			The distribution later than 1500 ns is still consistent with Michel electrons from muon decays and time independent activities estimated from Beam-off data even with the cosmic-muon veto.
			The activities before 1500 ns will be investigated with the measurement at the detector location.
			\begin{figure}[hbtp]
				\begin{center}
					\includegraphics[scale=0.6]{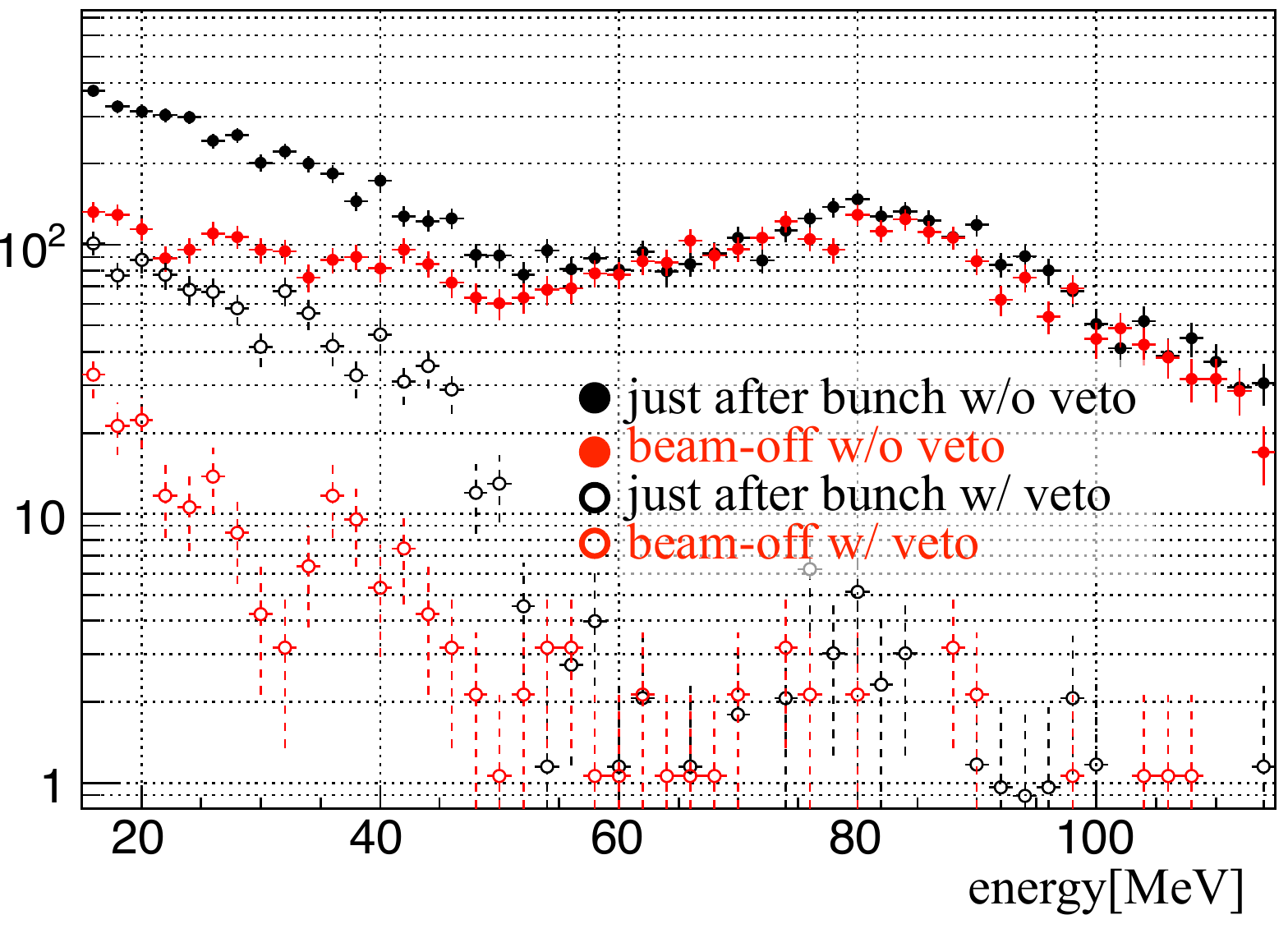}
					\caption{\setlength{\baselineskip}{4mm}
					Energy distributions of the activities just after beam bunches and during Beam-off with and without the cosmic-muon veto.
					The distributions above 60 MeV are consistent with each other between Beam-on and Beam-off,
					and activities below 60 MeV remain even with the cosmic-muon veto.}
					\label{prompt_energy}
				\end{center}
			\end{figure}
			\begin{figure}[hbtp]
				\begin{center}
					\includegraphics[scale=0.6]{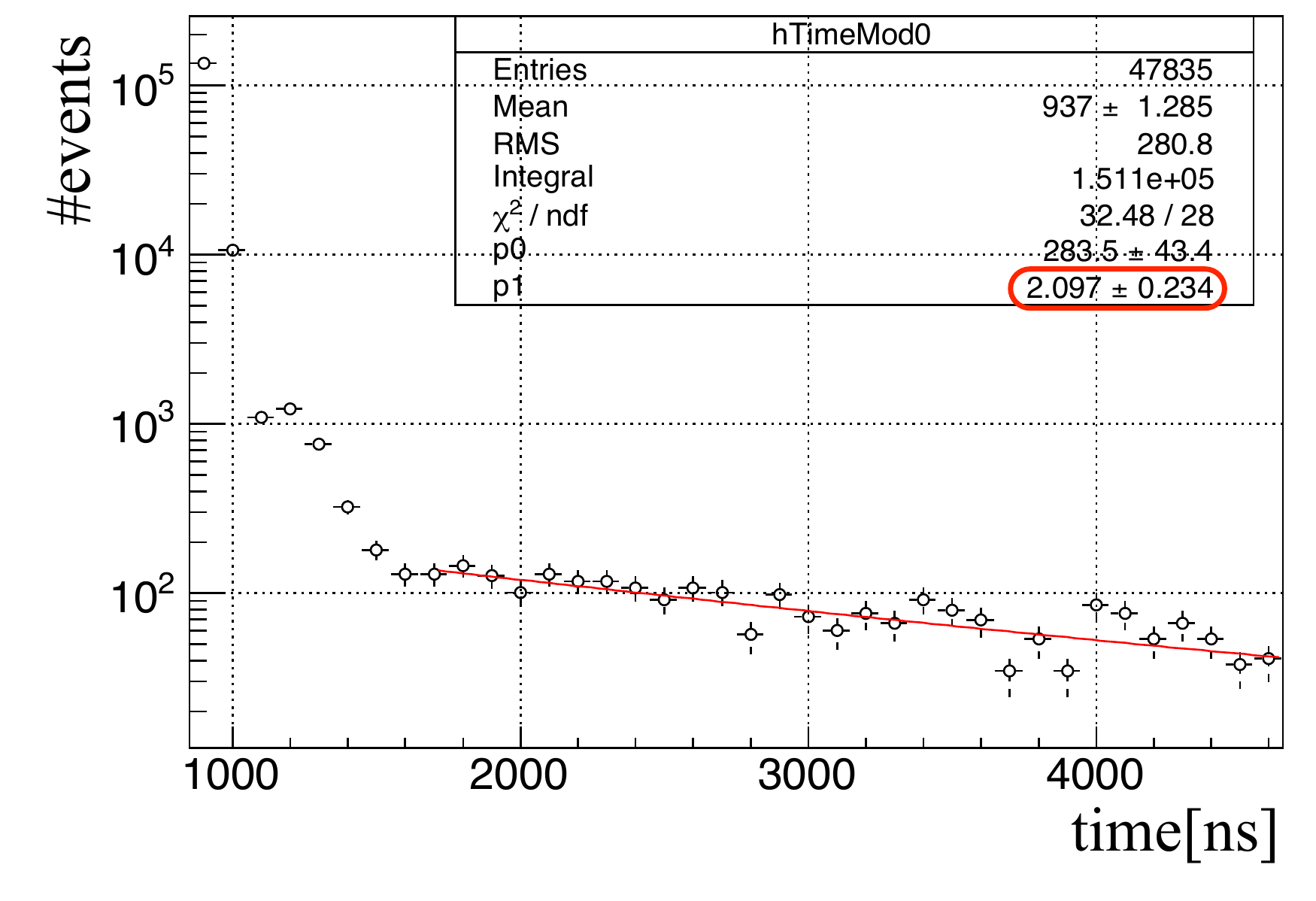}
					\caption{\setlength{\baselineskip}{4mm}
					Timing distribution from the 1st bunches for the activities: $20 < E [{\rm MeV}] < 60$ with the cosmic-muon veto.
					The distribution was fitted with the exponential plus constant (red line).
					The constant (time-independent) term was estimated with Beam-off data ($\sim 10$).
					The lifetime is also consistent with the muon lifetime.}
					\label{prompt_lifetime_wVeto}
				\end{center}
			\end{figure}
			
\begin{figure}[hbtp]
				\begin{center}
					\includegraphics[scale=0.6]{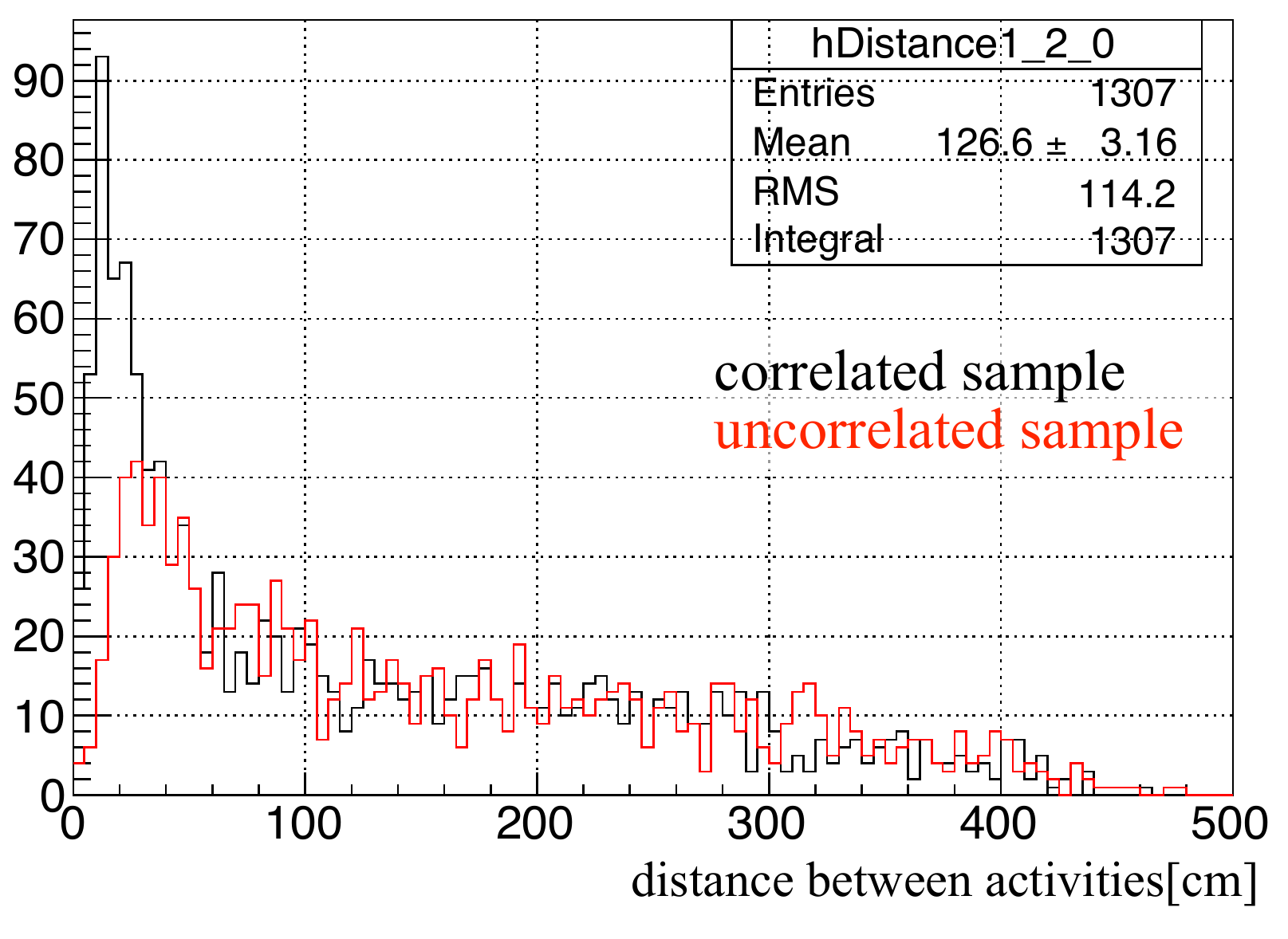}
					\caption{\setlength{\baselineskip}{4mm}
					Spatial distance distributions for both correlated (black) and uncorrelated (red) samples.
					A clear enhancement was seen within the distance of 30 cm for the correlated sample.}
					\label{prompt_distance}
				\end{center}
			\end{figure}

			We investigated the correlation between those 'prompt-like' activities and activities on bunch timings.
			For comparison, we first made two samples: correlated and uncorrelated samples.
			The first sample contains pairs of a 'prompt-like' activity and an activity on bunch timing at the same beam spill.
			The other one contains pairs of a 'prompt-like' activity and an activity on bunch timing at the next beam spill.
			We then calculated the spatial distance between the activities for each pair.
			Figure \ref{prompt_distance} shows the spatial distance distributions for both correlated and uncorrelated samples.
			A clear enhancement was seen within the distance of 30 cm for the correlated sample.
			By rejecting the prompt candidate which has an activity around itself on bunch timing,
			the Michel electron background is supposed to be suppressed.
			The rejection power of this cut will be discussed in Section \ref{BAFN}.

\begin{figure}[hbtp]
				\begin{center}
					\includegraphics[scale=0.6]{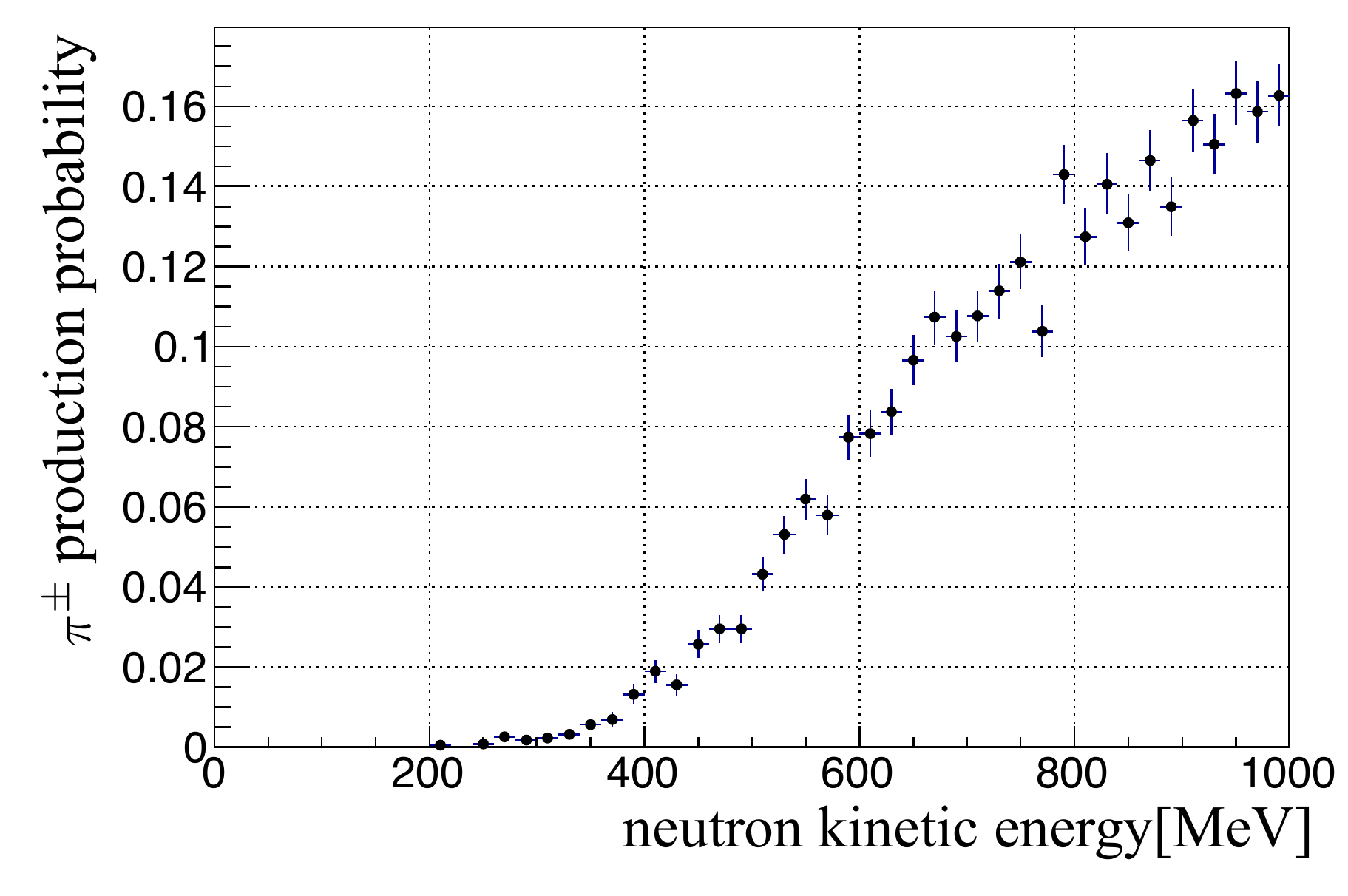}
					\caption{\setlength{\baselineskip}{4mm}
					Production probability of charged pions as a function of kinetic energy of incident neutrons.
					The production target was a plastic scintillator.
					The values were estimated with MC simulation.
					Neutrons whose kinetic energy are larger than about 200 MeV can produce charged pions.}
					\label{prompt_piproduct}
				\end{center}
			\end{figure}

			These results indicated that muons hit the scintillators on bunch timing and generated the Michel electrons.
			There are two possibilities of the muon source.
			\begin{itemize}
			\item Muons directly coming from outside of the scintillators.
			\item Fast neutrons coming on bunch timing hit scintillators and produced pions, and the muons were from those pion decays.
			Neutrons whose kinetic energy are larger than about 200 MeV can produce charged pions as shown in Figure~\ref{prompt_piproduct}.
			\end{itemize}
			If the first case, we can easily reject such events by surrounding the detector with veto counters.
			If the latter case, it is not obvious how well we can suppress such events.
			We thus supposed that the Michel-electrons are from $\pi \to \mu \to e$ decay induced by fast neutrons,
			and allocated the number of observed events itself to the background source for $\bar{\nu_e} + p \to e^+ + n$.
			
			The number of observed Michel electron events with the 300 kW beam at BL13 was equivalent to $5.6 \times 10^{-4}$ /spill within the time window, $1 < t [{\rm \mu s}]< 10$, without the spatial distance cut described above.

%%%%%%%%%%%%%%%%%%%
\subsubsection{Backgrounds for the delayed signal at BL13}
%%%%%%%%%%%%%%%%%%%
Backgrounds for delayed signal after 1 $\mu$s from the first beam bunch were also measured with the 1 ton scintillator detector.
Because of the 1 $\mu$s delay, the observed activities are dominated by neutron-capture related events.
			There are two types of backgrounds for delayed signal.

\begin{itemize}
\item Gammas by neutron captures in the materials outside of the detector.
 The gammas directly hit the detector from outside and mimic gammas by neutron captures in the detector.
\item Neutrons which are captured in the detector and emit gammas at the end.
\end{itemize}
			We measured each background level at BL13.\\
			
			\textbf{Gammas}\\
			
			Because the energy deposit for thermal neutron capture in the scintillators is 2.2 MeV,
			the activities with energy deposit $6<E{\rm [MeV]}<12$ were assumed to be gammas.
			
\begin{figure}[hbtp]
				\begin{center}
					\includegraphics[scale=0.6]{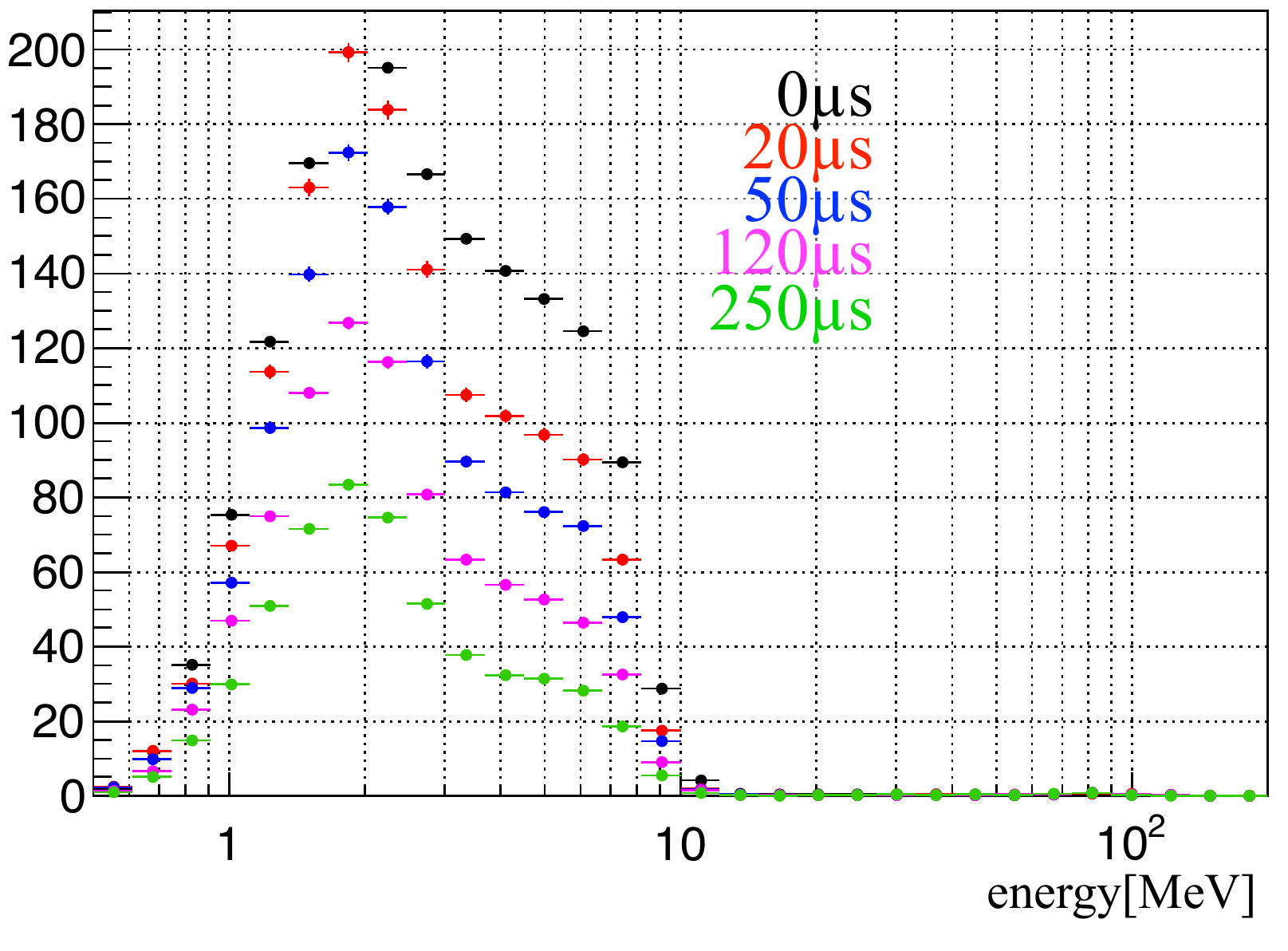}
					\caption{\setlength{\baselineskip}{4mm}
					Energy distributions as a function of the gate delay. (the length of the time window of the FADC is fixed though). For example, "100 $\mu$s" means the FADC gate is opened from 100$\mu$s to 105.5 $\mu$s after the first bunch starting time. The black, red, blue,
					magenta and green dots show the energy distributions after the beam
					bunch, 20 ${\rm \mu s}$, 50 ${\rm \mu s}$, 120 ${\rm \mu s}$ and 250
					${\rm \mu s}$ later, respectively. }
					\label{delayed_energy}
				\end{center}
			\end{figure}

\begin{figure}[hbtp]
				\begin{center}
					\includegraphics[scale=0.6]{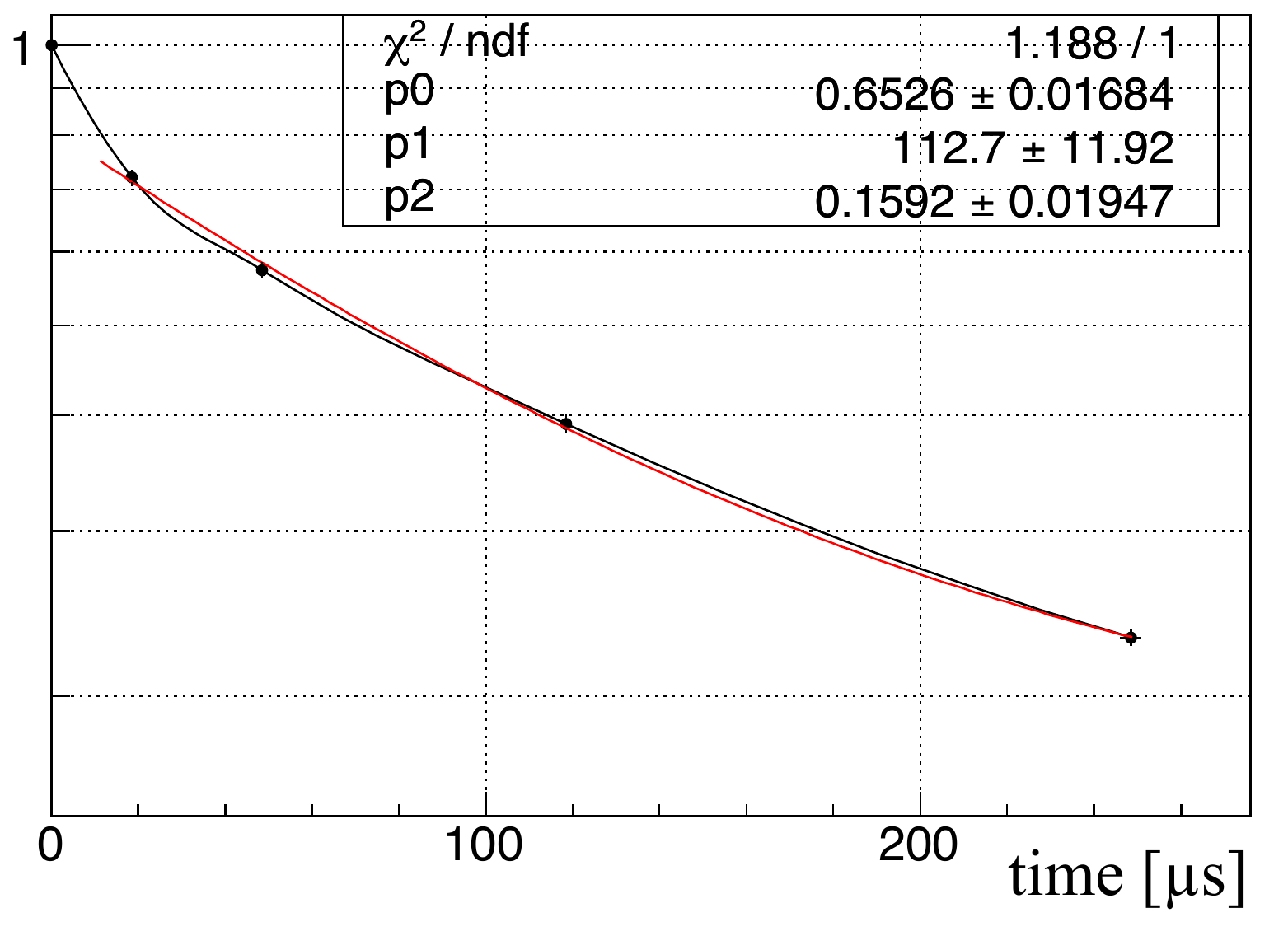}
					\caption{\setlength{\baselineskip}{4mm}
					Relative rate of gammas with energy deposit $6<E{\rm [MeV]}<12$ as a function of the FADC gate delay.
					The curve was fitted with the Equation \ref{eq_lifetime} (red line), which is , the exponential + constant term..}
					\label{delayed_lifetime}
				\end{center}
			\end{figure}

			We found the activities have some effective lifetime and thus measured the lifetime of those activities.
			Figure \ref{delayed_energy} shows the energy distributions for each period of time,
			and the relative rate of gammas for each period is shown in Figure~\ref{delayed_lifetime}.
			We fitted the curve with the Equation \ref{eq_lifetime} and obtained $\tau = 112.7 \pm 11.92\ \mu$s.
			
			The number of observed events with 300 kW beam at BL13 was equivalent to $7 \times 10^4$ /day within  5 $\mu$s window after the beam bunches.
			By considering the efficiency and the effective lifetime curve described above,
			the expected number of events within 100 $\mu$s window after the beam bunches was equivalent to 0.9 /spill with the 300 kW beam. \\

			\textbf{Neutrons}\\
			
			As described above, these neutrons were detected as thermal neutrons at the end.
			Because the energy deposit for thermal neutron capture in the scintillators is 2.2 MeV,
			we evaluated the amount of these neutrons by counting activities with the energy deposit $1<E{\rm [MeV]}<4$.
			
\begin{figure}[hbtp]
				\begin{center}
					\includegraphics[scale=0.6]{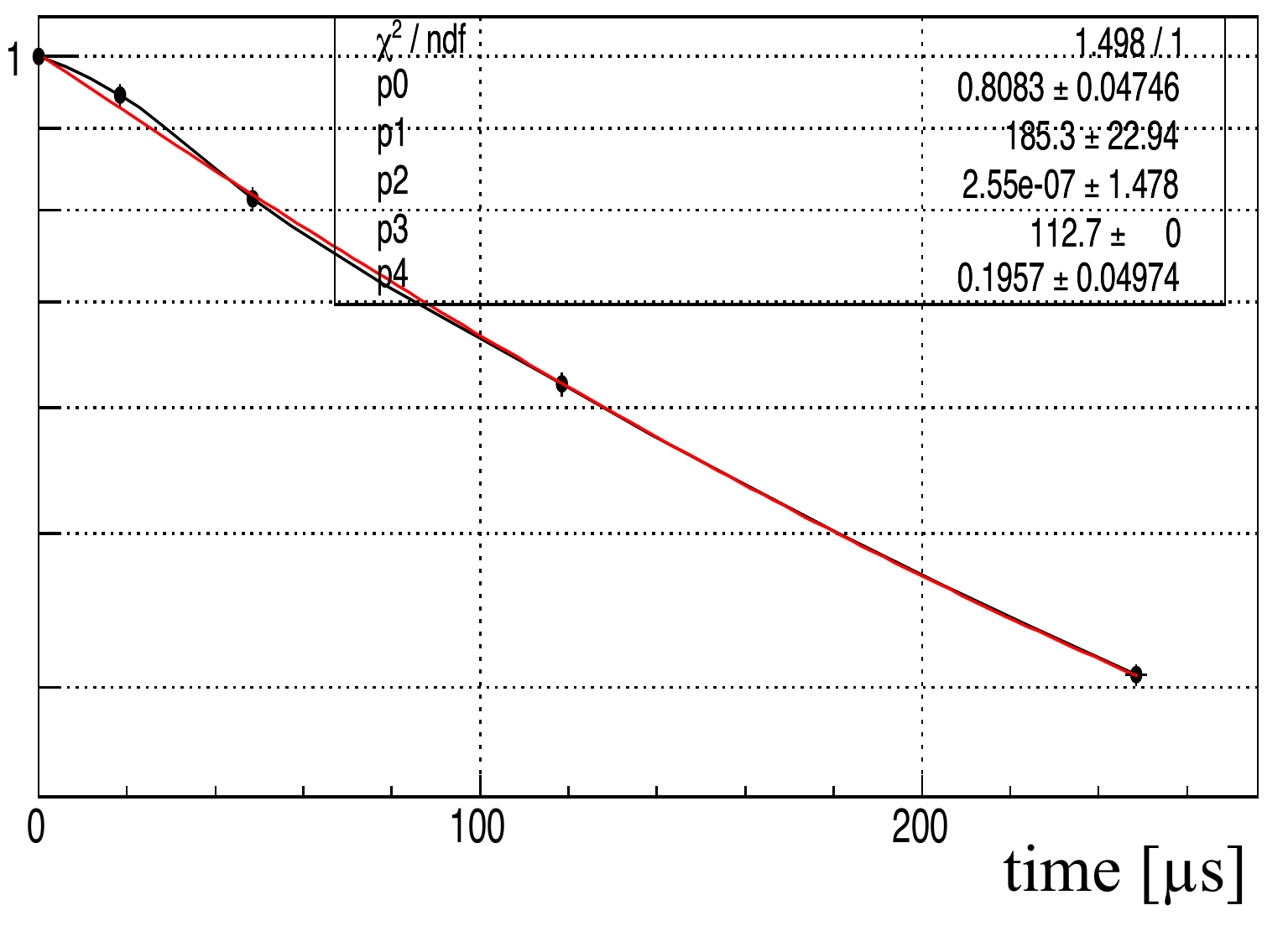}
					\caption{\setlength{\baselineskip}{4mm}
					Relative rate of the activities with energy deposit $1<E{\rm [MeV]}<4$ as a function of the FADC gate delay.
					The curve was fitted with the Equation \ref{eq_lifetime} (red line), the exponential + constant term. }
					\label{delayed_lifetime2}
				\end{center}
			\end{figure}

			We measured the effective lifetime of those activities.
			The relative event rate for each period of time is shown in Figure~\ref{delayed_lifetime2}.
			We fitted the curve with the Equation \ref{eq_lifetime} and obtained $\tau = 185 \pm 23\ \mu$s.
			
			The number of observed events with the 300 kW beam at BL13 was equivalent to $6 \times 10^5$ /day within  5 $\mu$s window after the beam bunches.
			By considering the efficiency and the effective lifetime curve described above,
			the expected number of events within 100 $\mu$s window after the beam bunches was estimated to be  equivalent to 14 /spill with the 300 kW beam for neutrons.
		
\
%%%%%%%%%%%%%%%%%%%%%%%
\subsubsection{Environmental gamma}
\begin{figure}[h]
\begin{center}
\includegraphics[width=100mm]{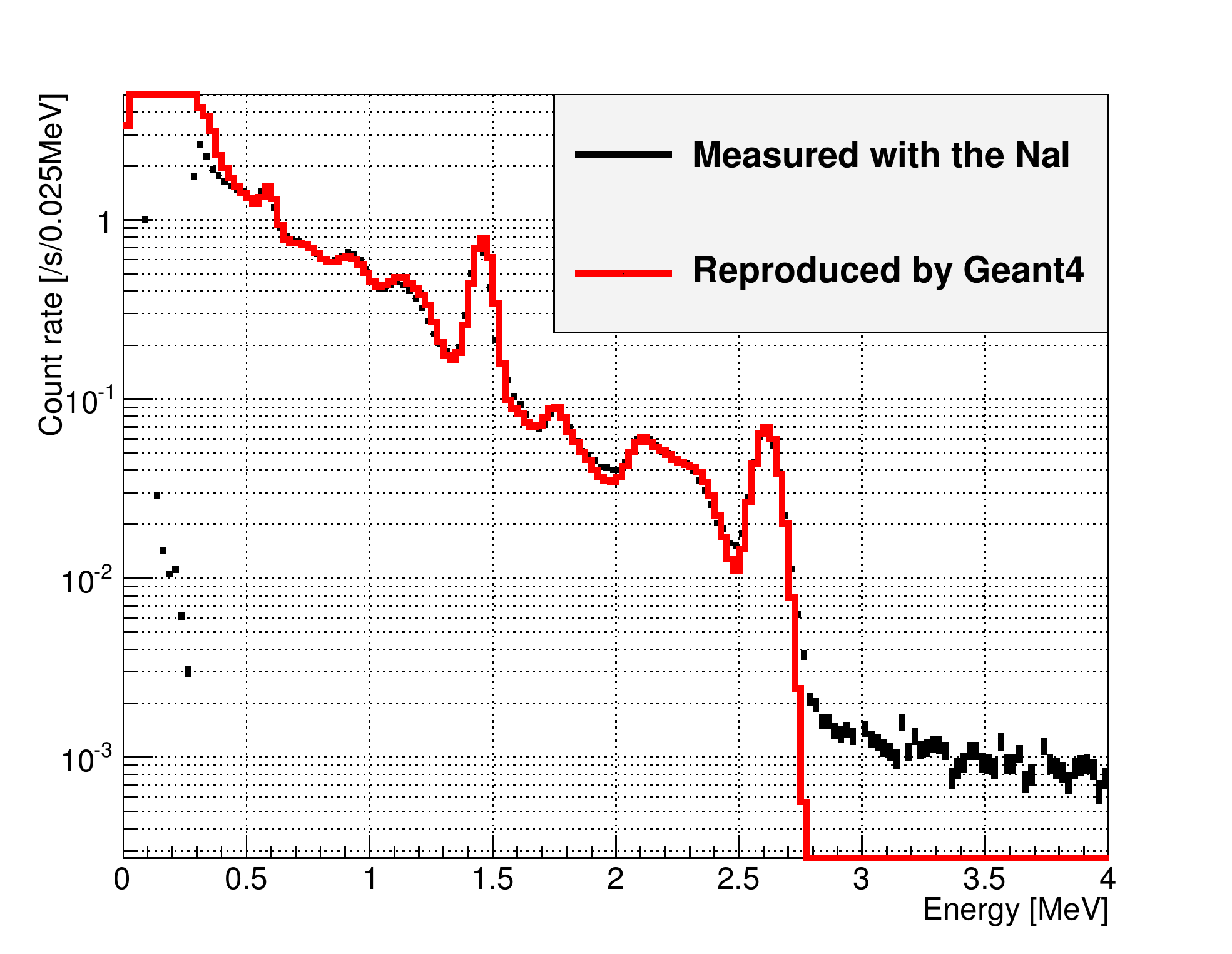}
\end{center}
\caption{\setlength{\baselineskip}{4mm}
The energy spectrum of environmental gammas at BL13 measured by the NaI counter. 
The black spectrum is measured data, while 
the red spectrum is MC result reproduced by Geant4. 
The effects of the cosmic ray events is observed above 2.6 MeV in data.
We can avoid this background with Gd-loaded scintillator.
}
\label{fig:Ev_gamma_BL13}
\end{figure}

%%%%%%%%%%%%%%%%%%%%%%%
~~
Environmental gamma was measured with the NaI counter at BL13 when the RCS was off.
The count rate between 1 and 3 MeV measured with the NaI counter is 10.13 Hz.
The activities above 3 MeV was induced by cosmic rays.
Using Geant4, the shape of spectrum can be reproduced from the combination of $^{238}$U, $^{232}$Th series and $^{40}$K, assuming 3\%/$\sqrt{E[\mathrm{MeV}]}$.
The spectra of measured and reproduced environmental gamma are shown in Figure \ref{fig:Ev_gamma_BL13}.
Estimated flux of environmental gamma is 4.3 /s/cm$^2$.
Similarly, environmental gamma was measured with other NaI counter at Tohoku University and estimated flux is 3.9 /s/cm$^2$.
Although amount of environmental gamma depends on surrounding materials such as concrete, soil and so on, the amount of environmental gamma at two different 
places are not much different.
Therefore the amount of environmental gamma at the 3rd floor is expected to 
be at the same level as above fluxes.
This measurement concludes that Gd-loaded scintillator is useful to reduce the background efficiently.

%%%%%%%%%%%%%%%%%%%%%%%
\subsection{Estimated background rates at the detector location}
\begin{figure}[hbtp]
	\begin{center}
		\includegraphics[scale=0.6]{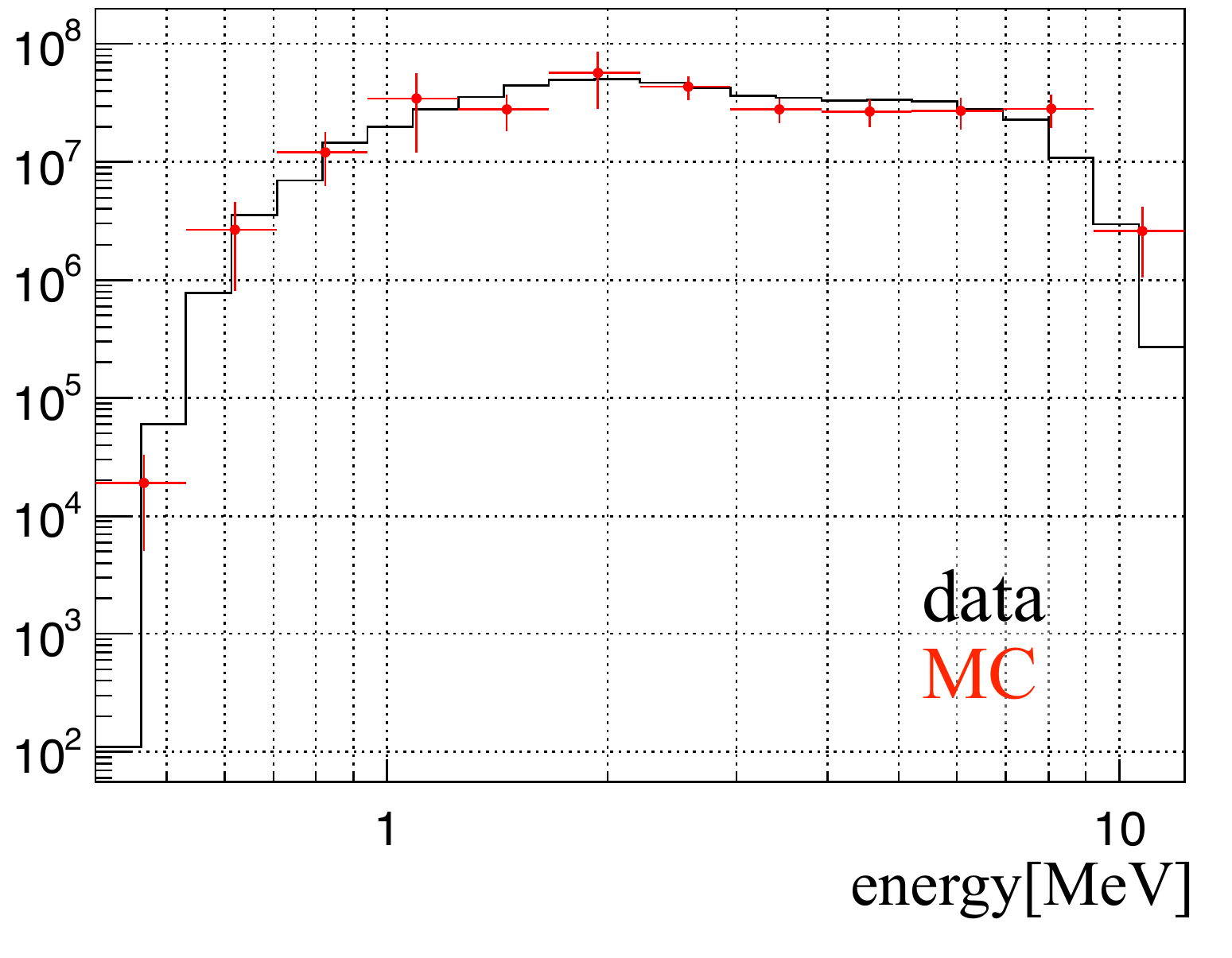}
		\caption{\setlength{\baselineskip}{4mm}
		Comparison of the energy spectrum below 12 MeV measured with the 1 ton detector and the estimation based on PHITS.
		The integrals of the histograms were normalized to each other. 
		The observed energy spectrum was reproduced by PHITS.}
		\label{1t_compare_phits}
	\end{center}
\end{figure}

%%%%%%%%%%%%%%%%%%%%%%%
The background rate at the detector location (the 3rd floor in MLF) was estimated by using Particle and Heavy Ion Transport code System (PHITS)\cite{PHITS}.
The PHITS is used for designing radiation shielding at MLF.
The outputs of PHITS and the radiation survey results at many places at MLF were compared and evaluated with each other.
The PHITS is suitable to estimate the background rate at MLF.
We first evaluated the background rate at BL13 by PHITS and check the validity of PHITS.
Figure \ref{1t_compare_phits} shows the comparison of the energy spectrum measured with the 1 ton detector, and the estimation based on PHITS.
The observed energy spectrum was reproduced.

We then estimated the rates of backgrounds hitting the detector at the 3rd floor by PHITS.
By taking the ratio of the estimated rates at BL13 and the 3rd floor by PHITS,
and multiplying the ratio to the observed background rate at BL13,
the expected background rate at the 3rd floor, $n_{\rm exp}$, is 
\begin{equation}
	n_{\rm exp} = n_{\rm BL13} \times \frac{N_{\rm candidate}}{N_{\rm BL13}}\ ,
\end{equation}
where $n_{\rm BL13}$ is the observed background rate at BL13, $N_i$ is the estimated background rate at the $i$-th location by PHITS.
By calculating the equation, the background rates at the 3rd floor at the surface of the supposed detector described in Section \ref{sec_detector} with 1 MW beam power were estimated to be 14 /spill/detector for gammas and 40 /spill/detector for neutrons.
By taking the probability of the backgrounds coming to the fiducial volume in the detector, into account,
these values are low enough to observe the $\bar{\nu_e}$ signals by IBD as described in Section \ref{ESBALL}.

\begin{figure}[hbtp]
	\begin{center}
		\includegraphics[scale=0.6]{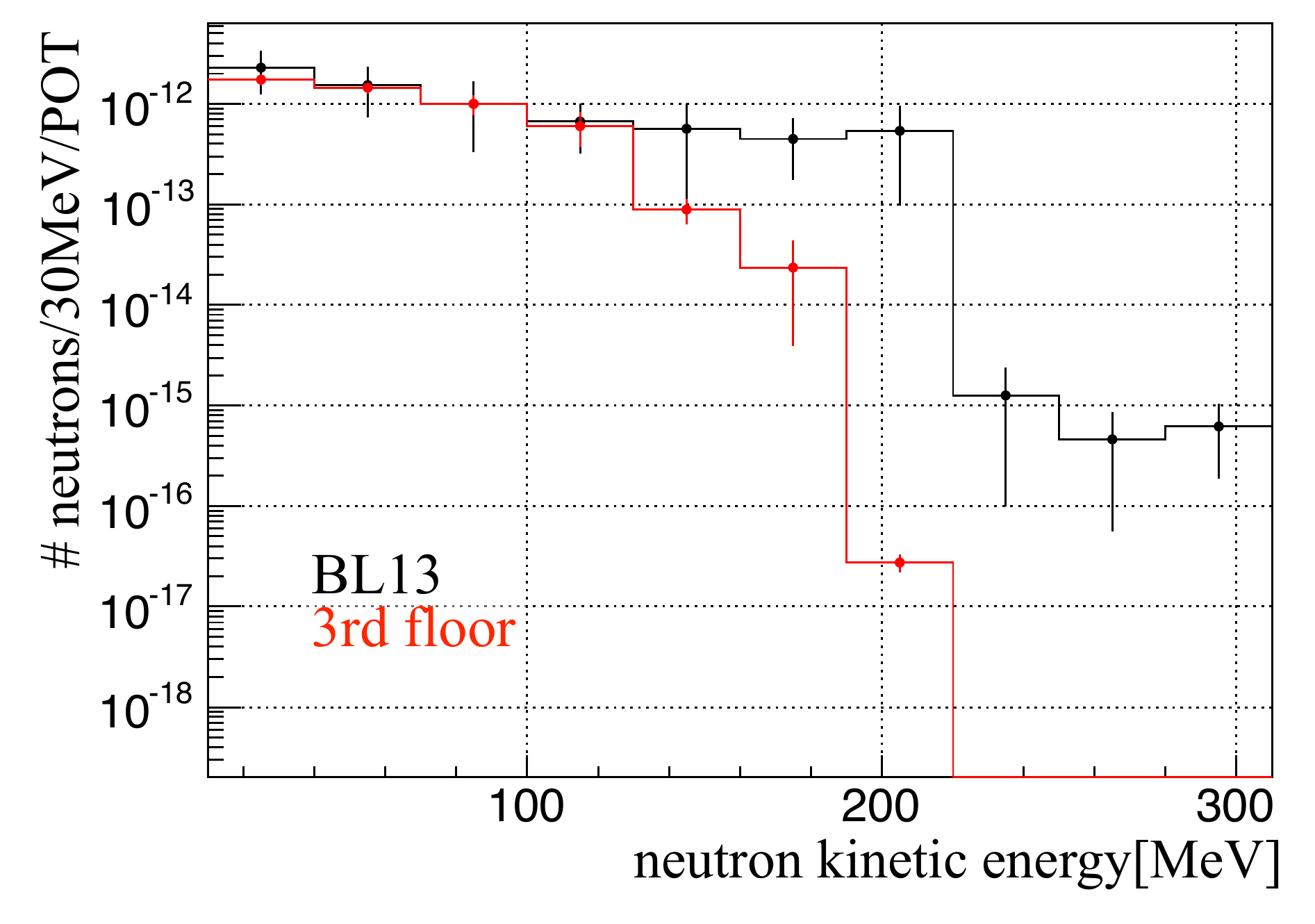}
		\caption{\setlength{\baselineskip}{4mm}
		Estimated rate of the neutrons hitting each detector at BL13 (black) and the 3rd floor (red) as a function of neutron kinetic energies. 
		The PHITS estimated more than 4 orders of magnitude smaller background rate at the 3rd floor than that at BL13 with respect to neutrons which can produce charged pions.}
		\label{phits_highEneutrons}
	\end{center}
\end{figure}

We also estimated the background rate of the Michel-electrons.
Because of the limitation of the Monte-Carlo statistics, we just compared the rates of high energy neutrons which can potentially produce charged pions ($E_{\rm neutron}>$200 MeV).
Figure \ref{phits_highEneutrons} shows the estimated rate of the high energy neutrons hitting each detector at BL13 and the 3rd floor.
The PHITS estimated more than 4 orders of magnitude smaller background rate at the 3rd floor than that at BL13.
By considering the beam power difference (300 kW $\to$ 1 MW),
we thus supposed the background rate of the Michel-electron at the 3rd floor was $2 \times 10^{-7}$ /spill/detector.

\pdfoutput=1
%%%%%%%%%%%%%%%%%%%%%%%%%%%%%%%%%%%
\section{The detector}
%%%%%%%%%%%%%%%%%%%%%%%%%%%%%%%%%%%
\label{sec_detector}

%%%%%%%%%%%%%%%%%%%%%%%%%%%%%%%%%%
\subsection{The detector site}
%%%%%%%%%%%%%%%%%%%%%%%%%%%%%%%%%%
\label{sec_detectorloc}

\begin{figure}
 \centering
 \includegraphics[width=1.0\textwidth]{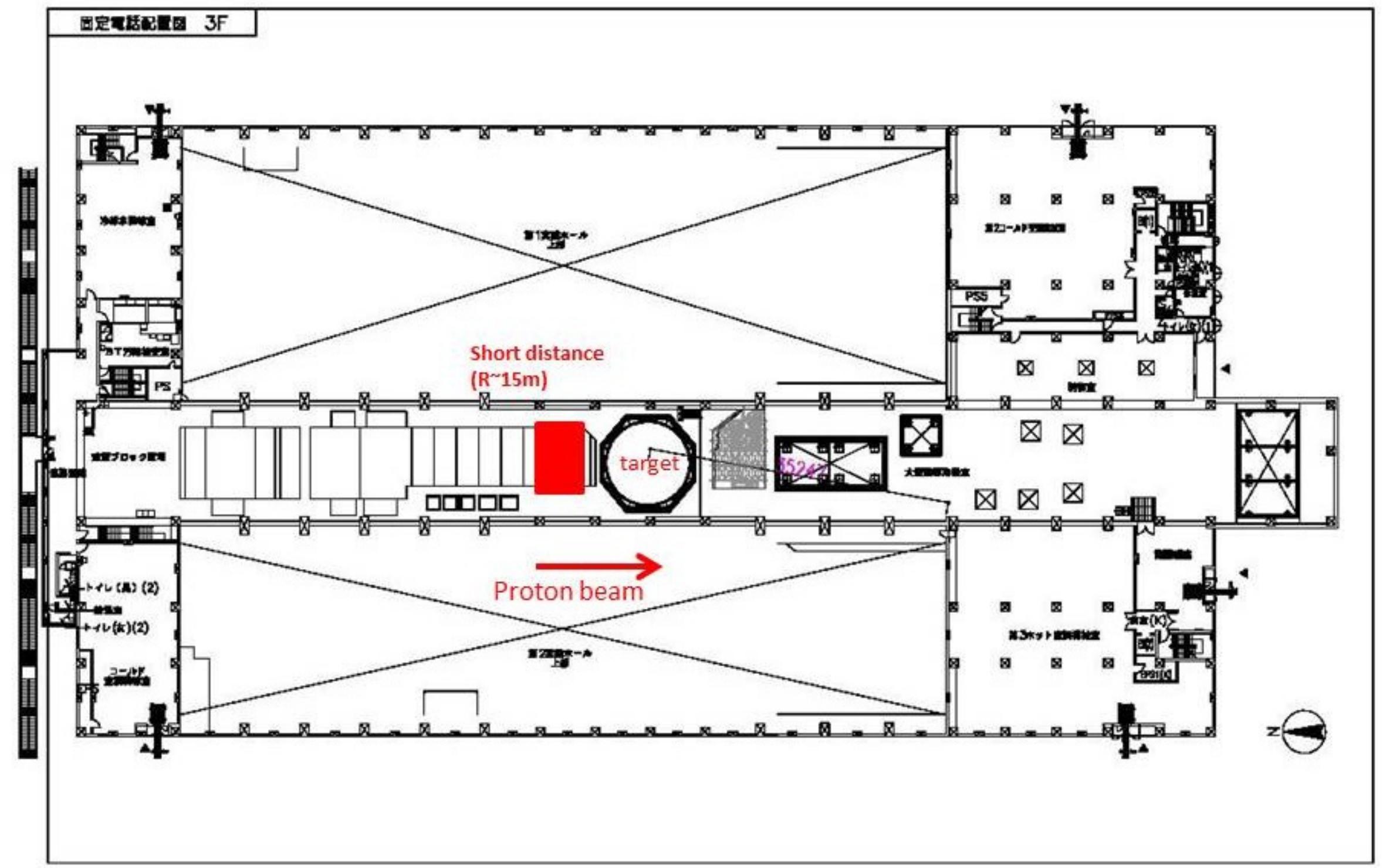}
 \caption{\setlength{\baselineskip}{4mm}
 Drawings of the 3rd floor of the MLF building (top view). Candidate site of the 50 ton detector is shown in the red box.   
 }
 \label{fig:location}
\end{figure}

\begin{figure}
 \centering
 \includegraphics[width=0.8\textwidth]{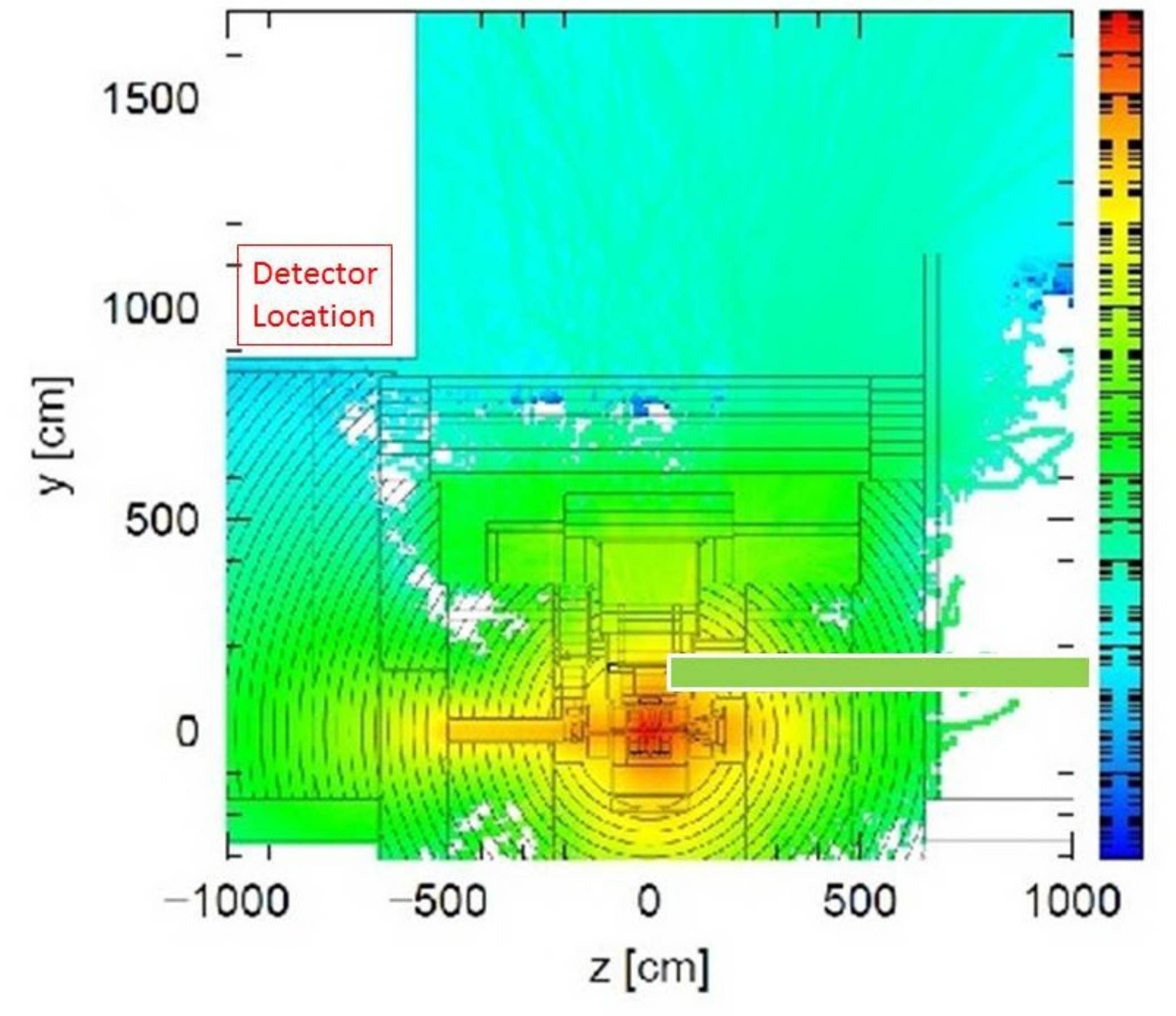}
 \caption{\setlength{\baselineskip}{4mm}
 Detector location of the 3rd floor of the MLF building (side view). 
 Color shows the radiation level, and the red part corresponds to 
 the mercury target.    
 }
 \label{fig:location2}
\end{figure}

The detector site is the third floor of the MLF, which is used 
for the maintenance of the mercury target. 
Figures~\ref{fig:location} and \ref{fig:location2} show the top and 
side view of the detector site.
Red box in Figure~\ref{fig:location} shows the detector site at the
3rd floor. It is put at 13 m above and 11 m upstream of the target.  

For the constraints from the entrance of the MLF building, two 
detectors are put in the red box area. A design of one detector
is shown in the next subsection.   

%%%%%%%%%%%%%%%%%%%%
\subsection{Detector structure}
%%%%%%%%%%%%%%%%%%%%

\begin{figure}[htbp]
   \begin{center}
    \includegraphics[keepaspectratio=true,height=90mm]{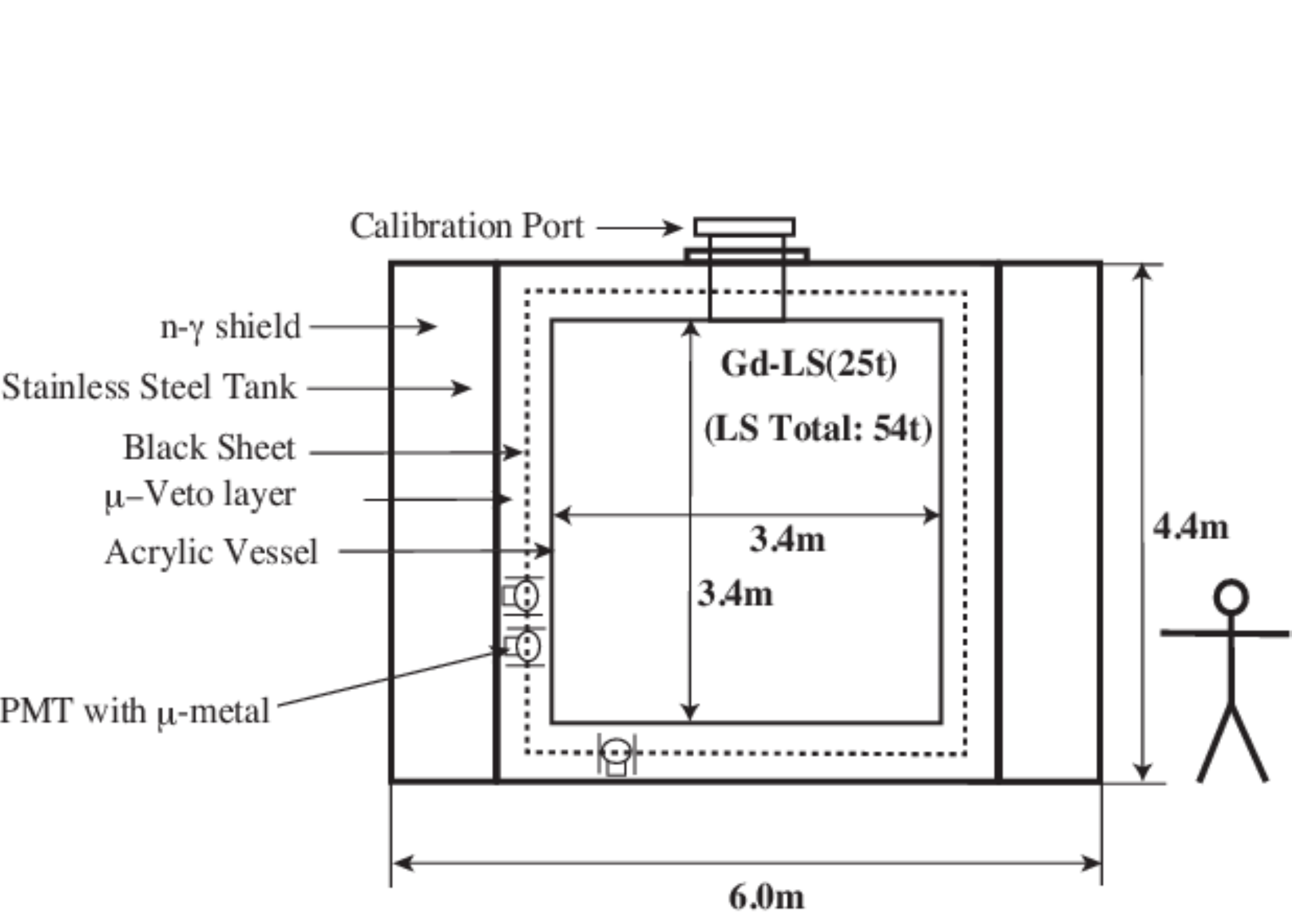}
    \caption{ \setlength{\baselineskip}{4mm}
Schematic drawing of the neutrino detector. (Note that the same two
detectors are put at the site, which have a total fiducial mass of 50 t consisting of Gd-loaded liquid scintillator.)
    }
    \label{fig:LSND_Type_Detector}
    \end{center}
\end{figure}

Figure~\ref{fig:LSND_Type_Detector} shows a schematic drawing 
of one of two proposed neutrino detectors.

 The outer tank is a stainless steel cylindrical tank with diameter of 4.4~m, height of 4.4~m and volume of 70~m$^3$.
 In the tank, there is an acrylic vessel of diameter of 3.4~m, height of 3.4~m and volume of 31~m$^3.$
  The acrylic vessel is filled with 25~tons of a gadolinium loaded liquid scintillator (Gd-LS) described in section \ref{sec:GdLS}.
  Signal is defined by the delayed Gd signal.
  In this way the detection efficiency of the prompt signal is independent of the energy.
 There is a calibration port on the top of the acrylic vessel which allows access of calibration devices from top of the detector.
 The region between the stainless steel tank and the acrylic vessel is filled with
 36~m$^3$ of gadolinium {\it unloaded} liquid scintillator.
 The scintillation yield of the LS and the specific gravity are matched to those of
 Gd-LS.
 This LS catches $\gamma$-rays which escape from the Gd-LS region and reproduces original
 neutron absorption energy.
  150 10-inch diameter PMTs (section \ref{sec:PMT}) are arranged such a way that their equator of the glass is 25~cm from the stainless steel wall.
   Black sheets equipped between PMTs optically separate the liquid scintillator layer into two.
 The LS can move freely through openings of the black sheets and there is no differential pressure between the two LS regions.
 The outer region is viewed by additional 5-inch PMTs and used as cosmic-ray anti counter.

 There are electronics racks near the tank.
  Most of the data processing is performed near the tank.
  Only a few network and power cables connect between the detector-electronics system and outer laboratory. 
The detector-electronics system can be moved by a crane at once after disconnecting such cables.

 The weight of the stainless steel tank is 5~tons.
 The weight of the liquid is 54~tons and electronics is 1~tons.
 The total weight of the detector is 60~tons and the weight per unit footprint
 is 4~ton/m$^2$.

   The whole detector structure is surrounded by neutron and $\gamma$-ray shields which is made of iron slab and boron loaded paraffin blocks or additional 
water tank or LS layer.
 These shields can be dismantled quickly.
 The detector has to be transferred to a different place at least once per year to make room for inspection of the MLF beam lines.
In such cases the radiation shield will be disassembled and the detector and electronics system will be moved by the crane of the lab.
  In case emergency access to the beamlines is necessary, the detector has to be quickly made ready for the transfer.

%=============================
\subsection{Liquid scintillator \cite{GdLS} \label{sec:GdLS}}
%-----------------

The detection of electron antineutrinos via the inverse beta-decay reaction is one of the main signals.
Various organic liquid scintillators (LS) have often been used as the detector medium because they produce relatively large numbers of photons at low energies of a few MeV.
The antineutrino signal is a delayed coincidence between the prompt positron and the capture of the neutron in an (n,$\gamma$) reaction after it has been thermalized in the LS.
This delayed coincidence tag serves as a powerful tool to reduce random backgrounds.

The neutron capture can occur on gadolinium, Gd.
The (n,$\gamma$) cross-section for natural Gd is high, 49,000 barns.
Because of this high cross section, only a small concentration of Gd, (0.1 $\sim $ 0.2 \% by weight), is needed in the LS.
The neutron-capture reaction by Gd releases a sum of 8-MeV energy in a cascade of 3-4
$\gamma$-rays.
The higher total energy release of the $\gamma$-rays and their enhanced isotropy help to exclude low-energy backgrounds from other sources, such as radioactive decay in the surrounding environment and materials.
The time delay for the neutron-capture is  $\sim$27$\mu$s in 0.1 \% Gd, and
this short delay time helps to reduce the accidental background rate significantly.

A multi-ton scintillation detector for an antineutrino oscillation experiment must satisfy a number of stringent requirements: The Gd-LS must be chemically stable 
which means no formation over time of any components
in the liquid that will absorb or scatter light, or change the concentration.
The Gd-LS must be optically transparent, have high light output, and pose low intrinsic radioactive background.
The Gd-LS must also be chemically compatible with the containment vessel.
For example, it is known that certain steels could leach metallic impurities into scintillator.
Several organic scintillation solvents were studied to test their feasibilities for the above-mentioned criteria.
 Linear alkylbenzene, LAB, first identified as a scintillation liquid from the
SNO+ R \& D, is composed of a linear alkyl chain of 10-13 carbons attached to a benzene ring;
it is commercially used primarily for the industrial production of biodegradable synthetic detergents.
LAB has a light yield comparable to that of PC and a high flash point, which significantly reduces the safety concerns.
These notable characteristics make it suitable for a large-scale neutrino experiment.
Current ongoing or proposed experiments for reactor electron antineutrinos, Daya Bay and RENO; double-beta decay, SNO+; and solar neutrinos, LENS, unanimously select LAB as their primary scintillation liquid.
Similarly, this proposed experiment will use LAB as the singular solvent for the Gd-loaded option; which has advantage of stability, high light-yield, and optical transmission over a binary solvent system (i.e. PC or LAB in dodecane or mineral oil).

It is commonly known in the community that the quality of Gd-LS is the key to the success of the LS experiments.
There is heightened concern for a new long-duration oscillation experiment such as we propose here.
The BNL neutrino and nuclear chemistry group has been involved in R\&D of chemical techniques for synthesizing metal-loaded organic liquid scintillators since 2000 and is currently a member of several liquid scintillator experiments.
A highly stable 0.1\% Gd-LS with attenuation length of $\sim$ 20m and $\sim$ 10,000 optical photons/MeV has been developed by the BNL group for the reactor antineutrino experiments. Indeed, Daya Bay has published its first observation of non-zero $\theta_{13}$ in 2012, based on the successful detection of the IBD reaction by Gd-loaded liquid scintillator.

%=============================
\subsection{PMT \cite{PMT} \label{sec:PMT}}
%----------------
150 low background 10 inch PMTs will be used to detect the scintillator photons produced by IBD reaction.
These kinds of PMTs have been used in paraffin oil at Double Chooz and RENO neutrino detectors.
\begin{figure}[htbp]
   \begin{center}
    \includegraphics[keepaspectratio=true,height=55mm]{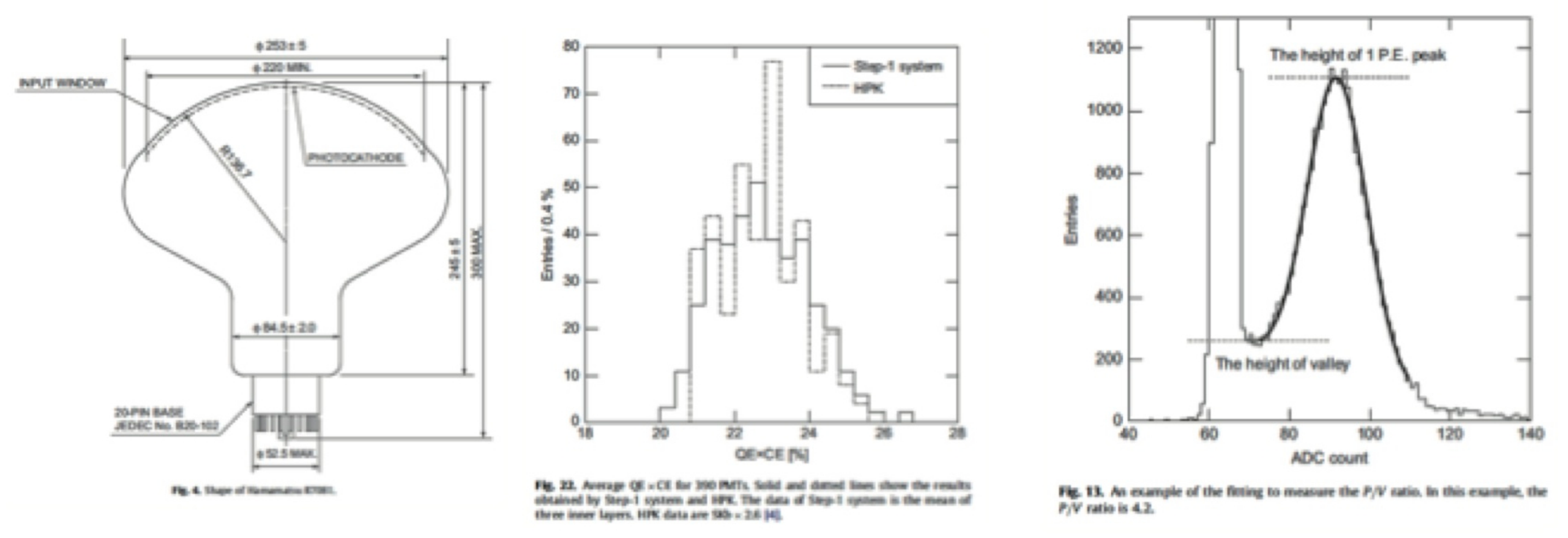}
    \caption{\setlength{\baselineskip}{4mm}
		HPK PMT properties.
    Left: Dimension.
    ~~Center: Quantum efficiency $\times$ collection efficiency.
    Right: Peak to Valley ratio for one photoelectron signal. }
    \label{fig:PMT}
    \end{center}
\end{figure}
For example, the PMT used in Double Chooz and RENO experiment is HPK-R7801MOD-ASSY which has the following properties.
The quantum efficiency times collection efficiency is 23\%.
TTS is 2.9~ns FWHM, dark rate for 1/4~p.e. threshold is 4~kHz (typical) and 
8~kHz (Max).
The peak to valley ratio is 2.8 and clear one photoelectron peak can be detected.
$10^7$ gain is obtained with 1500~V of high voltage and the power consumption is 0.2~W with
this voltage.
The PMT is operated with positive HV.
The base part of the PMT is molded by transparent epoxy to prevent the electric circuit material from directly touching the oil.
A Teflon jacket cable (RG303/U) is used to readout signal and supply HV.
There is a splitter circuit between HV and PMT to separate HV and signals.
There is a 50~$\Omega$ resister at the end of the signal cable to quickly dump the tail of large signals.
The PMT will be equipped with $\mu$-metal shield to reduce the effect of the earth magnetic field. 
The PMT base is thermally contacted to the stainless steel tank wall to dissipate the generated heat efficiently. 

The energy of neutron signal is 8~MeV and we require the energy resolution of
\begin{equation} 
 \frac{\delta E}{E} < \frac{15\%}{\sqrt{E({\rm MeV})}}.
\end{equation}

 4\% of photo coverage is necessary to obtain this energy resolution assuming there is 50\% of scintillation light inefficiency from various reasons, 
which corresponds to 150~PMTs. 
The glass is a low background type.
Low background sands are chosen for the glass material and they are melted in a platinum coated furnace to avoid contamination of radioactivities from the crust of the furnace wall.

The radioactive elements in the glass are U:13~ppb, Th:61~ppb and $^{40}$K:3.3~ppb and expected $\gamma$-ray rate ($E>1$~MeV) from the PMT glass is $\sim$ 400~Hz. 
 In addition to the PMTs for neutrino target, 50 additional 5~inch PMTs will be used for muon anti counter.

%=============================
\subsection{Liquid Operations}
%---------------------------
 %==========================
 \subsubsection{LS Filling Operation}
 %--------------------------
 The liquid scintillators will be delivered by using teflon coated iso-containers and
 lorry tracks from manufacturer.
 There will be temporary liquid handling hut with storage tanks and pumps at the unloading area of the MLF area.
 The delivered liquids are once stored in the storage tanks and then sent to the detector and carefully put in the detector by equalizing the liquid levels between inside and outside of the acrylic vessel.
  After the filling is complete, the oil filling pipes are disconnected from the detector to isolate it for the transfer.

%=========================
 \subsubsection{Detector Operation during data taking }
%--------------------------
 It is not necessary to circulate the liquid during data taking operation.
  So once the detector is filled by the liquids, it will be sealed. 
  On the other hand, the gas phase is filled by nitrogen from tanks.
 A small amount of nitrogen gas will be continuously supplied to the detector, which equalizes the differential pressure passively keeping up the changes of the atmospheric pressure.

%%%%%%%%%%%%%%%%%%%
\subsection{Electronics}
%%%%%%%%%%%%%%%%%%%
~~
\noindent

The requirements for the electronics are listed below:
\begin{enumerate}
\item The prompt signal must be recorded without dead time for 10 $\mu$s from the beam injection.
\item The delayed signal must be recorded with a time stamp for $\sim40$ ms from the beam injection.
\item The sampling rate must be high enough to achieve the timing resolution of 1 ns.
\item The dynamic range must cover from 1 MeV to 50 MeV with a resolution of 12 bits or more.
\end{enumerate}

\begin{figure}[htbp]
   \begin{center}
    \includegraphics[keepaspectratio=true,width=130mm]{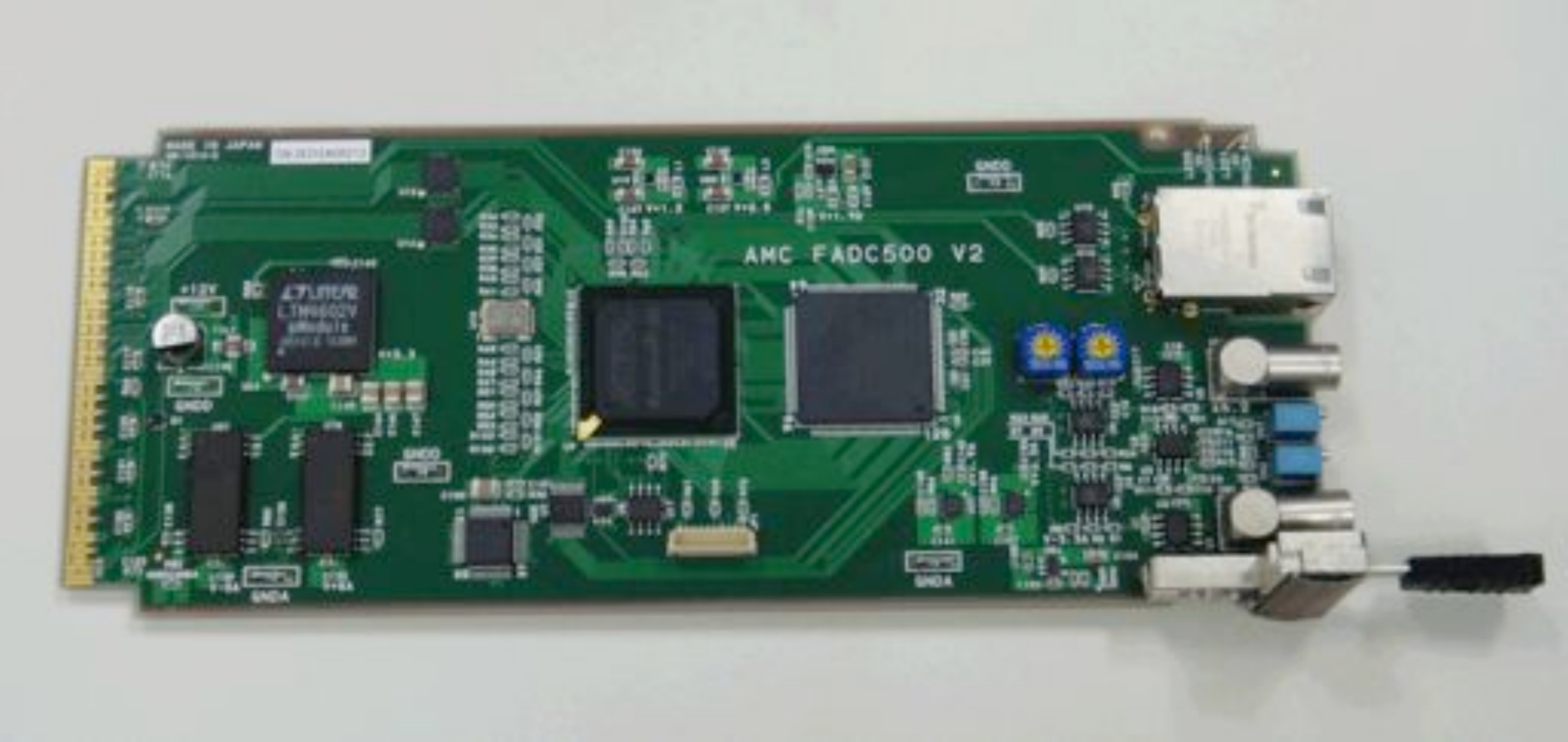}
    \caption{ \setlength{\baselineskip}{4mm}500-MHz flash ADC module used for the present CANDLES experiment.}
    \label{fig:fadc}
    \end{center}
\end{figure}

\begin{figure}[htbp]
   \begin{center}
    \includegraphics[keepaspectratio=true,width=130mm]{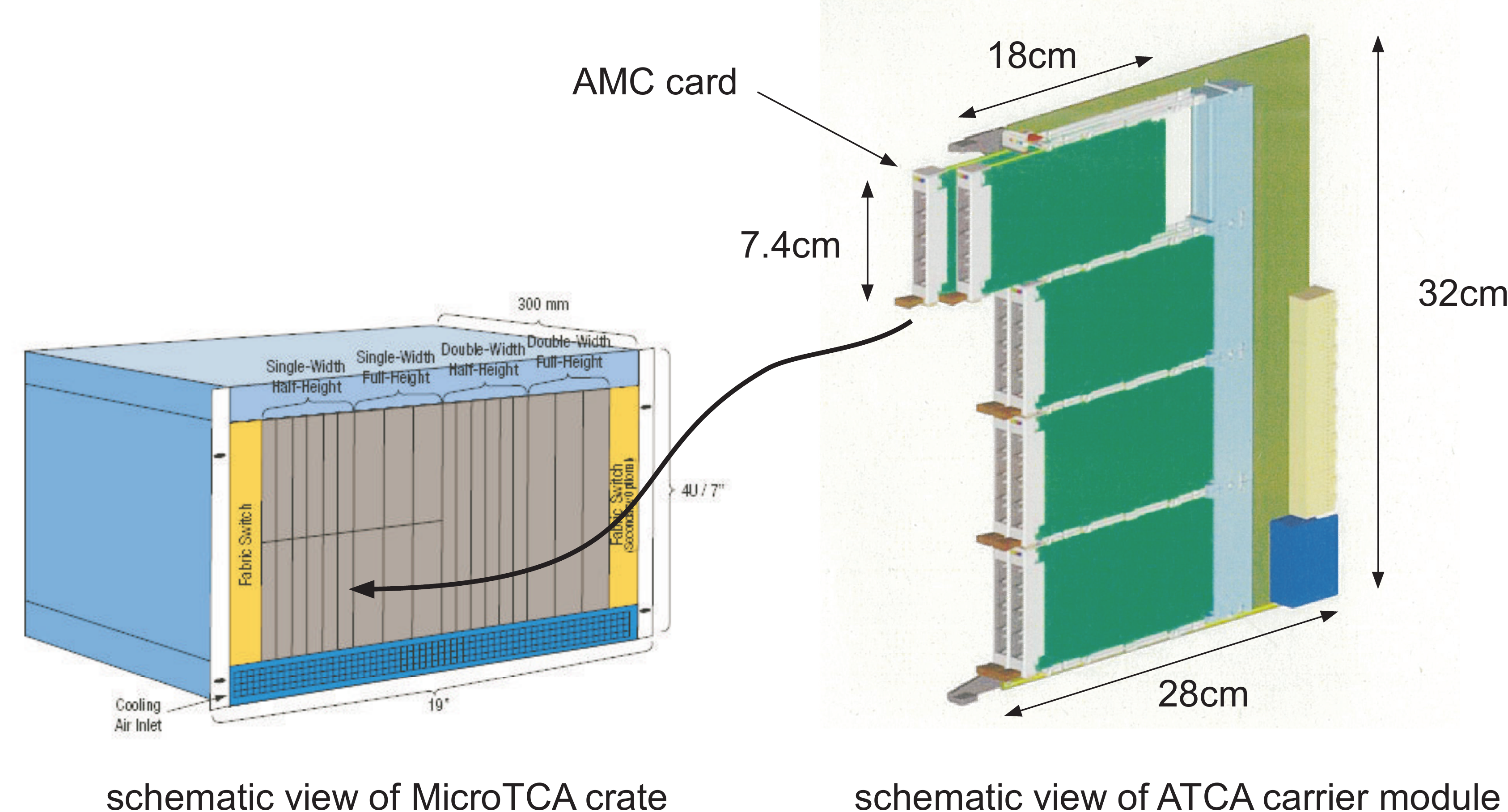}
    \caption{ \setlength{\baselineskip}{4mm}Schematic view of Micro TCA crate (left) and ATCA carrier module  (right).}
    \label{fig:amc}
    \end{center}
\end{figure}

\begin{figure}[htbp]
   \begin{center}
    \includegraphics[keepaspectratio=true,width=130mm]{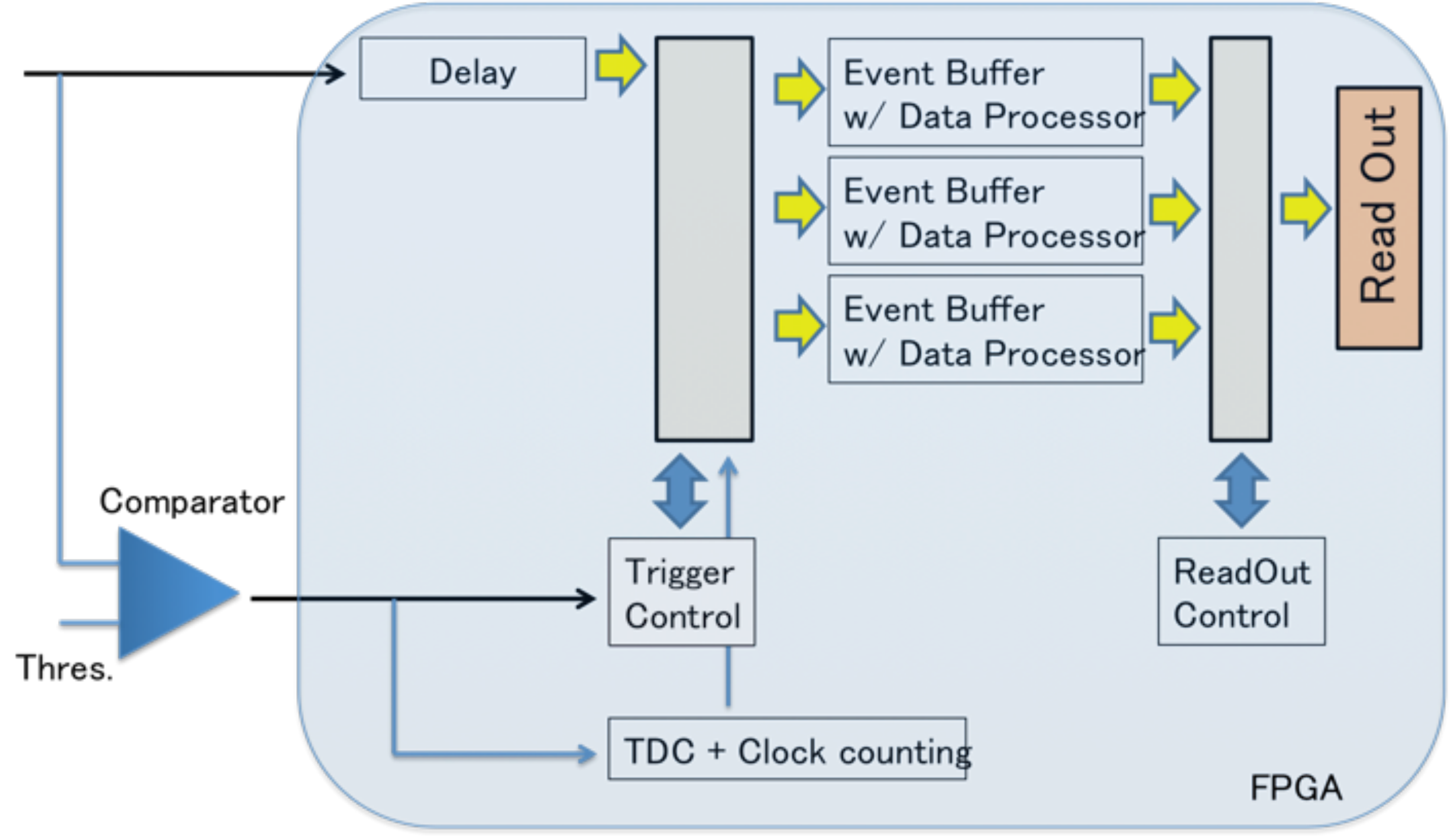}
    \caption{ \setlength{\baselineskip}{4mm}Concept of the electronics.}
    \label{fig:elec}
    \end{center}
\end{figure}

 In order to fulfill these requirements within a reasonable cost, we will modify a 500-MHz flash ADC and a Micro TCA readout system, which is designed at RCNP for the CANDLES experiment upgrade. The system is based on the existing a 500-MHz flash ADC (Figure~\ref{fig:fadc}) and an ATCA readout system~\cite{cite:noumachi} for the present CANDLES experiment~\cite{cite:umehara}. 
A Micro TCA system~\cite{cite:SpW2013} is described as the following. 
An AMC module has two 8-bit flash ADC with the 500 MHz sampling rate. 12 AMC modules are installed on a micro TCA sub-rack. Thus, one sub-rack accepts 24 analog signals. Each signal from the detector will be divided into low and high gain analog signals with a splitter at the input of the flash ADC. One AMC module handles one pair of signals. A total of 400 AMC modules will be used for the experiment. They are held on 34 sub-racks. Or using ATCA carrier, which holds 8 AMC modules, 5 ATCA sub-racks holds up to 480 AMC modules as shown in Figure~\ref{fig:amc}.
The dead-time-less signal processing for the trigger and the data readout is realized by pipeline data processing using a FPGA that implements delays, a trigger control, a clock, event buffers with a data processor and a read-out control (Figure~\ref{fig:elec}).
The concept of the pipeline processing is already applied to the CANDLES experiment. Performance of a prototype of the new flash ADC will be checked at RCNP in December 2013. 

Serial data link is used for the data collection from each FPGA chip. 
On board connections, backplane connections and inter-sub-rack connections 
use the same data link. The data link is extended by router chip of the data link. Therefore the system is highly scalable from small system to large system. Maximum data rate from each AMC module is 16 MB/s at the maximum. Collected data are read through several Gigabit Ethernet links by several PC. The number of PC and Ethernet link depends on the demands of the data rate. 

%%%%%%%%%%%%%%%%%%%%%%%%%%%%%
\subsection{Expected performance}
%%%%%%%%%%%%%%%%%%%%%%%%%%%%

%%%%%%%%%%%%%%%%%%%
\subsubsection{Vertex reconstruction and the position resolution}
\begin{figure}[htbp]
 \begin{center}
 \includegraphics[width=12cm]{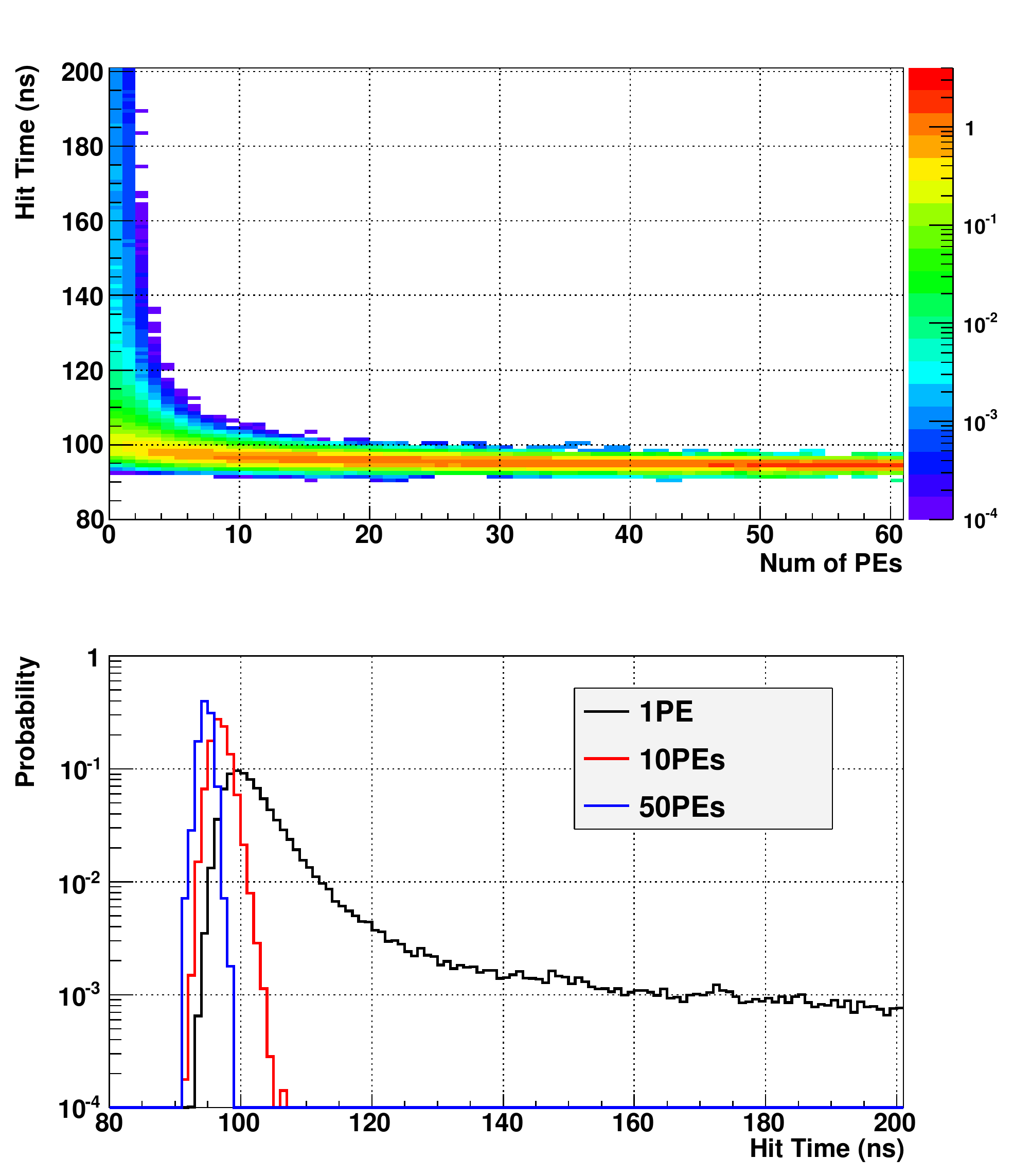}
 \end{center}
 \caption{\setlength{\baselineskip}{4mm}
 Hit timing delay as a function of the number of photo-electron received by PMTs. It is used for the probability density distribution for maximum likelihood method for the vertex reconstruction. Upper figure shows the probability density distribution. Lower figure shows the typical number of photo-electron dependence of the hit time. Black, red and blue lines show 1 photoelectron, 10 photoelectrons and 50 photoelectrons cases, respectively.}
 \label{VTXPDF}
\end{figure}

%%%%%%%%%%%%%%%%%%%
\indent
Vertex of an event is reconstructed by using hit times from the buffer PMTs. The hit time for each PMT is defined as time difference by when the pulse height exceeds 1 mV of the threshold from the trigger time. The hit time depends on time constant of light emissions in the liquid scintillator, time of flight of the 
light between the vertex and surface of each PMT, time shift due to time difference until when a signal is produced from when the event occurs, and 
number of photoelectrons hitting each PMT. So as free parameters in fit function for the vertex reconstruction, the vertex position (x,y,z) and the base shift (T0) are used. Here, corrected hit time of i$_{th}$ PMT (CorT$_{i}$)  is defined as follows.

\begin{equation}
CorT_{i}=HitT_{i}-T0-\frac{|\vec{P}_{VTX}-\vec{P}_{i}|}{C_{n}},
\end{equation}
where, $\vec{P}_{VTX}$ and $\vec{P}_{i}$ are the vertex and i$_{th}$ PMT positions, respectively, C$_{n}$ is light velocity in the liquid scintillator.
Upper figure in Figure~\ref{VTXPDF} shows probability density distribution($f_{pdf}$) of the corrected hit times depending on the number of photoelectrons($N^{PE}_{i}$), estimated by using MC samples of thermal neutron capture on Gd at various positions in the buffer tank. Lower figure shows the examples depending on number of photoelectrons. When the number of photoelectrons hitting 
a PMT is a few, the corrected hit times distribute following the time constant of light 
emission in the scintillator. When the number of photoelectrons increases, the time constant of the light emission does not affect the distribution of the corrected hit times. Then the distribution is close to Gaussian depending on TTS of the PMT. 
For the vertex reconstruction, maximum likelihood method with the probability density distribution. The likelihood($\mathcal{L}$) is defined as follows.

\begin{equation}
\mathcal{L}=\prod_{i}{f_{pdf}(CorT_{i}, N^{PE}_{i})},
\end{equation}
then, position with minimum value of $-ln\mathcal{L}$ is the reconstructed vertex. 

\begin{figure}[htbp]
 \begin{center}
 \includegraphics[width=16cm]{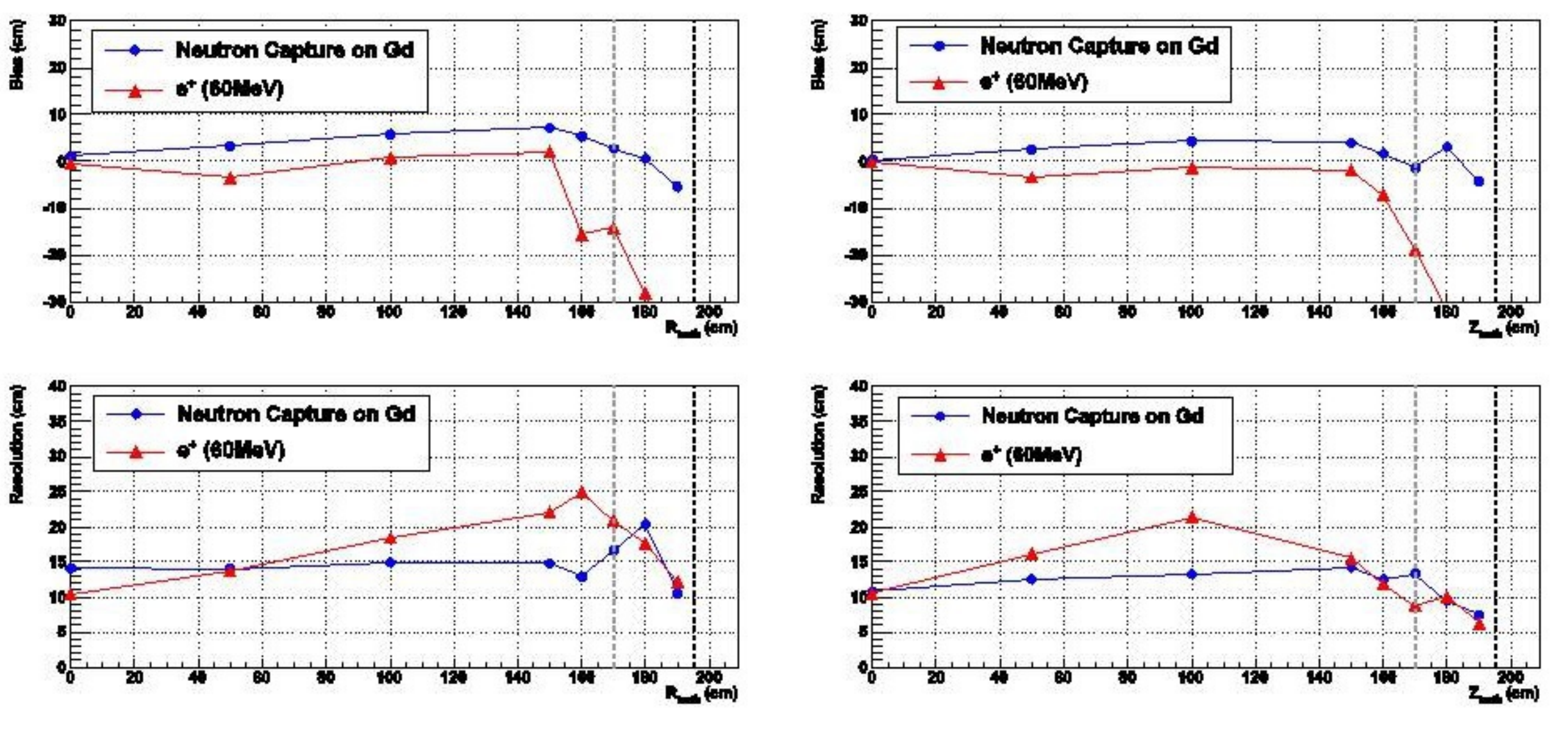}
 \end{center}
 \caption{\setlength{\baselineskip}{4mm}
 Reconstructed vertex bias(upper plot) and resolution(lower plot) on R direction (left) and Z direction (right). Blue and red lines show cases of neutron capture events on Gd (The true energy is 8 MeV) and positron events with 60MeV of the kinetic energy (The true energy is 61.022MeV due to annihilation gammas with electron), respectively. Gray and black lines show target and buffer walls, respectively.}
 \label{EVRVTX}
\end{figure}

Figure~\ref{EVRVTX} 
%and Fig.~\ref{EVZVTX}  
shows evaluations of the reconstructed position biases and resolutions in cases of thermal neutron capture on Gd (total 8 MeV gammas) and 60 MeV electron 
considering the e$^{+}$ energy of the IBD events for the $\overline{\nu}_{e}$ signals at various positions on R direction and Z axis in the cylindrical detector. The reconstructed position bias is estimated as mean value of a peak of distance distribution between true vertex position and the reconstructed one, and the reconstructed position resolution is estimated as the RMS. The reconstructed bias and position resolution in energy range of the IBD events for the $\overline{\nu}_{e}$ signals and whole target volume, are less than
$\pm$15 cm and 25cm, respectively.

%%%%%%%%%%%%%%%%%%%
\subsubsection{Energy reconstruction and the resolution}
%%%%%%%%%%%%%%%%%%%
\indent
Number of photoelectrons of i$_{th}$ PMT ($N^{PE}_{i}$) is calculated as follows considering attenuation length of the scintillation light ($L_{Att} \sim$ 10m) and solid angle to the PMT from the vertex position.
\begin{equation}
N^{PE}_{i}=N_{tot}\times \frac{S_{10inch}^{PMT}cos\theta_{i}}{4\pi L_{i}^{2}} \times exp(-\frac{L_{i}}{L_{Att}}),
\end{equation}
where $N_{tot}$ is total number of scintillation photons emitted at the vertex, $S_{10inch}^{PMT}$ is surface area of photocathode of 10~inch PMTs, $\theta_{i}$ is angle between the surface and incident direction of the light to the PMT, and $L_{i}$ is base line between the vertex and the PMT. Energy of the event (E) has a relation of $N_{tot}\propto E$, and  the energy is reconstructed as follows.
\begin{equation}
E=\frac{N_{tot}}{\alpha }=\frac{\sum N^{PE}_{i}/\alpha}{ \sum \left( \frac{ S_{10inch}^{PMT}cos\theta_{i}}{4\pi L_{i}^{2}} \times exp(-\frac{L_{i}}{L_{Att}}) \right) },
\end{equation}
where, $\alpha$ is number of scintillation photons emitted at the vertex per 1 MeV, and $N_{tot}$ is sum of numbers of photoelectrons of all PMTs. Then, the $\alpha$ is calculated by fitting with Gaussians to peaks in the $N_{tot}$ distribution of thermal neutron capture events on Gd at detector center.

\begin{figure}[htbp]
 \begin{center}
 \includegraphics[width=16cm]{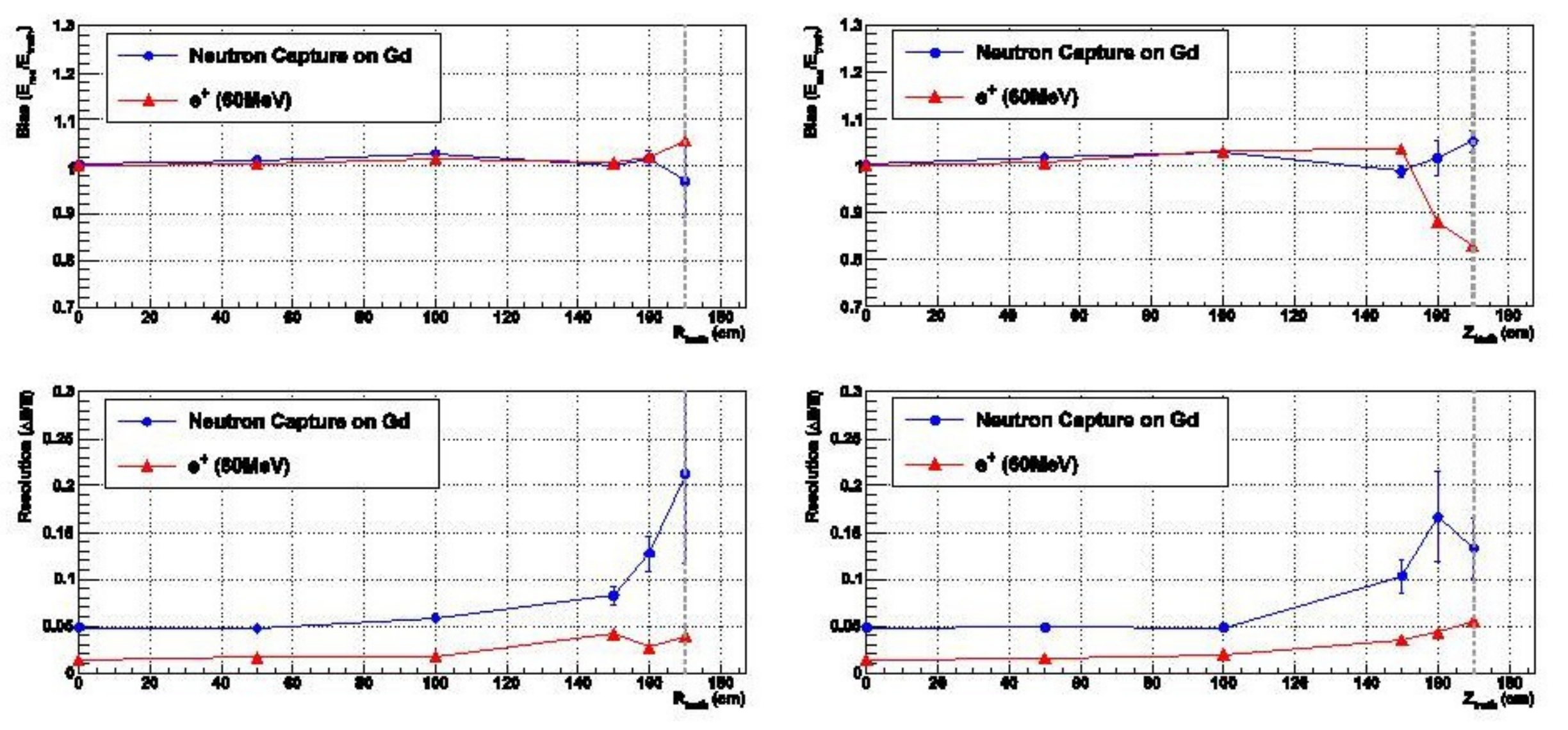}
 \end{center}
 \caption{ \setlength{\baselineskip}{4mm}
 Reconstructed energy bias(upper plot) and resolution(lower plot) on R direction (left) and Z direction (right). Blue and red lines show cases of neutron capture events on Gd (The true energy is 8 MeV) and positron events with 60MeV of the kinetic energy (The true energy is 61.022MeV due to annihilation gammas with electron), respectively. Gray and black lines show target and buffer walls, respectively.}
 \label{EVRE}
\end{figure}

Figure~\ref{EVRE} 
shows evaluations of the reconstructed energy biases and resolutions in cases of thermal neutron capture on Gd (total 8 MeV gammas) and $e^{+}$ with 60 MeV of 
the kinetic energy considering the e$^{+}$ energy of the IBD events for the 
$\overline{\nu}_{e}$ signals at various positions on R direction and Z axis in the cylindrical detector. The reconstructed energy bias is estimated as a ratio of true energy and mean value calculated by fitting with Gaussian to a peak of the reconstructed energy spectrum, and the reconstructed energy resolution is estimated as the sigma. The reconstructed bias in energy range of the IBD events for the $\overline{\nu}_{e}$ signals and whole target volume, are less than
5\% except for near the target wall.

\subsection{Neutron Selfshielding}
\begin{figure}[htbp]
 \begin{center}
 \includegraphics[width=11cm]{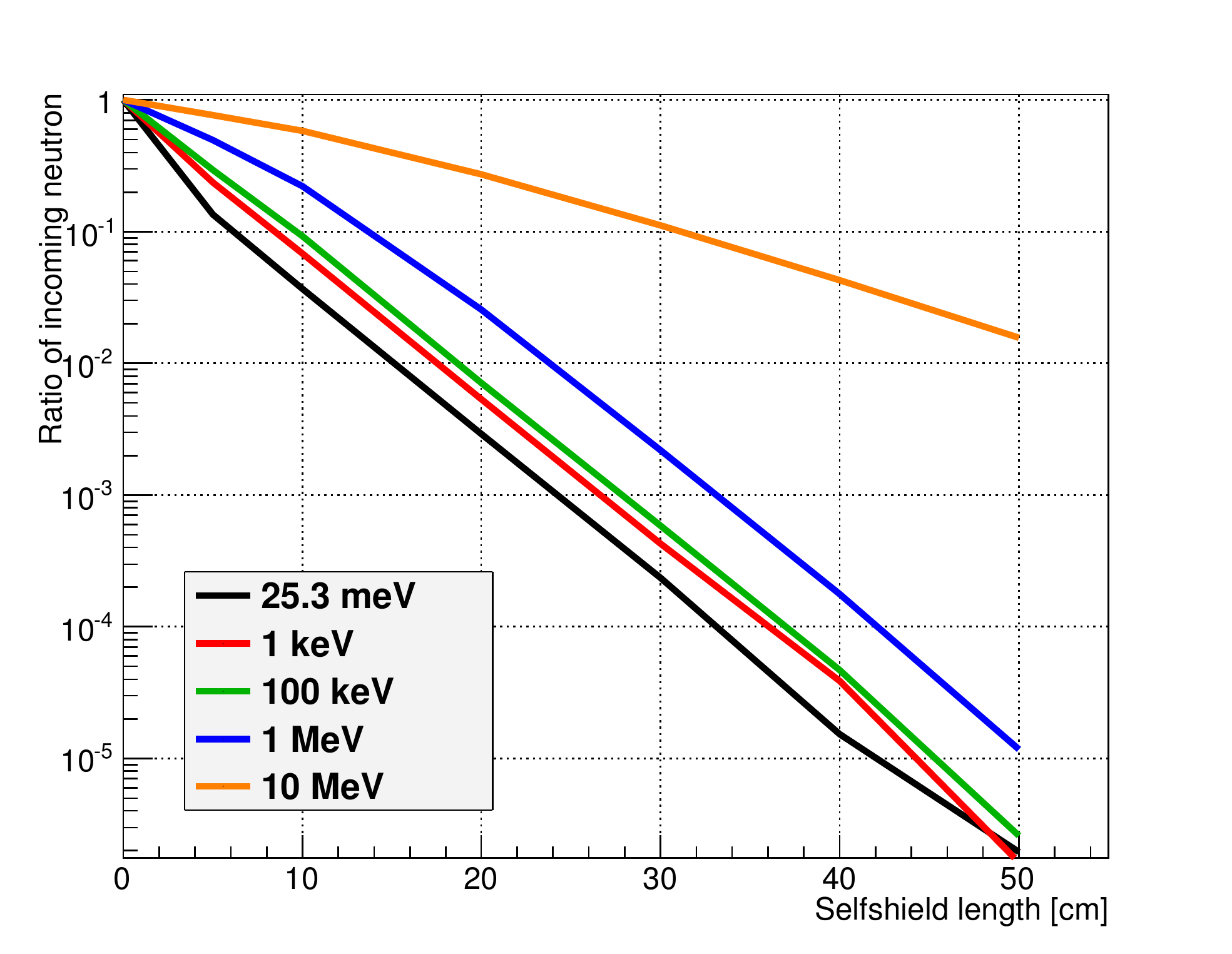}
 \end{center}
 \caption{\setlength{\baselineskip}{4mm}
Selfshield effect of the liquid scintillator as a function of the thickness on neutrons estimated with Geant4.}
 \label{self_shield_n}
\end{figure}

The detectors have selfshield regions with the 50-cm thick LS.
We estimated selfshield effect on neutrons with various energies using Geant4.
Figure~\ref{self_shield_n} shows ratio of neutrons which penetrate into the detector to incoming neutrons at the detector surface.
Neutrons below 100 keV of kinetic energy are reduced 5 orders of magnitude before they reach the surface of the acrylic tank.
On the other hand, the selfshield effect on neutrons above 1 MeV of kinetic 
energy is smaller than that of below 100 keV because the cross section on 
hydrogen 
of neutron elastic scattering decreases as kinetic energy of neutron increases.

\subsection{Particle Identification}

\begin{figure}[htbp]
\centering
\includegraphics[scale=0.650,angle=0]{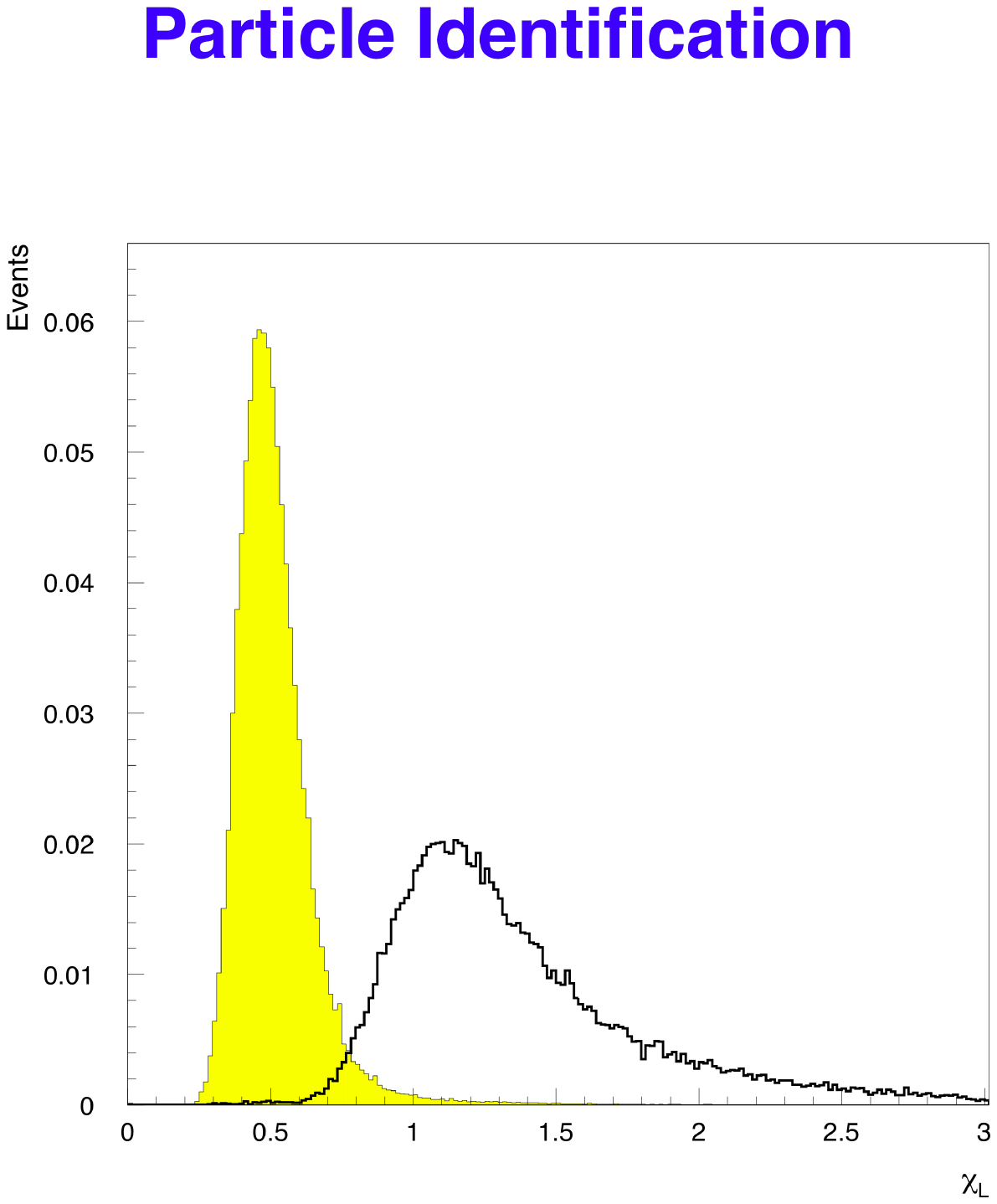}
\caption{\setlength{\baselineskip}{4mm}
The separation of low-energy electrons from low-energy
neutrons in the LSND experiment. The low-energy electrons come from stopped 
muon decay (shaded yellow), while the low-energy neutrons are cosmic-ray induced (blank).}
\label{PID}
\end{figure}

As shown in Figure \ref{PID}, the LSND experiment was able to reject 
low-energy ($< 100$ MeV) neutrons relative to electrons by a factor 
of more than 100. This rejection factor was obtained by combining the fit 
to the Cherenkov cone with the fraction of hit phototubes that were 
late ($> 10$ ns) relative to the fitted event time. The most powerful
particle identification parameter was the fraction of hit phototubes
that were late, due to the fact that low-energy electrons produce 
both prompt Cherenkov light and delayed scintillation light, while 
low-energy neutrons only produce scintillation light with time constants 
of $\sim 2$ ns and $\sim 20$ ns. The LSND mineral
oil was doped with a low concentration (0.031 g/l) of b-PBD, so that
electrons produced approximately 10 photoelectrons per MeV of Cherenkov
light and 20 photoelectrons per MeV of scintillation light at the
LSND case, while
neutrons produced only scintillation light. 

Improved neutron rejection factors are possible in the future with
better electronics and with longer time constant scintillation fluors. 
The LSND electronics only recorded the time of the first photoelectron
on a given phototube. However, better electronics could record the
time of all photoelectrons, which would increase the photoelectron 
statistics and improve the separation between electrons and neutrons. 
In addition, the use of longer time constant scintillation fluors
would further improve the identification of Cherenkov light and 
scintillation light and the rejection of neutrons. 

In the MLF experiment, 
the expected numbers of photoelectrons (pe) from Cherenkov and 
scintillation
light for a 30 MeV electron are approximately 100 pe and 1000 pe,
respectively. The prompt Cherenkov light can be distinguished from 
the
delayed scintillation light with timing, as the time constant for 
scintillation light is
$>2$ ns. Cherenkov light is very important both for particle 
identification (30 MeV neutrons, for example, produce no Cherenkov 
light) and for angular reconstruction.

\pdfoutput=1
%%%%%%%%%%%%%%%%%%%%%%%%%%%%%%
\section{Event selection}
%%%%%%%%%%%%%%%%%%%%%%%%%%%%%%
%%%%%%%%%%%%%%%%%%%%%%%%

\subsection{Summary of this section}
\indent

In this section, cuts for the IBD event selection and remaining 
backgrounds after
the selection and efficiency of the cuts are described. 
Table~\ref{tab:efficiency} summarizes the selected number of events for 
each event category with 50 tons fiducial mass and 1MW times 4-year 
exposure (assuming 4000 hours operation per year).

The accidental background is expected to be 
well suppressed with good shielding 
at the detector site and after the various cuts. Therefore, 
the dominant background is $\bar{\nu}_{e}$ from $\mu^{-}$. 

\begin{table}[htb]
\begin{center}
\begin{tabular}{|c|c|c|c|}\hline
&Contents&/4years/50tons&Comment\\\hline
&&&$\Delta m^2=3.0eV^{2}$,\\
Signal&$\overline{\nu}_{\mu}\to \overline{\nu}_{e}$ &811&$sin^22\theta=3.0\times10^{-3}$\\
&&&(Best $\Delta m^{2}$ for MLF exp.)\\\hline
&&&$\Delta m^2=1.2eV^{2}$,\\
&&337&$sin^22\theta=3.0\times10^{-3}$\\
&&&(Best fit values of LSND)\\\hline
&$\overline{\nu}_{e}$ from $\mu^{-}$&377& FLUKA (Table~\ref{tab:pionprod})\\
&$^{12}C(\nu_{e},e^{-})^{12}N_{g.s.}$&38& see ~\ref{NUCNGS}\\
Backgrounds&beam associated fast neutron&0.3& see ~\ref{BAFN}\\
&Cosmic ray induced fast neutron&42& see ~\ref{COSMUEVE2}\\
&Total accidental events &37& see \ref{ACCBG}\\\hline
\end{tabular}
\caption{\setlength{\baselineskip}{4mm}
	Numbers of events of the signal and backgrounds with total fiducial mass of 50 tons after applying IBD selection criteria shown in Table~\ref{SC} for 4~years measurement.
}
\label{tab:efficiency}
\end{center}
\end{table}

%%%%%%%%%%%%%%%%%%%%%%%%
\subsection{Selection criteria}
%%%%%%%%%%%%%%%%%%%%%%%
\label{NuSC}
Before applying IBD event selection, events in time window from 1~$\mu$s to 10~$\mu$s after the beam trigger are selected as the prompt signal in order to reject the on bunch events ($\Delta$t$_{prompt}$ cut). 
Variables for the IBD event selection are energies of prompt (E$_{prompt}$) and delayed (E$_{delayed}$) signals, $\Delta$t$_{delayed}$, which is time difference between the prompt and delayed signals, and $\Delta$VTX, which is distance between reconstructed positions for the prompt and delayed signals. 
Cut condition and the efficiency of $\overline{\nu}_{e}$ signals for each variable are shown in Table~\ref{SC}. This efficiency is for
oscillated signals for high $\Delta  m^{2}$ region (e.g.; 
$\Delta m^{2} >$ 100 eV$^{2}$).

Cut condition of the prompt energy is decided considering with the spectra of oscillated $\overline{\nu}_{e}$ signals and other neutrino backgrounds. 
Cut conditions of the delayed energy and $\Delta$t$_{delayed}$ are decided 
considering the total 8 MeV gammas via thermal neutron capture on Gd in the 
target volume and the 30 $\mu$s of capture time. 
The $\Delta$VTX cut is powerful for reducing the accidental background, because $\Delta$VTX of correlated events like $\overline{\nu}_{e}$ signals are
usually distributed within 1 m, while the $\Delta$VTX of accidental events are 
distributed over several meters depending on the target and buffer volumes.
Total cut efficiency of $\overline{\nu}_{e}$ from $\overline{\nu}_{\mu}$ is 48\%. Number of $\overline{\nu}_{e}$s (50~tons of total fiducial mass, 4~years measurement) in case of $\Delta$m$^2$=3.0eV$^2$ and sin$^2$2$\theta$=3.0$\times$10$^{-3}$ after applying the selection criteria, is 811~events.

\begin{table}[htb]
\begin{center}
\begin{tabular}{|c|c|} \hline
Cut Condition&Cut Efficiency \\\hline
1.0$\le \Delta$t$_{prompt}\le$10$\mu$s&74\% \\\hline
6$\le$E$_{delayed}\le$12MeV&78\% \\\hline
20$\le$E$_{prompt}\le$60MeV&92\% \\\hline
$\Delta$t$_{delayed}\le$100$\mu$s&93\% \\\hline
$\Delta$VTX$\le$60cm&96\% \\\hline\hline
Total&48\% \\\hline
\end{tabular}
\caption{\setlength{\baselineskip}{4mm}
IBD Selection criteria and efficiencies for the oscillated signals for high $\Delta m^{2}$ case.}
\label{SC}
\end{center}
\end{table}

\begin{figure}[htbp]
 \begin{center}
 \includegraphics[width=10cm]{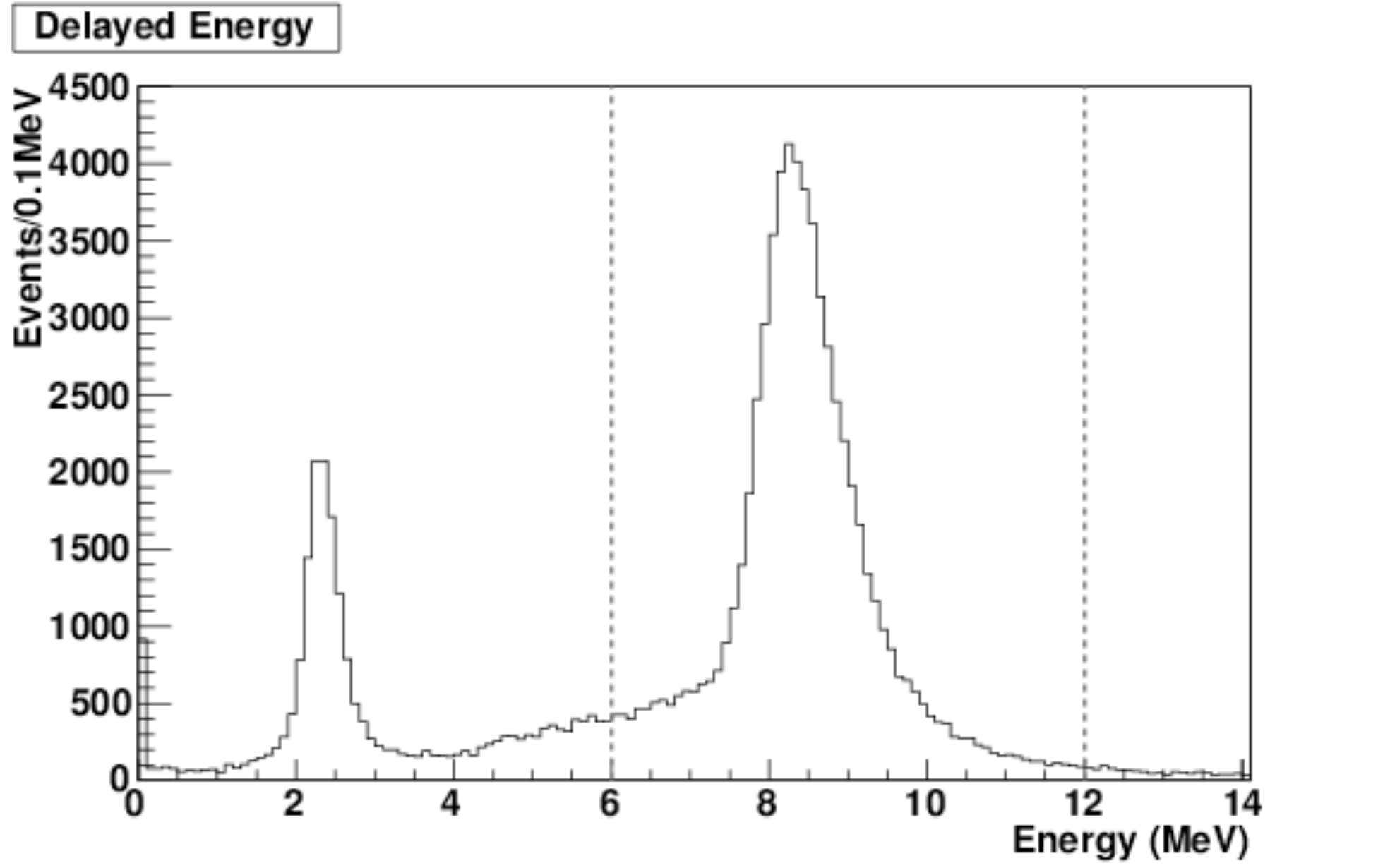}
 \end{center}
 \caption{\setlength{\baselineskip}{4mm}
 Energy spectrum of delayed IBD signal is shown. Even using Gd-LS, there 
is some amount of H capture inside the detector, which
create capture gammas with 2.2 MeV. It reduces the detection efficiency.}
 \label{Edelayed}
\end{figure}

Energy spectrum of delayed IBD signal is shown in Figure~\ref{Edelayed}. 
Even using Gd-LS, there is some amount of H capture inside the detector, which
creates capture gammas with 2.2 MeV. It reduces the detection efficiency.

%%%%%%%%%%%%%%%%%%%%%%%%%%%%%%%
\subsection{Backgrounds}
%%%%%%%%%%%%%%%%%%%%%%%%%
\label{ESBALL}
\subsubsection{Backgrounds for IBD events} 
After the IBD selections, there are remaining backgrounds in the IBD 
candidates. The backgrounds are classified into accidental and correlated backgrounds for the IBD events. 
The accidental background is induced by two independent events entering in the 
detector in $\Delta$t$_{delayed}$ for the IBD coincidence accidentally (see~\ref{ACCBG}).
The correlated background is induced by sequential signals caused by one event. 
Main sources of correlated backgrounds are beam associated fast neutrons 
accompanied by Michel electrons and thermal neutron capture on 
Gd (see \ref{BAFN}), $\overline{\nu}_{e}$ from $\mu^{-}$ (see \ref{NUMUM}), $^{12}C(\nu_{e},e^{-})^{12}N_{g.s.}$ reaction(see \ref{NUCNGS}), and cosmic muon induced fast neutron from outside of the detector(see ~\ref{COSMUEVE2}). 

In order to achieve the current design sensitivity for sterile neutrino search,
we need to reduce the total number of background events for the IBD candidates, 
except for the $\overline{\nu}_{e}$s from $\mu^{-}$, 
to several events/year per one detector, compared with several tens 
events/year/detector of $\overline{\nu}_{e}$s from $\mu^{-}$. The detail of each background is described in each subsection.

%%%%%%%%%%%%%%%%%
\subsubsection{Beam associated fast neutron}
\label{BAFN}
%%%%%%%%%%%%%%%%%
Fast neutrons are induced by beam protons interacting in the target, and
these beam neutrons contribute to the background and are classified into 
the 3 types below: 
\begin{enumerate}
\item Neutrons below several tens MeV.
\item High energy neutrons with several hundreds MeV entering in the detector. 
\item Gamma rays generated via thermal neutron capture with several $\mu$s of 
capture time outside of the detector.
\end{enumerate}

\paragraph{Neutrons below several tens MeV}

The neutron flux entering the detector at the candidate site is 41/spill per one detector (see section 4). 
The remaining beam neutron rate after applying the IBD selections for delayed signals (delayed energy between 6 and 12MeV, and $\Delta$t$_{delayed}<$100$\mu$s) is 1.9/spill. In order to reduce more neutrons, number of hit PMTs cut (NHIT cut) is very effective in addition to the IBD selections. Because most remaining events after applying the IBD selections are ones captured on H in the buffer scintillator region near PMT surfaces, the visible energy becomes larger when
comparing with the true energy, but the number of hits is not large when
comparing with events captured on Gd. Actually, the neutrino signals after applying the IBD selections are not reduced by the NHIT cut (the cut efficiency 
is above 99~\%). The neutron rate reduced by $\sim$10$^{-3}$ by applying the NHIT cut. 
Finally, the remaining neutron rate after applying both NHIT and IBD selections for delayed signals is expected to be 2.4$\times$10$^{-3}$/spill per one detector.

\paragraph{High energy neutrons with several hundred MeV entering in the detector}

\begin{figure}[htbp]
 \begin{center}
 \includegraphics[width=10cm]{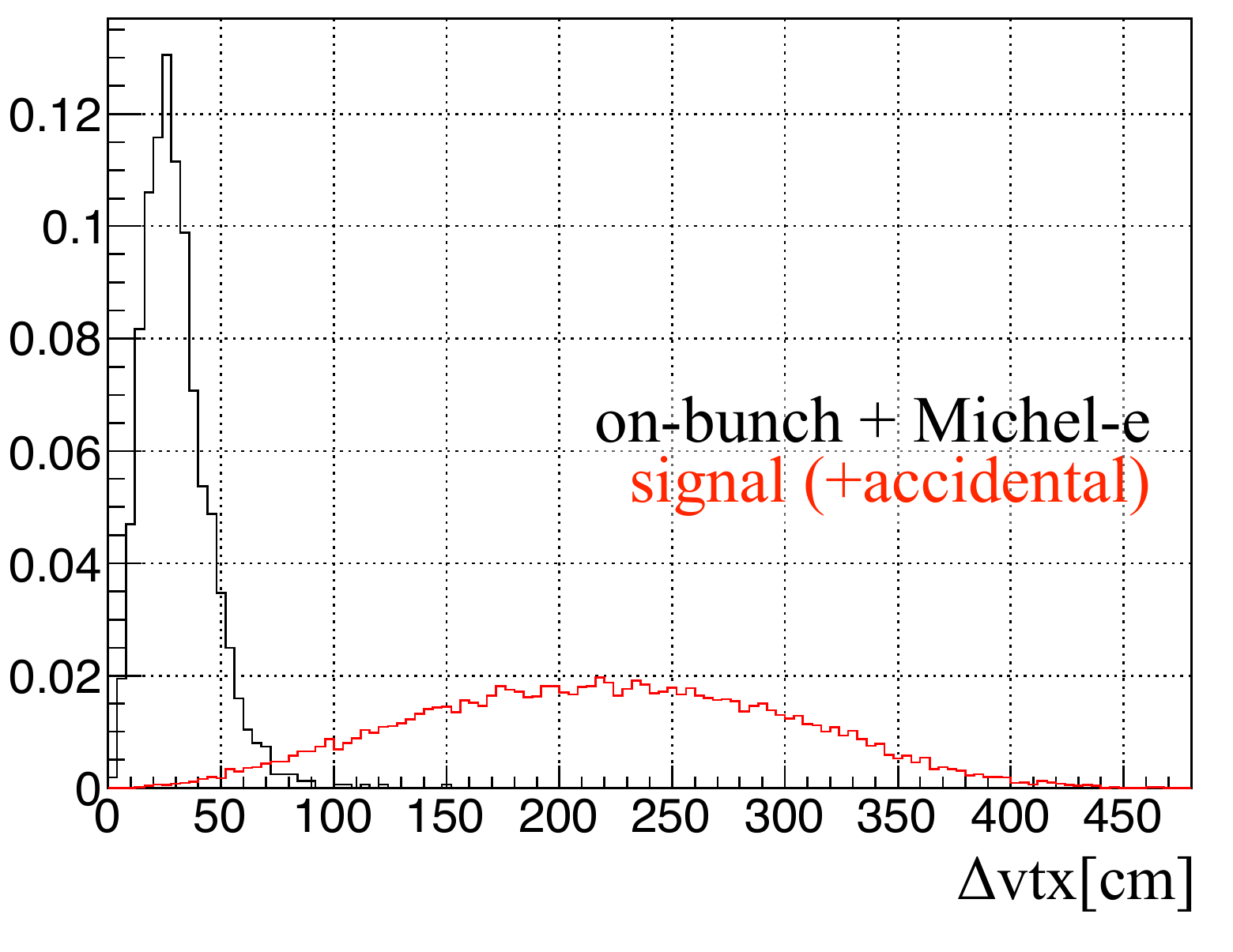}
 \end{center}
 \caption{\setlength{\baselineskip}{4mm}
 	The $\Delta$VTX distributions between Michel electrons and on-bunch signals (black line), and between the neutrino prompt signals and accidental on-bunch signals (red line), calculated by the MC simulation.}
 \label{BAMDVTX}
\end{figure}

\begin{figure}[htbp]
 \begin{center}
 \includegraphics[width=0.49\textwidth]{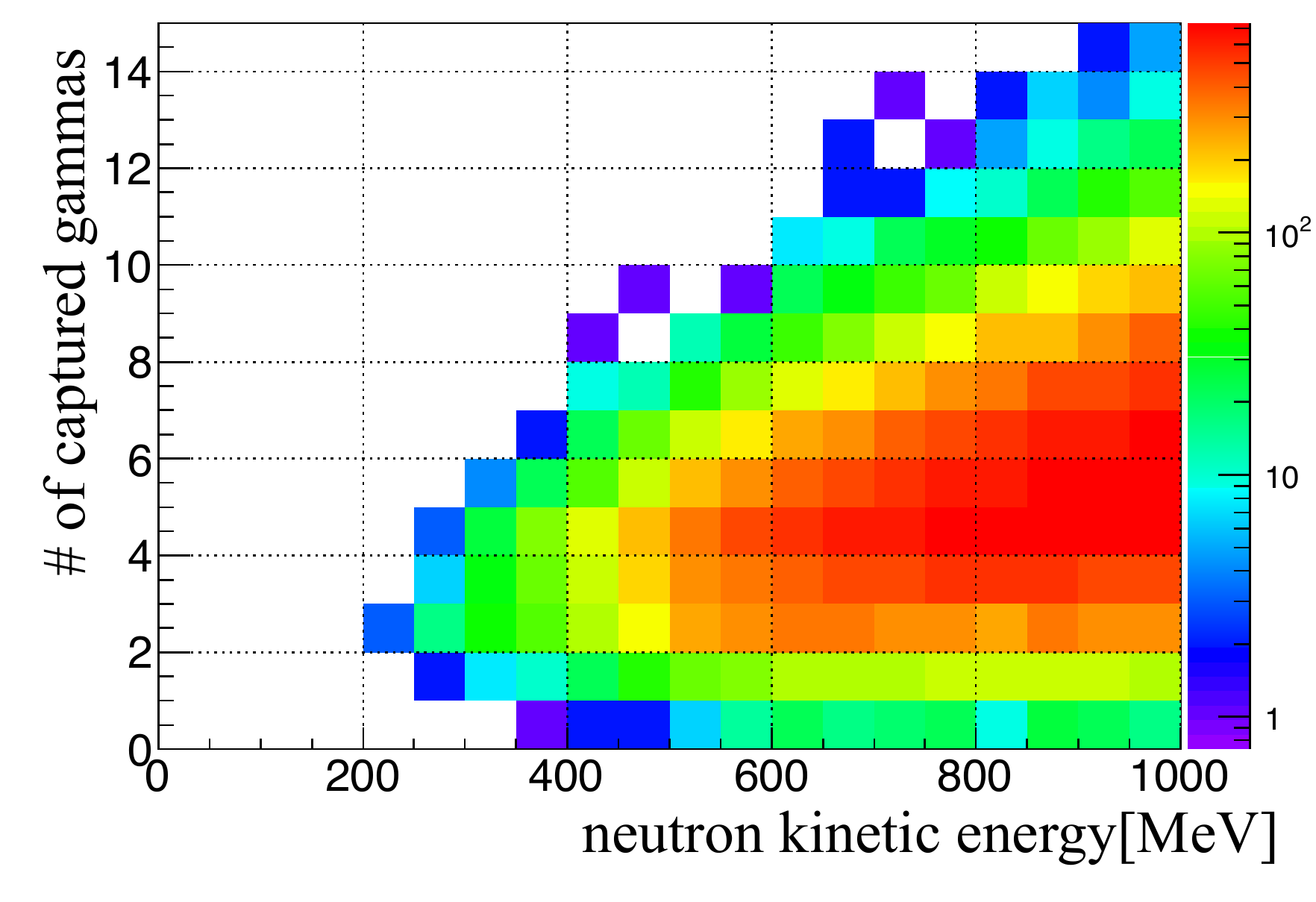}
 \includegraphics[width=0.49\textwidth]{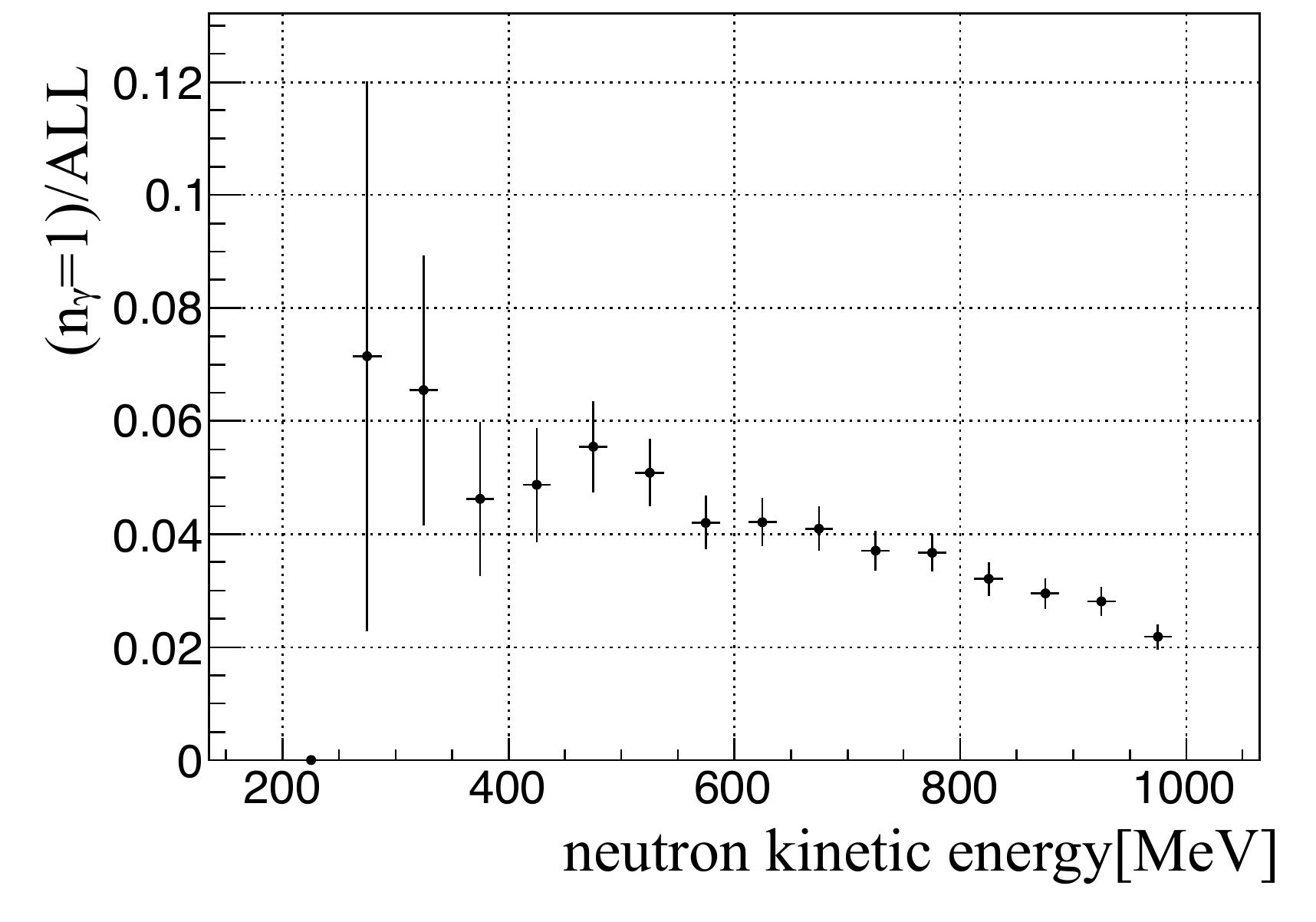}
 \end{center}
 \caption{\setlength{\baselineskip}{4mm}
 	The relation between energy of the initial neutron and the multiplicity of neutrons induced by the hadronic interactions (left figure), and the 
rejection power applying the multiplicity=1 selection.}
 \label{BAMMP}
\end{figure}

The Michel electron rate in the prompt energy range entering in the detector
at the candidate site is estimated to be 2$\times$10$^{-7}$/spill per one 
detector for 1MW. The number of Michel electrons per year is then
72 per one detector.
For reducing the number of Michel electrons more, the $\Delta$VTX cut between 
prompt-like signal due to Michel electron and  on-bunch signal due to protons 
recoiled by the neutrons and rejection of the multi-neutron captures are 
expected to be effective. Figure~\ref{BAMDVTX} shows the $\Delta$VTX distributions between Michel electrons and on-bunch signals, and between the neutrino prompt signals and accidental on-bunch signals, calculated by the MC simulation. There is good separation between both distributions. Thus, the rejection power is 
expected to be 100. 
Figure~\ref{BAMMP} shows the relation between energy of the initial neutron and 
the multiplicity of neutrons induced by the hadronic interactions (left 
figure), and the rejection power applying the multiplicity=1 selection
(right figure). The mean rejection power is 20. So by applying both cuts, the
number of Michel electrons contributing as the correlated backgrounds for the 
IBD candidates is reduced by 1/2000. It is expected to be 0.3~events considering total fiducial mass of 50 tons for 4~years measurement.


\paragraph{Gamma rays generated via thermal neutron capture}

The gamma flux entering the detector at the candidate site is 14/spill per one detector (see section 4). 
Many remaining events after applying the IBD selections for delayed signals 
deposit energy in the buffer scintillator region near the PMT surfaces, so the NHIT 
cut, in addition to the IBD selections, is also effective for more gamma 
reduction. Finally, the remaining gamma rate after applying both NHIT cut and IBD selections for delayed signals is reduced by 4.7$\times$10$^{-2}$/spill per one detector from 0.8/spill in case of only applying the IBD cuts. 

\subsubsection{$\overline{\nu}_{e}$ from $\mu^{-}$}
\label{NUMUM}
$\overline{\nu}_{e}$ from $\mu^{-}$ decay is detected with same sequence of the IBD reaction as $\overline{\nu}_{e}$ from $\overline{\nu}_{\mu}$. 
The total cut efficiency is 47\%. Assuming the rate is suppressed as FLUKA case, it is expected to be 803~events per 50~tons of total fiducial mass 
for 4~years measurement. The remaining rate after applying the IBD selection 
criteria is then 377~events.

%%%%%%%%%%%%%%%%%%%
\subsubsection{$^{12}C(\nu_{e},e^{-})^{12}N_{g.s.}$}
%%%%%%%%%%%%%%%%%%
\label{NUCNGS}
$^{12}C(\nu_{e},e^{-})^{12}N_{g.s.}$ reaction accompanies subsequent 
$\beta$ decay of $^{12}N_{g.s.}$ with 15.9 ms life time as follows:
\begin{equation}
^{12}C+\nu_{e} \to e^{-} + ^{12}N_{g.s.};  ^{12}N_{g.s.} \to ^{12}C + e^{+} + \nu_{e}
\end{equation}
Then, e$^{-}$ events contribute to the prompt-like signal and e$^{+}$ events 
decaying within time window for the IBD selection contribute to the delayed-like signal. 
The total cut efficiency is 0.17\%.
The rate after applying the selection criteria is expected to be 38~events 
for 4~years measurement. 

The e$^{-}$ and e$^{+}$ events contribute to prompt and delayed like signals for the accidental background, respectively. A rate for both prompt and delayed like signals is 8.0$\times$10$^{-6}$/spill per one detector.
Meanwhile, there is also the $^{12}C(\nu_{e},e^{-})^{12}N^{*}$ mode, which
is a similar reaction. The rate contributing to prompt-like signals for the 
accidental background was estimated 
to be approximately  3.8$\times$10$^{-6}$/spill.

%%%%%%%%%%%%%%%%%%%%
\subsubsection{Cosmic muon induced events}
%%%%%%%%%%%%%%%%%%%%
\label{COSMUEVE2}
There are some important backgrounds induced by cosmic rays; fast neutrons and
spallation products. Fast neutrons, which are created by the concrete or 
the iron shield located outside of the detector, enter to the 
detector, and they recoil protons inside the detector and
then are thermalized and create captured gammas. Thus, one neutron can produce 
both a faked ``IBD prompt'' and  ``IBD delayed'' signal at the same time. 
As shown in section~\ref{FASTN} in detail, the 
estimated number of events with 50 
tons fiducial mass times 4 years is 42 events assuming factor 100 rejection 
using sheilds or PID ability such as in the LSND experiment.  

On the other hand, the spallation products are expected to be negligible, 
as shown in section~\ref{CSPP}.

\subsubsection{Accidental backgrounds after applying IBD selection criteria}
\begin{figure}[htbp]
\begin{center}
\includegraphics[width=14cm]{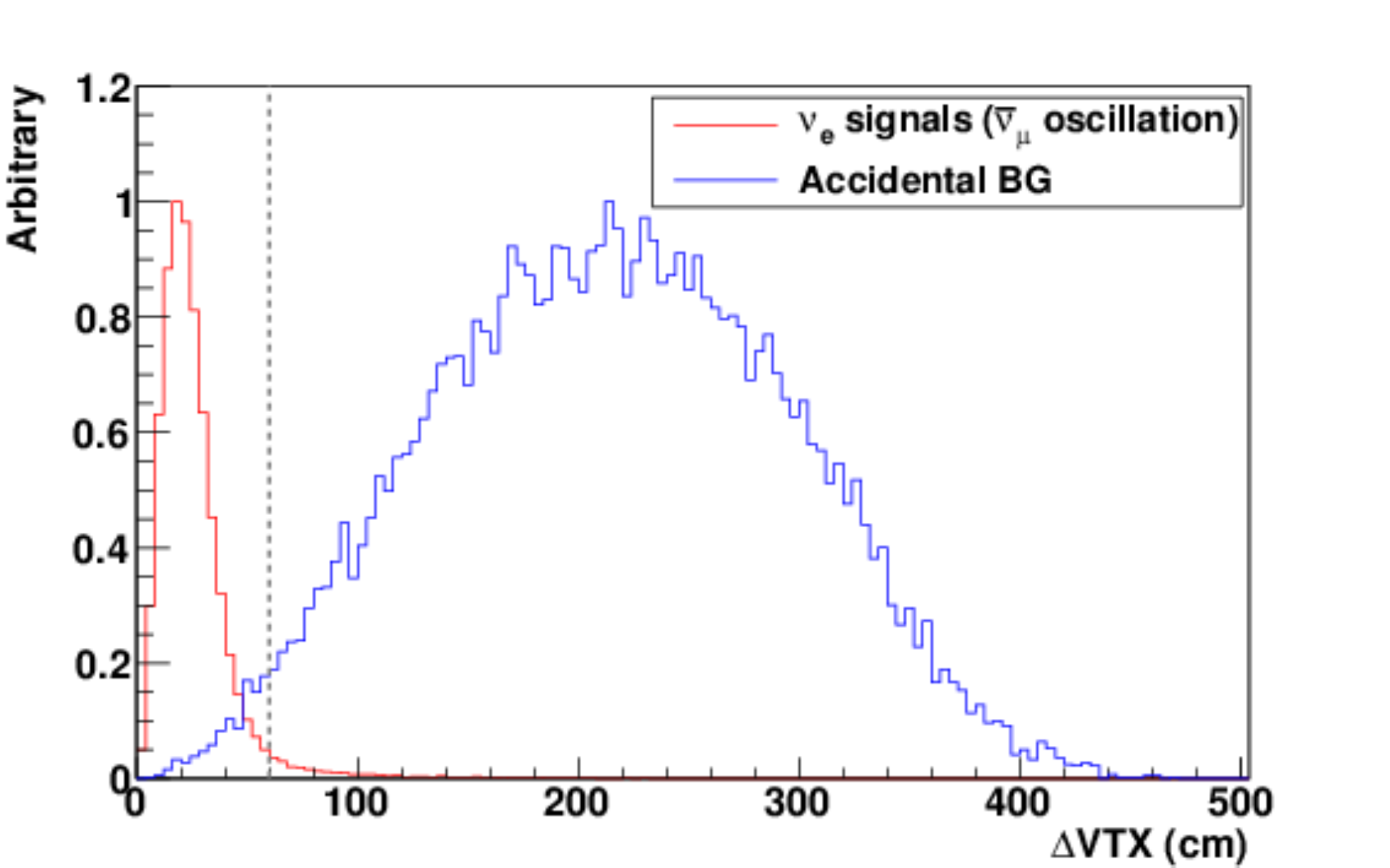}
\end{center}
\caption{\setlength{\baselineskip}{4mm}$\Delta$VTX distributions of 
$\overline{\nu}_{\mu} \rightarrow \overline \nu_e$ oscillation events
(red line) and the accidental events (blue line). Gray line shows $\Delta$VTX cut condition.}
\label{DVTXACC}
\end{figure}

%%%%%%%%%%%%%%%%%%%%%%%%%
\label{ACCBG}
There is a possibility that background events in energy cut range and time window for the prompt signals contribute to the prompt like signals of accidental backgrounds. Meanwhile, there is a possibility that background events in the energy cut range and time window for the delayed signals contribute to the delayed like signals of accidental backgrounds. In this proposal, the accidental rates
(R$_{acc}$) per one detector were calculated by multiplication of number of spills per year~(3.6$\times$10$^{8}$~spills/year), single rates per spill for the prompt and delayed like signals (R$_{prompt}$, R$_{delayed}$) considering the
time windows and the cut efficiencies, and the $\Delta$VTX cut efficiency 
($\epsilon_{VTX}$) as follows:

\begin{equation}
R_{prompt}=\sum (R^{prompt}_{i} \times \epsilon ^{prompt}_{i})
\end{equation}
\begin{equation}
R_{delayed}=\sum (R^{delayed}_{i} \times \epsilon ^{delayed}_{i})
\end{equation}
\begin{equation}
R_{acc}=R_{prompt} \times R_{delayed} \times \epsilon_{VTX} \times 3.6\times10^{8}spills/year
\end{equation}
where $i$ is the type of event contributing to the accidental backgrounds, 
R$_{prompt}\\ $(R$_{delayed}$) is calculated as sum rate of all types
of the prompt(delayed) like signal considering each cut efficiency($\epsilon ^{prompt}_{i}$, $\epsilon ^{delayed}_{i}$). Actually, the $\Delta$VTX cut efficiency depends on each combination of prompt and delayed like signals, 
but the $\Delta$VTX cut efficiency for combinations of the $\overline{\nu}_{e}$ prompt signals and delayed like signals of the neutron captured on Gd distributing uniformly in the target volume, was used approximately for all components in this proposal. Figure~\ref{DVTXACC} shows distributions of the $\Delta$VTX cut.
The cut efficiency below 60 cm is 2.3$\pm$0.1\%.
Table~\ref{ACTB} shows a summary of values for calculation of the accidental background rate. Finally, total number of accidental events is 36.8~events considering 50 tons of total fiducial mass 
%of two detectors 
for 4~years measurement.

\begin{table}[htb]
\begin{center}
\begin{tabular}{|c|l|c|} \hline
&background & R$^{prompt}_{i}$ (/spill) \\\hline
Prompt&e$^{-}$ ($^{12}C(\nu_{e},e^{-})^{12}N_{g.s.}$)&8.0$\times$10$^{-6}$  \\\cline{2-3}
like&e$^{-}$ ($^{12}C(\nu_{e},e^{-})^{12}N^{*}$)&3.8$\times$10$^{-6}$ \\\cline{2-3}
signals&e$^{+}$ ($\overline{\nu}_{e}$ from $\mu^{-}$)&$<$10$^{-6}$ \\\cline{2-3}
&R$_{prompt}$&1.3$\times$10$^{-5}$/spill \\\hline \hline
&background & R$^{delayed}_{i}$ (/spill) \\\hline
&Gamma (Beam associated)&4.7$\times$10$^{-2}$ \\\cline{2-3}
Delayed&Neutron (Beam associated)&2.4$\times$10$^{-3}$\\\cline{2-3}
like&e$^{+}$ ($^{12}C(\nu_{e},e^{-})^{12}N_{g.s.}$)&1.2$\times$10$^{-5}$ \\\cline{2-3}
signals&Spallation products&$\sim$10$^{-4}$ \\\cline{2-3}
&R$_{delayed}$&4.9$\times$10$^{-2}$/spill \\\hline \hline
\multicolumn{2}{|c}{R$_{acc}$}&\multicolumn{1}{|c|}{4.8/year}\\\hline
\end{tabular}
\caption{Values for calculation of the accidental background rate per one detector.}
\label{ACTB}
\end{center}
\end{table}

%%%%%%%%%%%%%%%%%%%
\subsection{Selection summary}
%%%%%%%%%%%%%%%%%%%
\indent
Table~\ref{tab:efficiency} summarizes numbers of signal and background events 
with 50 tons of total fiducial mass from two detectors after applying IBD selection criteria for 4~years. Finally, $\overline{\nu}_{e}$ from $\mu^{-}$ events are expected to be the dominant background for the IBD candidates. The 
number of $^{12}C(\nu_{e},e^{-})^{12}N_{g.s.}$ events is expected to be around 
10\% of the neutrino signal, and  it is expected that other backgrounds can also be reduced to around the 10 \% level for each in this proposal. Assuming a 
4 year measurement with 2 detectors, the number of signal events in the case 
of $\Delta m^2=3 eV^{2}$ and $sin^2(2\theta)=3.0\times10^{-3}$ and the 
background from $\overline{\nu}_{e}$s from $\mu^{-}$ decay
are 811 and 377 events, respectively.
However, the backgrounds at the candidate site can be measured with some 
prototype detector like the 200 L Gd-loaded liquid scintillator detector. 
In order to confirm the reduction of background, the components below
should be measured at the candidate site.
\begin{enumerate}
\item The beam associated neutron rate below several tens MeV and correlated
gammas. The on-bunch rate is especially important.
\item Michel electron events induced by the high energy neutrons above 200 MeV.
\item Cosmic ray induced fast neutrons.
\end{enumerate}

\pdfoutput=1
%%%%%%%%%%%%%%%%%%%%%%%%%%%%%%%%%%%%%%%%%%%%
\section{Neutrino Oscillation Sensitivity}
%%%%%%%%%%%%%%%%%%%%%%%%%%%%%%%%%%%%%%%%%%%%
\indent

As discussed in previous section, the dominant background is 
$\bar{\nu}_{e}$ from $\mu^{-}$ decay in the MLF experiment.
Others are estimated carefully, e.g. from the neutrinos,
however those backgrounds can be neglected for the sensitivity fit 
study at this stage since the fraction of each background component
is less than 10$\%$ of the $\bar{\nu}_{e}$ from $\mu^{-}$ decay background.  

For the fit of the oscillation parameters, $\Delta m^{2}$ and $\sin^2(2\theta)$, constraints of the background normalization are 
important. However, $\bar{\nu}_{e}$ from $\mu^{-}$ has a
very poor normalization constraint from the external information 
since the stopping point of $\mu^{-}$ decays and pion production are not known
well, even when using MC simulation. 
Therefore, the uncertainty of the normalization factor for this background 
is taken to be 50$\%$. (See section~\ref{sec_beam} in details)
 
On the other hand, the cross section for the $\nu_e + ^{12}C \rightarrow e + ^{12}N_{gs} $ reaction is known at the 2$\%$ level~\cite{cite:XSEC12C}. The 
lifetime of $N_{gs} \beta$ decay and the $e^{-}$ energy spectrum are also 
well known, as shown before. The measurement of the reaction provides the 
normalization factor for the oscillated signal ($\bar{\nu_{e}} + p \rightarrow e^{+} + n$) since the parent particle for the oscillated signal is 
$\bar{\nu_{\mu}}$ from $\mu^{+}$ decays ($\mu^{+} \rightarrow e^{+} + \bar{\nu_{\mu}} + \nu_{e}$).
Note that the determination of the normalization factor can be done at the
10$\%$ level; even disappearance oscillations, if they occur, should be
small (less than 10$\%$).

%%%%%%%%%%%%%%%%%%%%%%%%%%%%%%%%%%
\subsection{Fit method}
%%%%%%%%%%%%%%%%%%%%%%%%%%%%%%%%%%
\indent

The binned maximum likelihood method is used for the analysis. The 
method fully utilizes the energy spectrum of each background and 
signal components, thus the amount of the signal components can be 
estimated efficiently.

\begin{figure}
\centering
\includegraphics[width=1.1\textwidth]{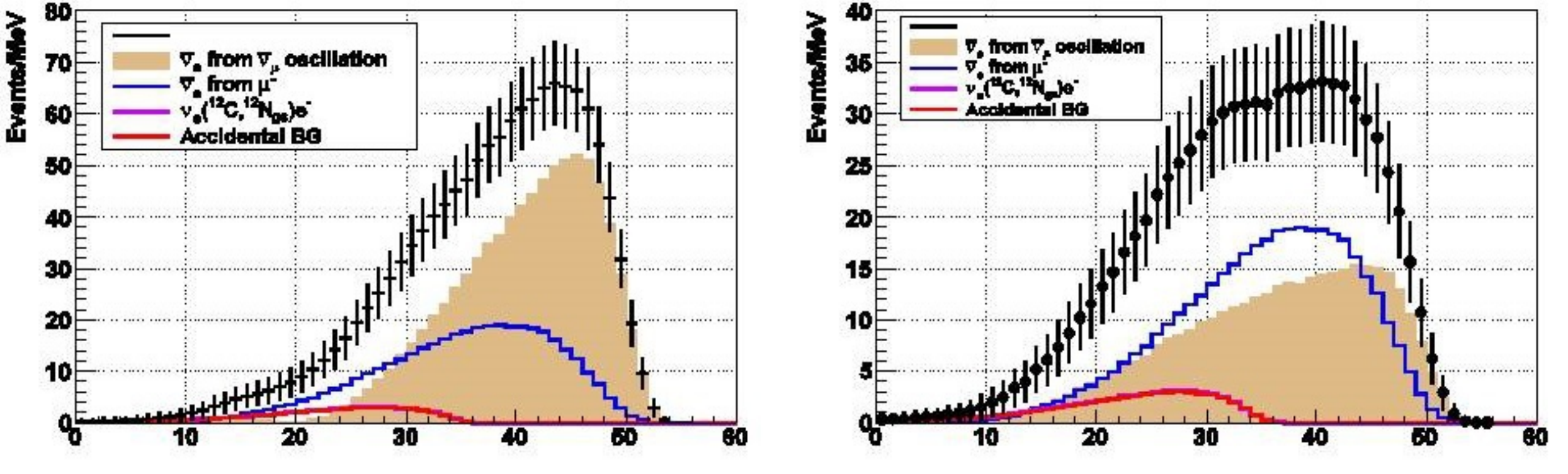}
\caption{\setlength{\baselineskip}{4mm}
The typical energy spectrum from $\mu^{-}$ (blue), the oscillated signal with ($\Delta m^{2}$, $\sin^2 2\theta$) = (3.0, 0.003) (brown shaded; best $\Delta m^{2}$ for the MLF experiment (left)),
(1.2, 0.003)(LSND best fit point (right)), 
$^{12}C (\nu_{e},e^{-}) ^{12} N_{gs}$(red)   
are shown. Black points with error bar  
correspond to the sum of the all components. Positron energy is 
smeared by 15.0$\% /\sqrt{E}$ for the detector effect. }
\label{fig:tenergy}
\end{figure}

The typical energy spectrum from $\mu^{-}$ (blue), the oscillated signal with ($\Delta m^{2}$, $\sin^2 2\theta$) = (3.0, 0.003) (brown shaded; best $\Delta m^{2}$ case (left)) and (1.2, 0.003) LSND best fit case (right)),
$^{12}C (\nu_{e},e^{-}) ^{12} N_{gs}$ (red) are shown in
Figure~\ref{fig:tenergy}. 
Here we assume the fiducial mass of the detector is 50 tons,
1MW beam power at MLF, and four years operation with 4000 hours 
in the beam exposure time during each year. 
The signal detection efficiency is assumed to be 48$\%$. 
The detector is put at the distance 17 m from the
target. 
The number of events at each energy bin is still 
statistically small, therefore we use maximum likelihood instead of the usual 
minimum $\chi^{2}$ method. The fitter estimates 
the oscillation parameters by varying the size and shape of the brown part
to best reproduce the energy distribution of the black points. 

For this purpose, the following equation is used.
\begin{eqnarray}
\label{Eq:likelihood}
L &=& \displaystyle \Pi_{i} P(N_{exp} | N_{obs})_{i} \\
P(N_{exp} | N_{obs}) &=& \frac{e^{-N_{exp}} \cdot (N_{exp})^{N_{obs}}}{N_{obs}!}
\end{eqnarray}
where, $i$ corresponds to $i$-th energy bin, $N_{exp}$ is expected 
number of events in $i$-th bin, $N_{obs}$ is number of observed
events in $i$-th bin. $i$ is starting from 20 MeV and ends at 60 MeV
because the energy cut above 20 MeV is applied for the primary signal
as explained before. Note that $N_{exp} =  N_{sig}(\Delta m^{2}, \sin^{2}2\theta) + \displaystyle\sum N_{bkg}$, and $N_{sig}(\Delta m^{2}, \sin^{2}2\theta)$ is 
calculated 
by the two flavor neutrino oscillation equation as shown before, 
$P(\bar{\nu_{\mu}} \rightarrow \bar{\nu_{e}}) = \sin^{2}2\theta \sin^{2}(\frac{1.27 \cdot \Delta m^{2} (eV^{2}) \cdot L (m)}{E_{\nu} (MeV)})$.

The maximum likelihood point gives the best fit parameters, and 2$\Delta lnL$ provides 
the uncertainty of the fit parameters. As shown in the 
PDG~\cite{cite:PDG}, we have
to use the 2$\Delta lnL$ for 2 parameter fits to determine the uncertainties from
the fit.   
 
%%%%%%%%%%%%%%%%%%%%%%%%%%%%%%%%%%
\subsection{Systematic uncertainties}
%%%%%%%%%%%%%%%%%%%%%%%%%%%%%%%%%%
\indent

Equation~\ref{Eq:likelihood} takes only statistical 
uncertainty into account, therefore the systematic uncertainties
should be incorporated in the likelihood.  
Fortunately, energy spectrum of the oscillated signal and background
components are well known, thus the error (covariance) matrix of 
energy is not needed.
In this case, uncertainties of the overall normalization
of each component have to be taken into account, and the assumption
is a good approximation at this stage. 

In order to incorporate the systematic uncertainties, the constraint terms should be
added to Equation~\ref{Eq:likelihood} and the equation is changed as follows.

\begin{eqnarray}
L &=& [\Pi_{i} P( N_{exp}^{'} | N_{obs})_{i}] \times e^{- \frac{(1-f_{1})^{2}}{2 \Delta \sigma_{1}^{2}}} \times e^{- \frac{(1-f_{2})^{2}}{2 \Delta \sigma_{2}^{2}}} 
\label{Eq:likelihood2}
\end{eqnarray}\
where $f_{j}$ are nuisance parameters to give the constraint term on the overall
normalization factors. $N_{exp}^{'} = f_{1} \cdot N_{sig} (\Delta m^{2}, \sin^{2}2\theta) + f_{2} \cdot N_{bkg}$. 
$\Delta \sigma_{i}$ gives the uncertainties on the normalization factors of each 
components.
In this proposal, the profiling fitting method is used to treat the systematic 
uncertainties. The method is widely known as the correct fitting method as well as the marginalizing method.
The profiling method fits all nuisance parameters as well as 
oscillation parameters. 

As mentioned above, 
the flux of the $\bar{\nu}_{e}$ from $\mu^{-}$ decays around 
the mercury target has very poor constraints from the external information.
For this situation, the uncertainty of this background
component is assigned to be 50$\%$.

Table~\ref{tab:constraints} shows the summary of the 
uncertainty of the normalization factors for the signal and background 
components. They are regarded as inputs of $\Delta \sigma$ although only 
$\bar{\nu}_{e}$ from $\mu^{-}$ is used in this proposal as mentioned above.

\begin{table}[h]
\begin{center}
	\begin{tabular}{|l|c|c|}
	\hline
	components  & uncertainty & comments \\ \hline \hline
	signal      & 10$\%$ & normalized by $\nu_e$ from $\mu^{+}$ \\ \hline
	$\bar{\nu_{e}}$ from $\mu^{-}$  & 50$\%$ &   \\ \hline
	$\nu_e + ^{12}C \rightarrow e + ^{12}N_{gs} $  & $<<$10$\%$ & for correlated BKG \\ \hline
	$\nu_e + ^{12}C \rightarrow e + ^{12}N_{gs}$ + delayed acc.  & $<<$10$\%$ & if  delayed BKG is known within 10$\%$ \\ \hline
	$\nu_e + ^{12}C \rightarrow e + ^{12}N^{*}$ & 20$\%$ & \\  
	\hline
	cosmic / beam & 5$\%$ & well known from calibration source \\ \hline
	\end{tabular}
	\caption{\setlength{\baselineskip}{4mm}
	Summary of uncertainties on the normalization factors. Note that only $\bar{\nu}_{e}$ from $\mu^{-}$ are used in the fitting in this proposal since it is dominant one.}
	\label{tab:constraints}
\end{center}
\end{table}

%%%%%%%%%%%%%%%%%%%%%%%%%%%%%%%%%%
\subsection{Sensitivity}
%%%%%%%%%%%%%%%%%%%%%%%%%%%%%%%%%%
\indent

In order to obtain the experiment's sensitivity, we assume there is no 
oscillation signal in the pseudo-data, and then calculate 2$\Delta ln(L)$ 
from the maximum $2ln(L)$ points. In this case, the maximum $2ln(L)$ 
point stays at $\sin^{2} 2\theta$ = 0.000. The points, which 
have 2$\Delta ln(L)$ to be 11.83 in ($\sin^{2} 2\theta, \Delta m^{2})$ 
2D plane correspond to the 3 $\sigma$ case. (28.76 for 5 $\sigma$ case)

\begin{figure}
\centering
\includegraphics[width=0.5\textwidth]{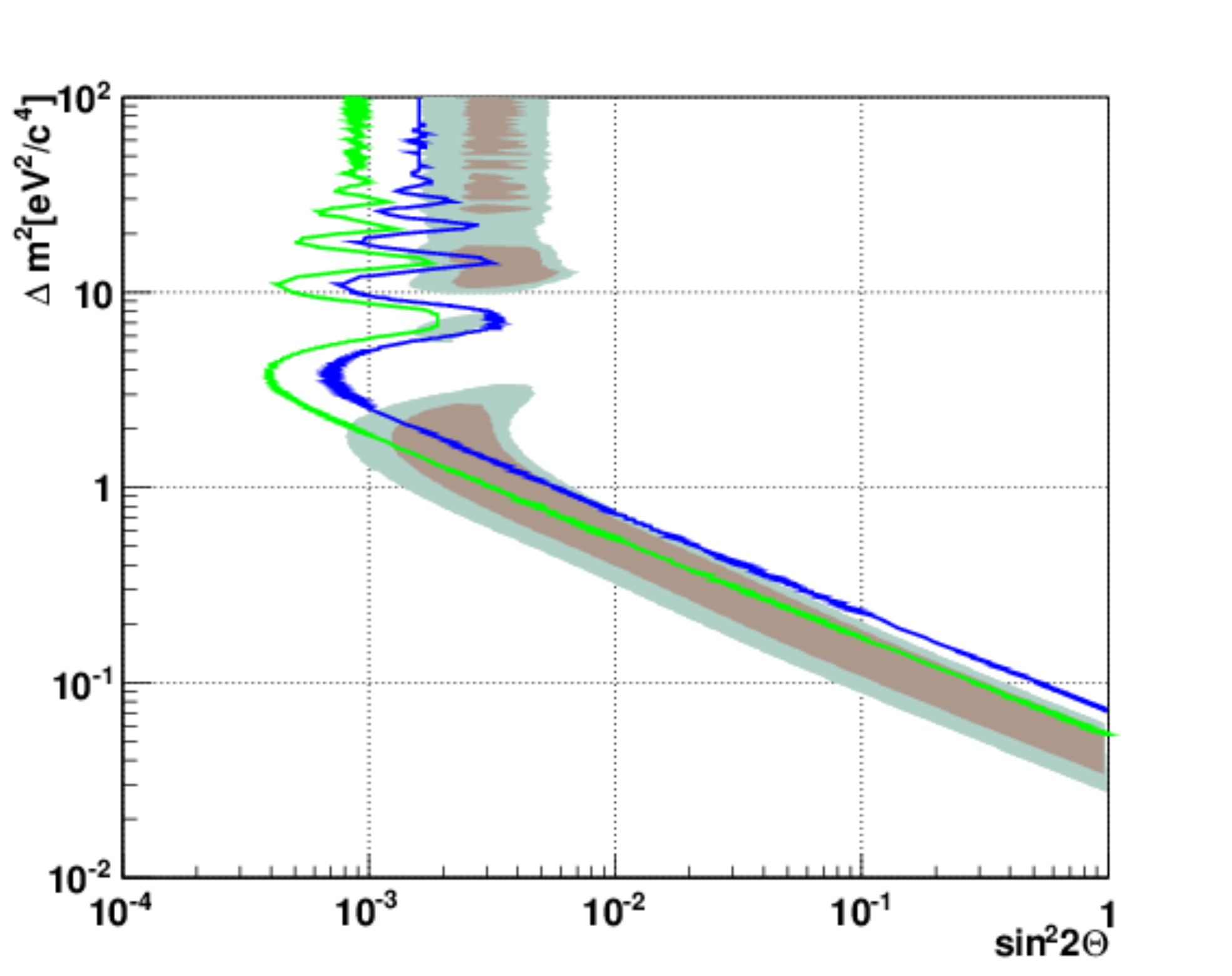}
\includegraphics[width=0.9\textwidth]{Near_A17_4y.pdf}
\caption{\setlength{\baselineskip}{4mm}
The sensitivity of the MLF experiment assuming 2 and 4 years 
operations (4000 hours / year) assuming $\sim$50$\%$ 
detection efficiency and 17 m baseline. The blue line shows the 5 $\sigma$ C.L., while green one corresponds to 3 $\sigma$.
Top plot shows the case with 2 years exposure, while the bottom plot 
shows that with 4 years. 
}
\label{fig:sensitivity}
\end{figure}

Figure~\ref{fig:sensitivity} shows the sensitivity of our experiment 
with 3 (green) and 5 $\sigma$ (blue), respectively. 
A top plot shows the case with
2 years exposure, while the bottom plot shows the that with 4 years.

If no definitive positive signal is found by this experiment, 
a future option exists to cover small $\Delta m^2$ region. 
This needs a relatively long baseline and requires a large detector to compensate for the reduced neutrino flux. (See appendix)
 
\pdfoutput=1
%%%%%%%%%%%%%%%%%%%%%%%%%%%%%%%%%%%%%%%%%%%%%%%
\section{Milestone}
%%%%%%%%%%%%%%%%%%%%%%%%%%%%%%%%%%%%%%%%%%%%%%%

The following measurements at the candidate site (the 3rd floor of the MLF building) are planned as a part of the feasibility test of the experiment. 
\begin{enumerate}
\item Total amount of beam associated gammas and neutrons for the "IBD delayed" background. 
\item Confirmation of reduction of Michel electrons compared to BL13.
\item Total amount of cosmic induced fast neutrons for the IBD correlated background.

\end{enumerate}
According to PHITS simulation, the gamma and neutron rates are $\sim$10$^{6}$/day/ton/1MW/100$\mu$s each, while the Michel electron rate is $\sim$10$^{-2}$/day/ton/1MW/9$\mu$s compared to $\sim$10$^{3}$ at BL13. The cosmic induced fast neutrons are measurable with self trigger, then the flux is expected to be $\sim$10Hz/m$^2$. All materials are measureable using a 1 ton level detector. 
The measurement provides confidence for the experiment if the measured rate is consistent with the PHITS simulation.

\pdfoutput=1
%=============================
\section{Cost estimation}
%----------------

Here is the cost estimation. 

\begin{table}[htbp]
 \begin{center}
  \begin{tabular}{lrrr} 
  \hline
  \hline
    Item                             &Unit price  & Quantity  & Total    \\
    \hline
* PMTs \& Electronics system :        & 500Ky/ch  &400~ch  & 200My     \\
* Tanks \& Acrylic Vessels :          & 50My/set  & 2~sets & 100My     \\
* Gd-LS, Buffer-LS	                  &           &        & 100My      \\
* Fluid handling and infrastructure   & 50My/set  & 1~set  &  50My     \\
* Miscellaneous                       &           &        &  50My   \\
\hline
\hline
	Grand Total	                     &            &        & 500My \\
	\end{tabular}
 \caption{Cost estimation (2 detectors)}
 \end{center}
\end{table}

\newpage
\appendix
\pdfoutput=1
\section{Test measurements of backgrounds at the MLF}

	\subsection{1 ton plastic scintillation counters}
		\label{appendix_1t}
		A plastic scintillator detector was placed at BL13 in MLF to measure the background level of neutrons and photons as was shown in Figure~\ref{1t_view}.
		We will describe the setup, the calibration and the obtained resolution of the 1 ton scintillator detector in the following subsections.		
		
		\subsubsection{setup}
\begin{figure}[hbtp]
				\begin{center}
					\includegraphics[scale=0.8]{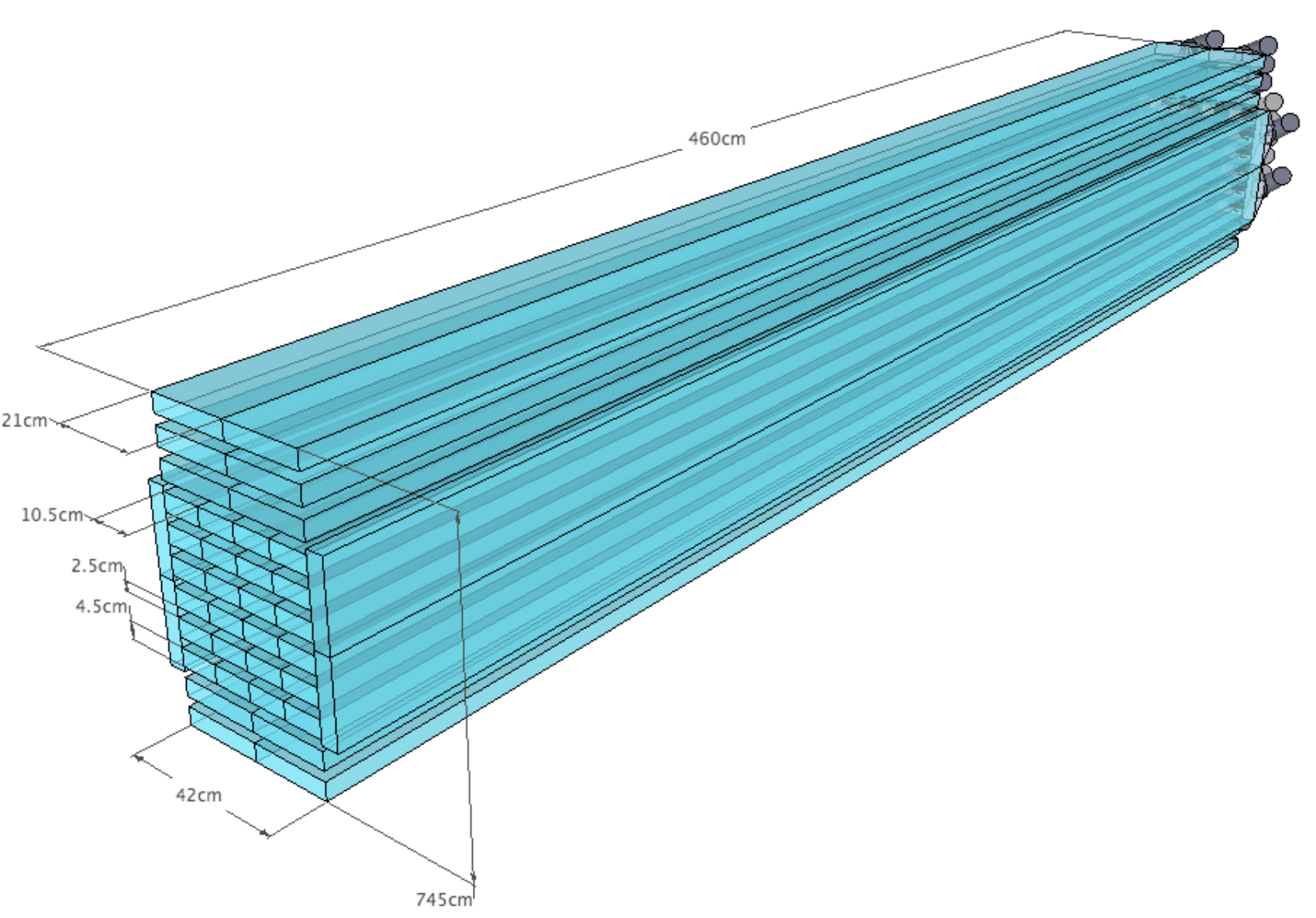}
					\caption{\setlength{\baselineskip}{4mm}Schematic view of the 1 ton plastic scintillator counters.
					It consists of 24 pieces of 10.5 cm width and 14 pieces of 21.0 cm width scintillators.
					All of them are 460 cm long.
					Both ends of each scintillator were viewed by PMTs.}
					\label{1t_setup}
				\end{center}
			\end{figure}

			Figure \ref{1t_setup} shows a schematic view of the 1 ton plastic scintillator counters placed at BL13.
			It consists of 24 pieces of $10.5 \times 4.5 \times 460\ {\rm cm^3}$ scintillator and 14 pieces of $21.0 \times 4.5 \times 460\ {\rm cm^3}$ scintillator.
			Each end of the scintillators were viewed by PMTs.
			Signals from each PMT were recorded by 65 MHz FADC with 50 ns RC-filter.
		\subsubsection{calibration}
\begin{figure}[hbtp]
				\begin{center}
					\includegraphics[scale=0.6]{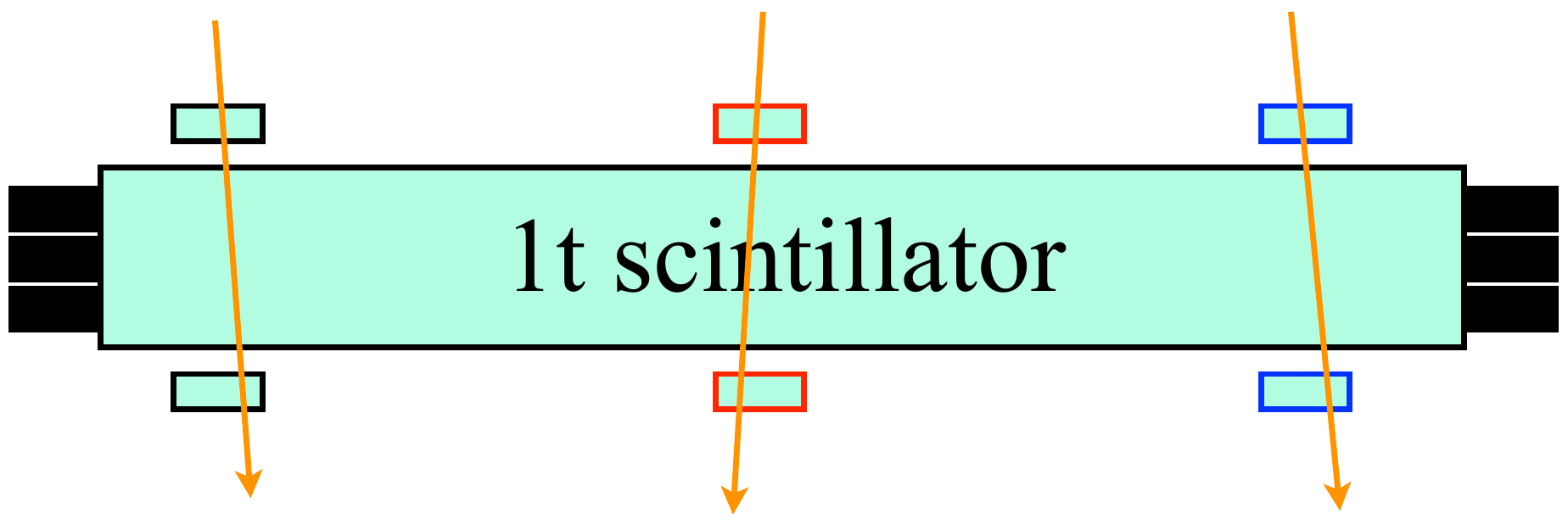}
					\caption{\setlength{\baselineskip}{4mm}Schematic view of the cosmic muon trigger counters.
					We prepared 3 pairs of plastic scintillators to trigger cosmic muons.}
					\label{1t_setup2}
				\end{center}
			\end{figure}

			We used cosmic muons to measure the attenuation length of the scintillator, to calibrate energy and timing.
			We prepared 3 pairs of scintillator counters to trigger the cosmic rays.
			Figure \ref{1t_setup2} shows a schematic view of the cosmic muon trigger counters.
			
\begin{figure}[hbtp]
				\begin{center}
					\includegraphics[scale=0.6]{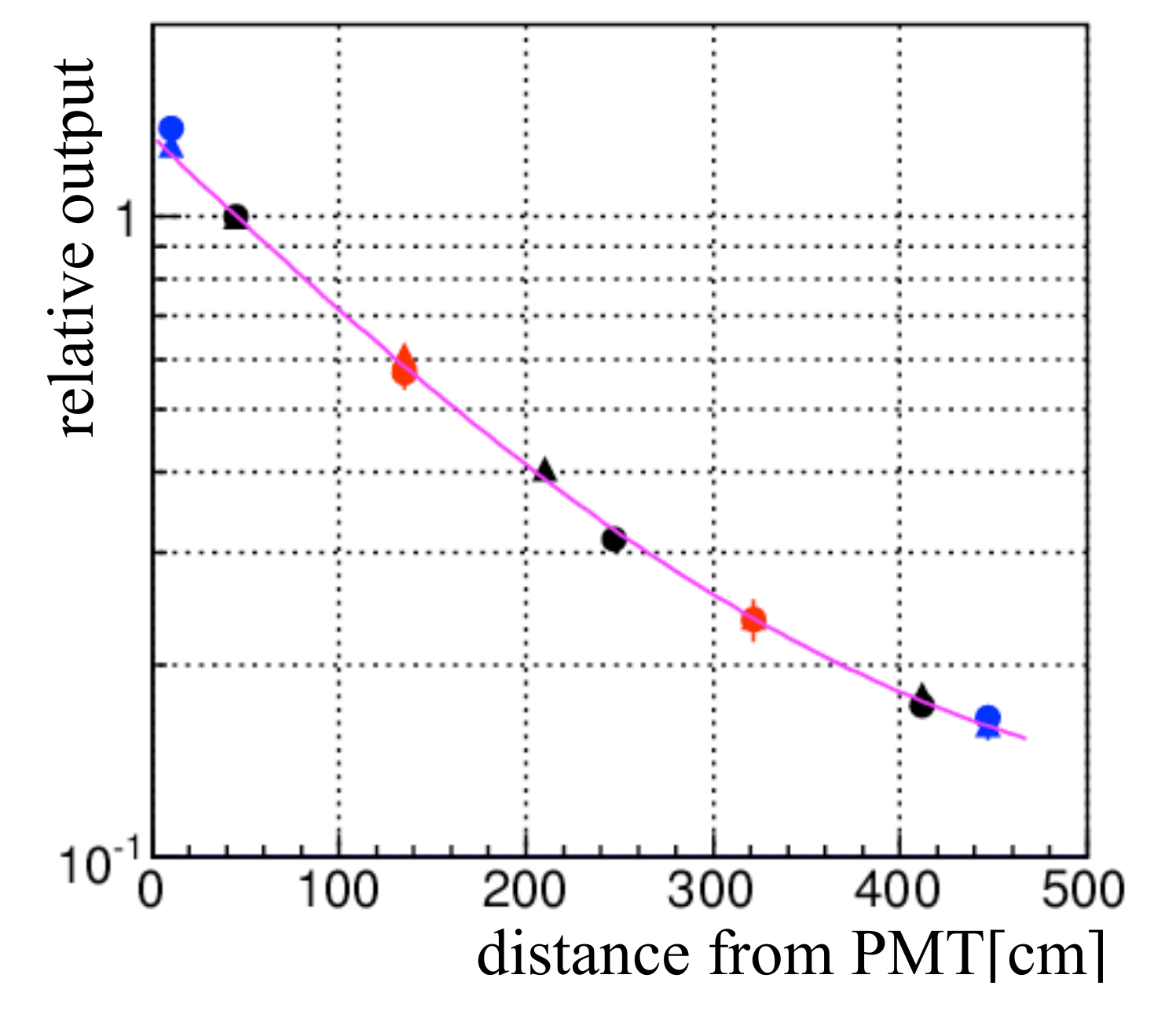}
					\caption{\setlength{\baselineskip}{4mm}Typical measured attenuation curve for a certain scintillator.
					We measured the curve and parameterized one for each scintillator (magenta line).}
					\label{1t_attenuation}
				\end{center}
			\end{figure}

\begin{figure}[hbtp]
				\begin{center}
					\includegraphics[scale=0.6]{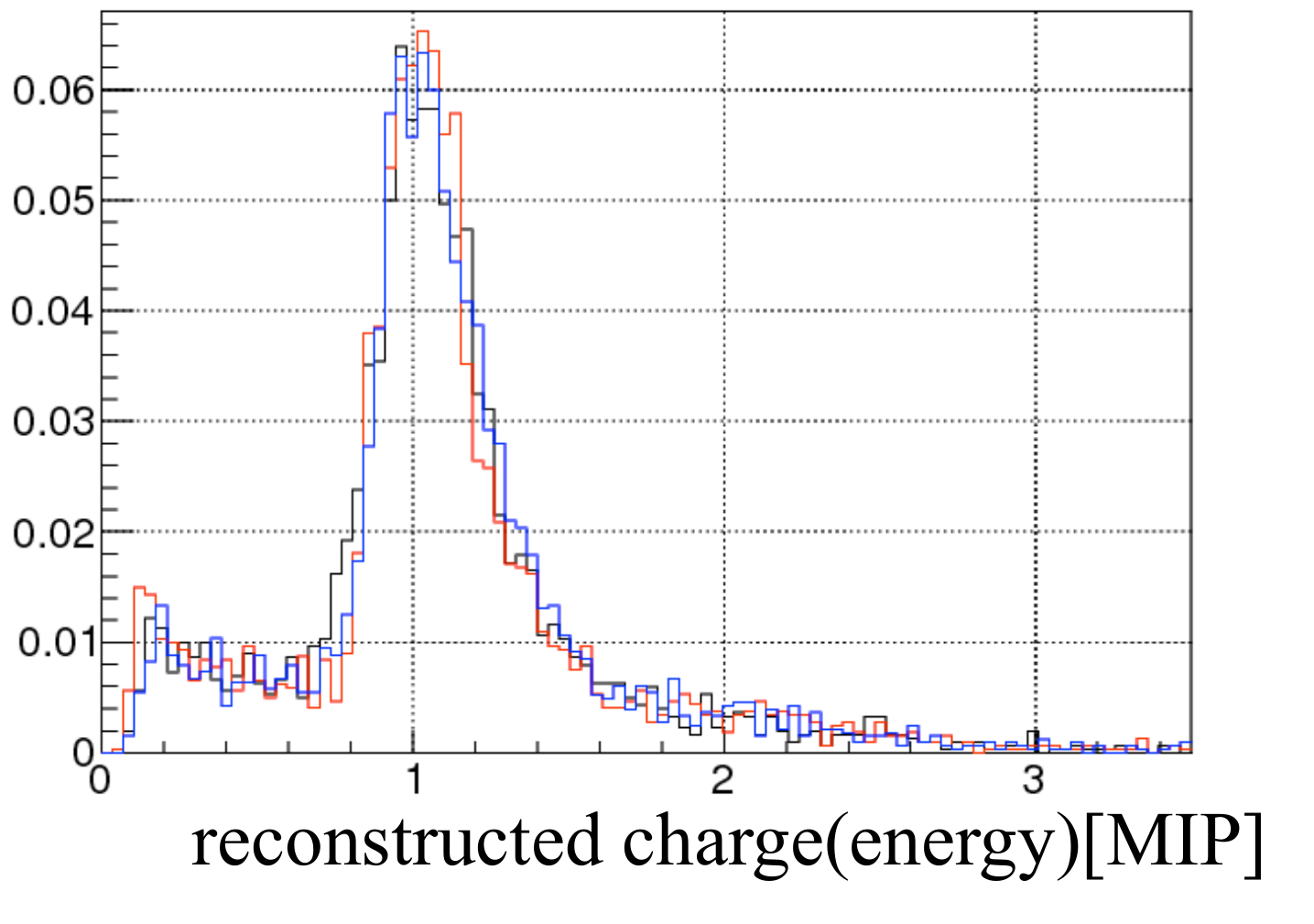}
					\caption{\setlength{\baselineskip}{4mm}Reconstructed charge for different incident positions by considering the measured attenuation length.
					The MIP energy for the 4.5 cm thick scintillator is 8 MeV.
					The colors correspond to the events triggered by the cosmic muon counters shown in Figure~\ref{1t_setup2}.}
					\label{1t_reccharge}
				\end{center}
			\end{figure}

			We made some dedicated runs to measure the attenuation length of the scintillators.
			In some of the dedicated runs, we changed the position of the trigger counters along the beam axis.
			Figure \ref{1t_attenuation} shows a measured typical attenuation curve.
			We measured the curve and parameterized one for each scintillator.
			By considering the attenuation length, the reconstructed charge was independent from the incident position as shown in Figure~\ref{1t_reccharge}.
			
\begin{figure}[hbtp]
				\begin{center}
					\includegraphics[scale=0.6]{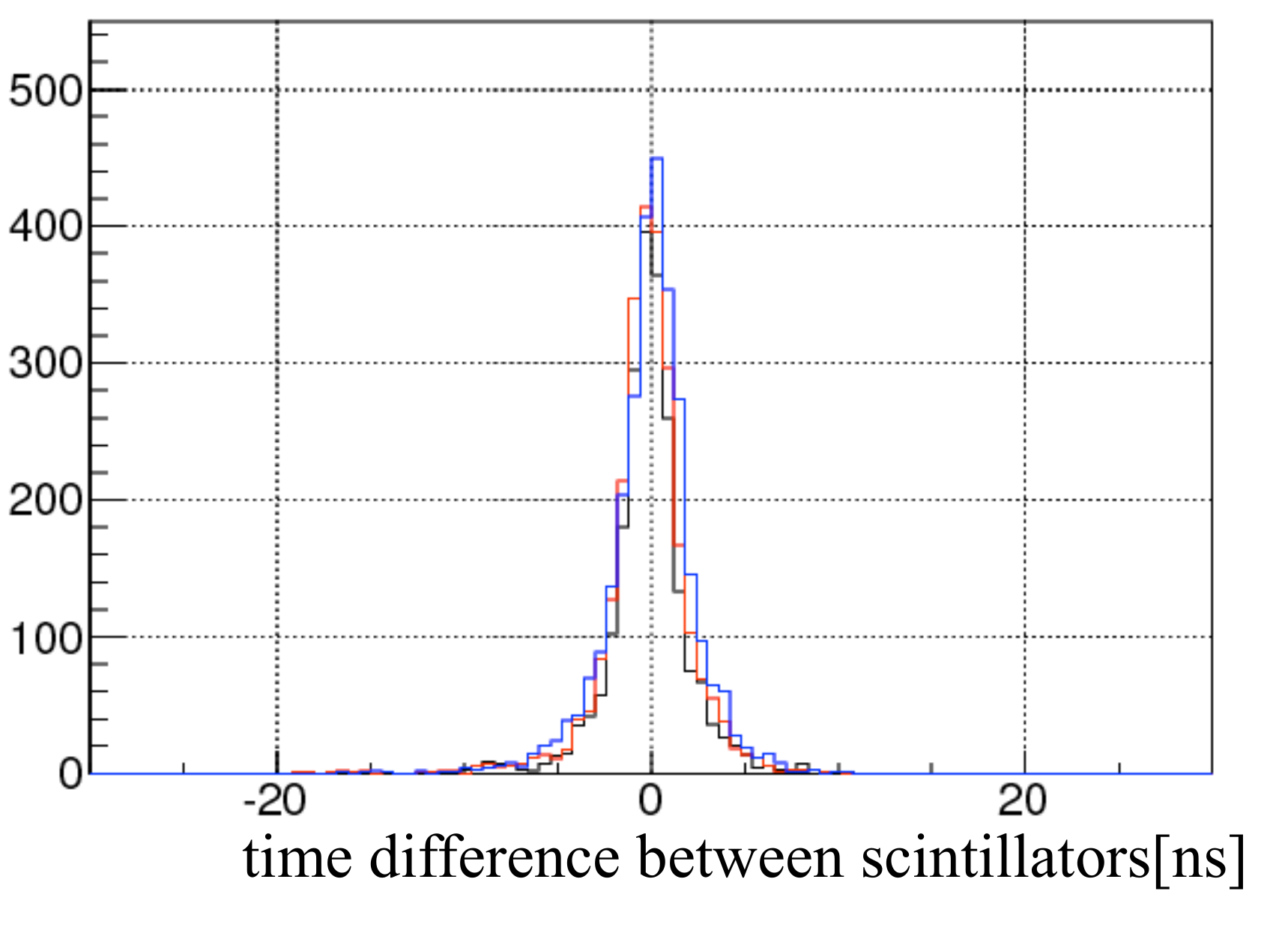}
					\caption{\setlength{\baselineskip}{4mm}Time differences between scintillators of cosmic ray events for different incident positions.
					The colors correspond to the events triggered by the cosmic muon counters shown in Figure~\ref{1t_setup2}.}
					\label{1t_deltat}
				\end{center}
			\end{figure}

			We also aligned the timing of each PMT.
			Time offsets were determined to minimize the time difference between each pair of PMTs on each end.
			The velocity of cosmic muons passing through the detector was considered.
			Figure \ref{1t_deltat} shows the time difference between scintillators for cosmic ray events.
			The timing of each PMT and scintillator was well calibrated.
			The light velocity in each scintillator was also measured.
			The typical velocity was 15 cm/ns.
			
		\subsubsection{Energy and Timing Resolution}
\begin{figure}[hbtp]
				\begin{center}
					\includegraphics[scale=0.42]{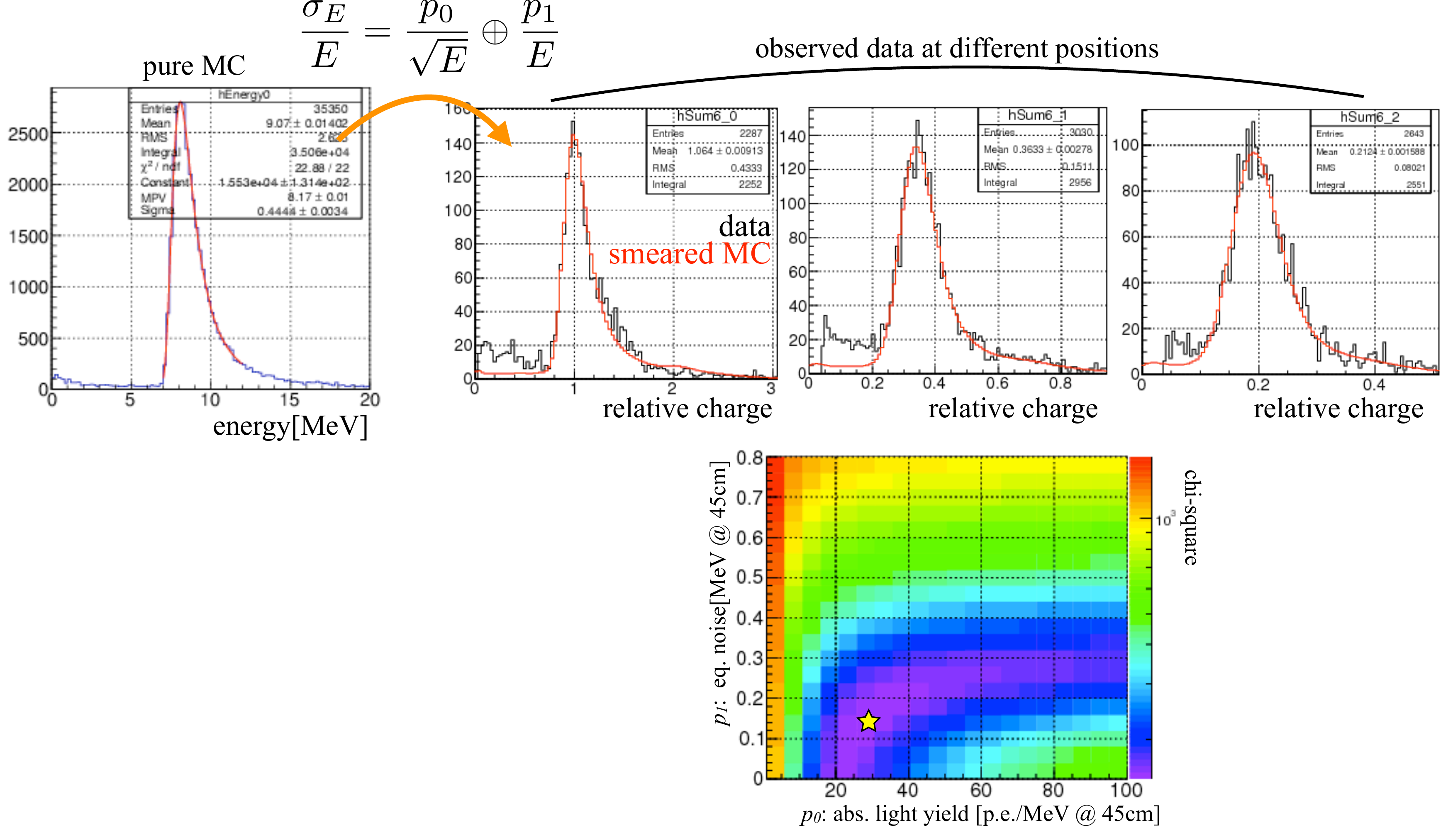}
					\caption{\setlength{\baselineskip}{4mm}Schematic view of the estimation procedure of the energy resolution.
					The charge distribution at 3 different positions along the beam axis for the cosmic muon events were compared with the smeared MC outputs.}
					\label{1t_eres_concept}
				\end{center}
			\end{figure}

			We evaluated the obtained energy resolution by smearing the output of the pure Monte-Carlo simulation, and comparing it with data.
			Figure \ref{1t_eres_concept} shows a schematic view of the estimation procedure of the energy resolution.
			We parameterized the energy resolution as follows:
			\begin{equation}
				\frac{\sigma_E}{E} = \frac{p_0}{\sqrt{E}} \oplus \frac{p_1}{E} \oplus p_2,
			\end{equation}
			where $p_0$ represents the photo-statistics term, $p_1$ represents the noise contribution and $p_2$ represents the calibration precision.
			The charge distribution at 3 different position along the beam axis were compared for each PMT.
			Because only one PMT was in interest at a time, we set $p_2=0$.
			We fitted the charge distributions by changing the rest of parameters,
			and obtained the light yield $n_{p.e.}\sim$ 30 p.e./MeV and the equivalent noise level $\sigma_{\rm noise}\sim$ 0.15 MeV at 45 cm from the PMT (typical values).
			
\begin{figure}[hbtp]
				\begin{center}
					\includegraphics[scale=0.4]{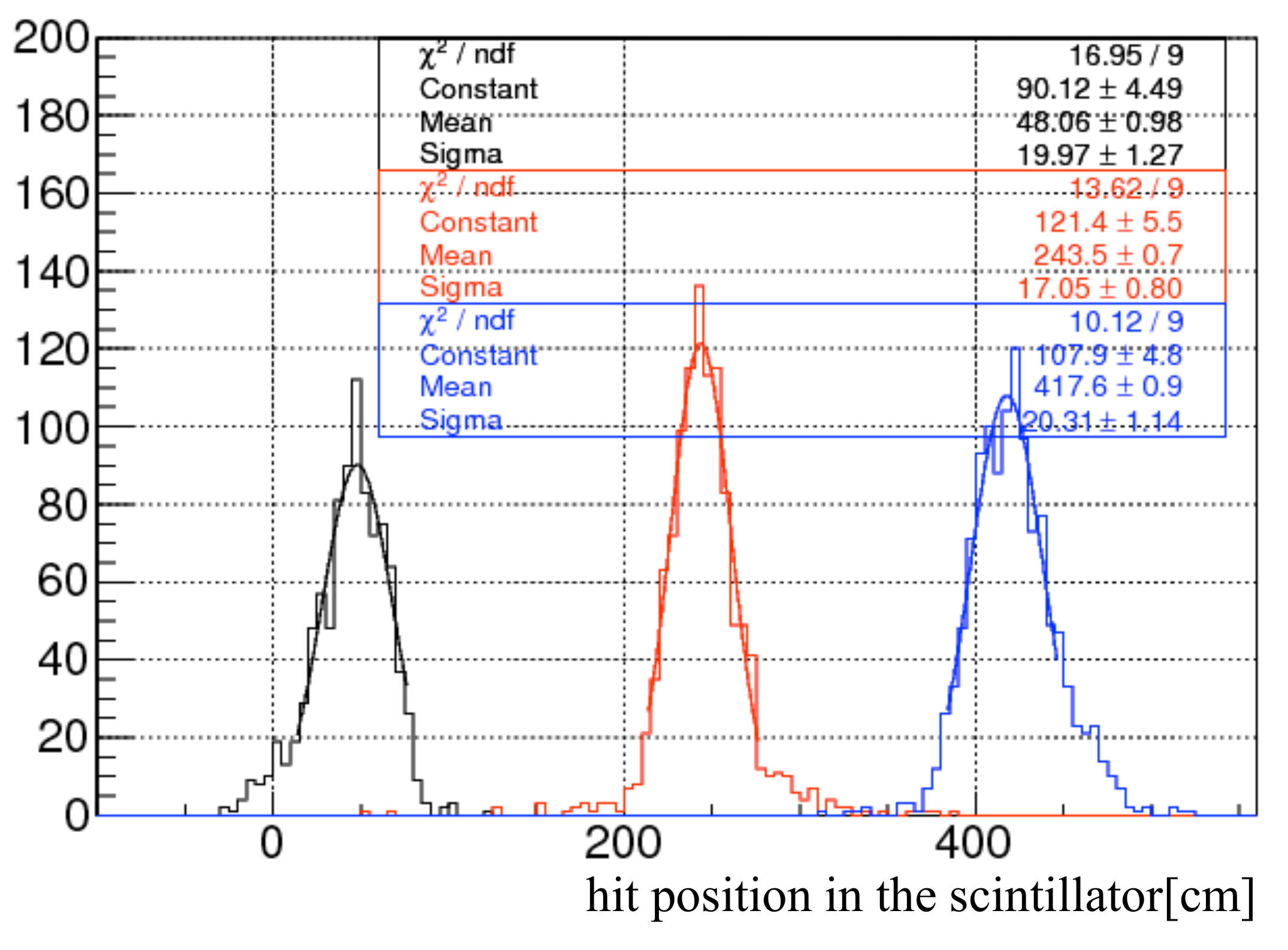}
					\caption{\setlength{\baselineskip}{4mm}Reconstructed hit positions along beam direction of cosmic ray events for different incident positions.
					The colors correspond to the events triggered by the cosmic muon counters shown in Figure~\ref{1t_setup2}.}
					\label{1t_recz}
				\end{center}
			\end{figure}

			We also evaluated the obtained position resolution and timing resolution.
			The hit position in a scintillator along the beam direction was calculated by using the time difference between PMTs on both ends of the scintillator.
			Figure \ref{1t_recz} shows the reconstructed hit position in the scintillators of cosmic ray events for different incident positions.
			The obtained position resolution has slight position dependence because of the light attenuation in the scintillators.
			The obtained position resolution around the middle of the scintillator for MIP energy was $\sigma_x = 17\ {\rm cm}$.
			By considering the observed light velocity in the scintillator 15 cm/ns, the timing resolution of the scintillator for MIP energy is $\sigma_t=1.1\ {\rm ns}$.
	\subsection{NaI counter (370 g)}
		\label{appendix_NaI}
		The NaI counter measured environmental $\gamma$ at BL13.
		The NaI counter was located on the floor of BL13 near downstream side of the 1 ton detector with respect to the beam and the opposite side of the 1 ton detector with respect to BL14.
		The NaI crystal is a cylinder with diameter of 2 inch and thickness of 2 inch.
		Pulse heights of signals from a PMT are measured by a multi channel analyzer (MCA) through a preamplifier, a spectroscopy amplifier and a linear gate and stretcher module.
		Based on the pulse heights, a histogram of the pulse heights with 1024 bins is made with a PC.
		Energy calibration was done using peaks of environmental $\gamma$, such as 1.461 MeV $\gamma$ of $^{40}$K and 2.615 MeV $\gamma$ of $^{208}$Tl.
		Energy resolution is about 3\%/$\sqrt{E[\mathrm{MeV}]}$, estimated using Geant4.

\pdfoutput=1
%%%%%%%%%%%%%%%%%%%%%%%%%%%%%%%%%%%%%%%%%
\section{Event generator and detector simulation}
%%%%%%%%%%%%%%%%%%%%%%%%%%%%%%%%%%%%%%%%%

%%%%%%%%%%%%%%%
\subsection{Event generator}
\label{EVENTGEN}
%%%%%%%%%%%%%%%%
Event generators are tools built to select the energy and momentum of the particles produced in the reactions expected in the detectors, which will be the entry for the MC simulations. The selection of the particle's energy and momentum method follows the flux distribution and cross-sections values, found in many different references. The tools also generate a random interaction point in the detector, based on its geometry, which is used to calculate the baseline value and momentum direction. All of them are build using the ROOT libraries, and some details of each is defined in the following subsections. 

%%%%%%%%%%%%%%%%%%%%%%%%
\subsubsection{Signal}
%%%%%%%%%%%%%%%%%%%%%%%%

\begin{figure}[htbp]
 \centering
 \includegraphics[width=0.45\textwidth]{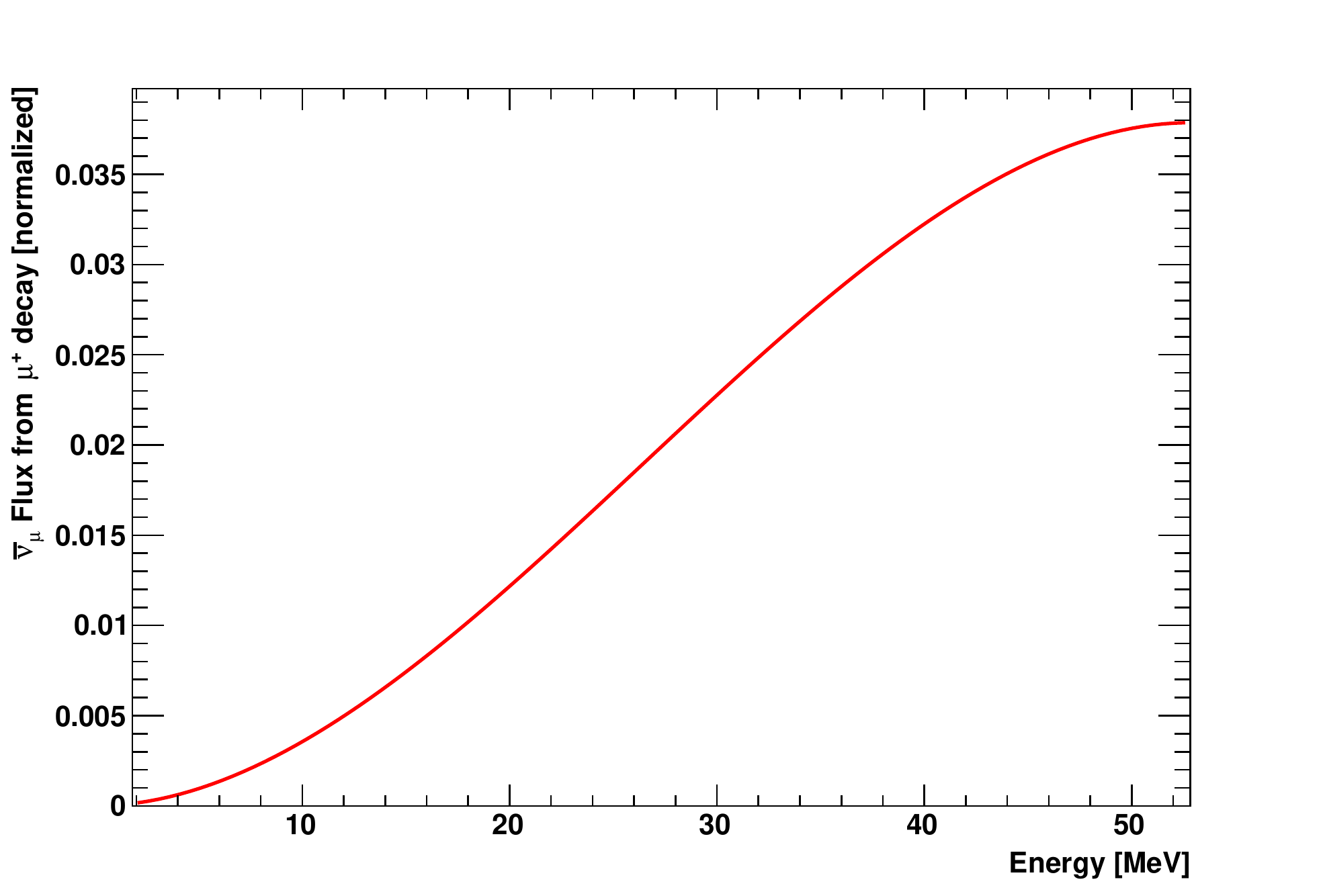}
 \includegraphics[width=0.45\textwidth]{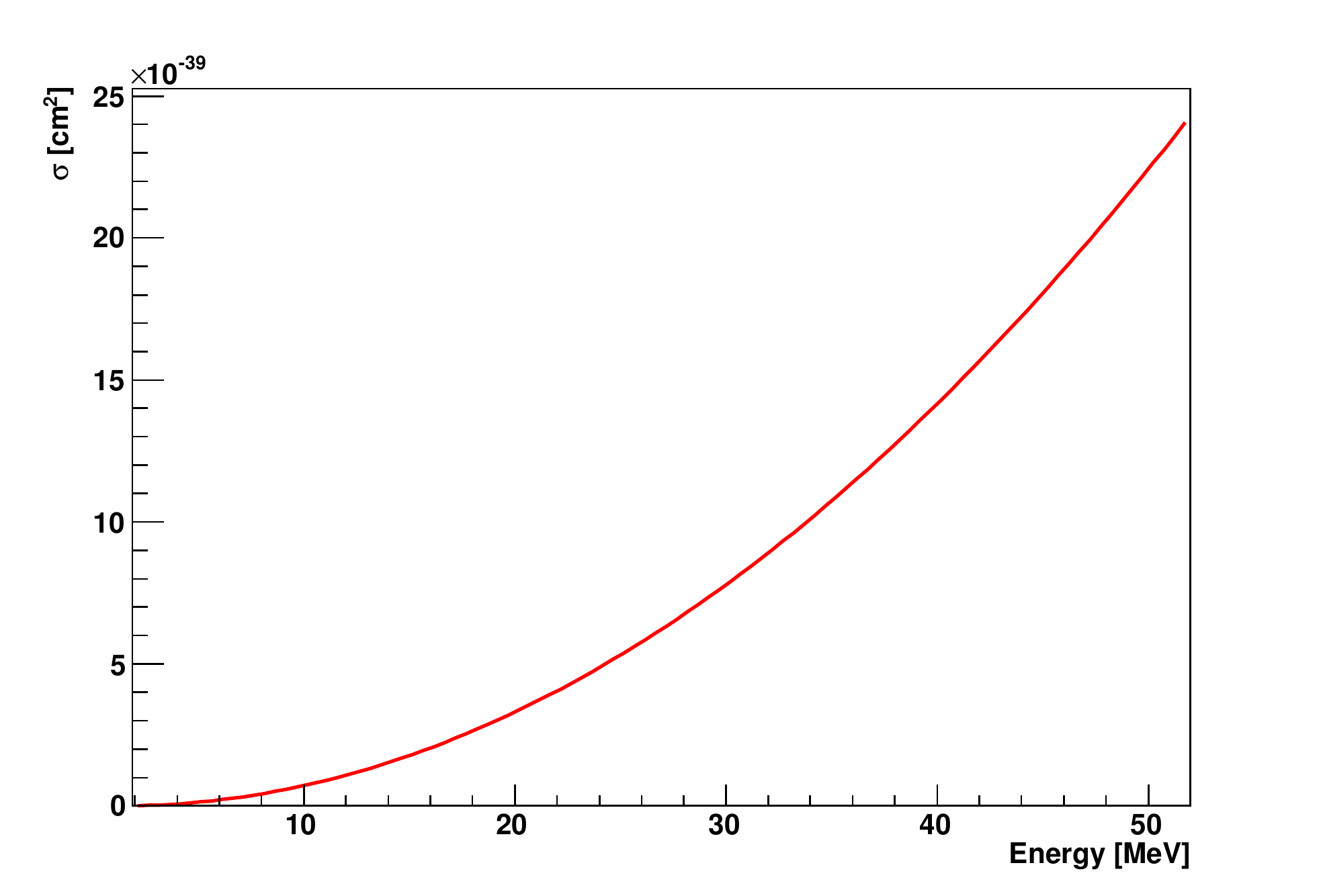}
 \caption{\setlength{\baselineskip}{4mm}
	Flux shape of the $\bar{\nu}_\mu$ produced in the $\mu^+$ decay, on the left, and the IBD energy dependence of the cross-section, on the right.}
 \label{fig:IBD_nu_and_CS}
\end{figure}

We search for the oscillated $\bar{\nu}_{e}$ signal from $\bar{\nu}_{\mu}$,
therefore, its spectral shape will be the same as the one for $\bar{\nu}_\mu$, from a $\mu^+$ decay at rest, described by:
\begin{equation}
\label{eq:nu_mu_flux}
\frac{d \Gamma}{dE_{\bar{\nu}_\mu}} \sim E^2_{\bar{\nu}_\mu} \left( 3 - 4 \frac{E_{\bar{\nu}_\mu}}{m_\mu} \right)
\end{equation}

which comes from the weak theory calculations of the $\mu$ decay, and it is represented in left of Figure~\ref{fig:IBD_nu_and_CS}.

\begin{figure}[htbp]
 \centering
 \includegraphics[width=0.95\textwidth]{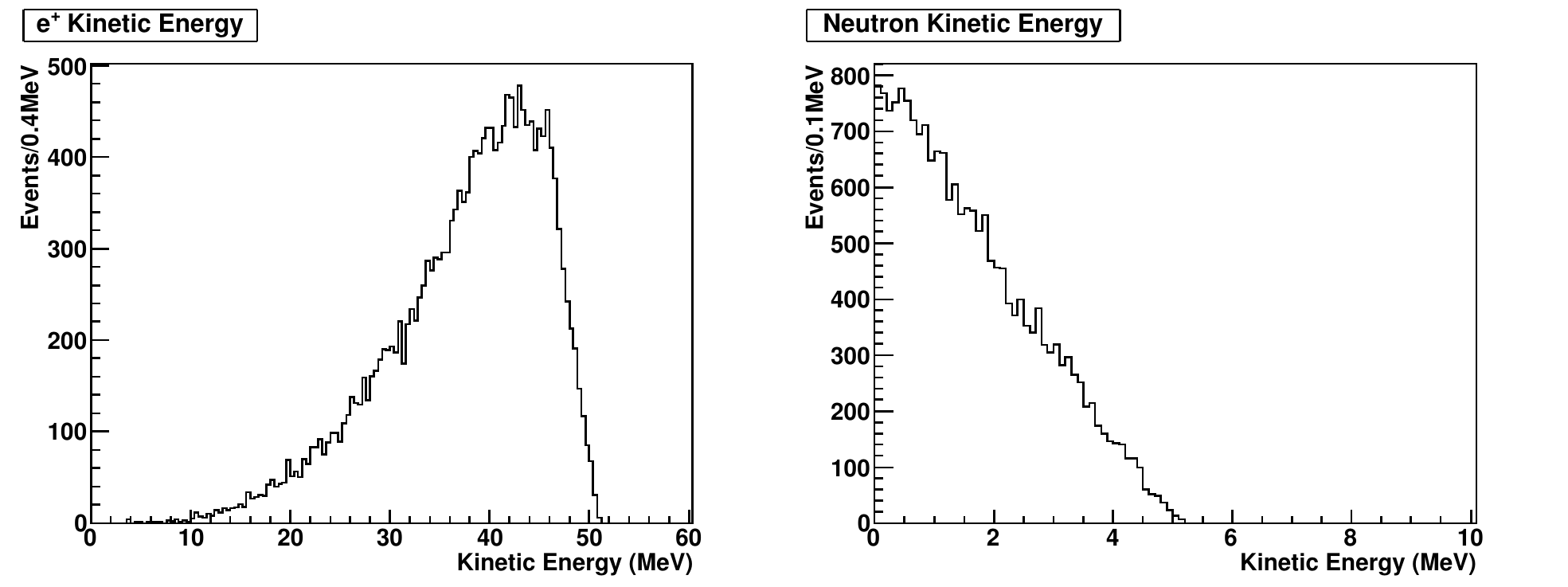}
 \caption{\setlength{\baselineskip}{4mm}
 	Positron (left) and neutron (right) energy distribution after IBD interactions, where the $\bar{\nu}_e$ comes from an oscillated $\bar{\nu}_\mu$.}
 \label{fig:IBD_pos_neu_energy}
\end{figure}

The interaction cross-section of the Inverse Beta Decay (IBD) is shown as:

\begin{equation}
\sigma_{\rm IBD} = \frac{2\pi^2/m_e}{f^{\rm R}_{p.s.}\tau_n}E_ep_e
\end{equation}
where $\tau_n$ is the measured neutron lifetime, $f^{\rm R}_{p.s.} = 1.7152$ is the phase space factor, including the Coulomb, weak magnetism, recoil, and outer radiative corrections, being depicted in right plot of Figure~\ref{fig:IBD_nu_and_CS}. An expression of the cross section depending also on the angle between the neutrino and the emitted positron can be found in~\cite{cite:IBD}. 
Figure~\ref{fig:IBD_pos_neu_energy} shows the energy distributions of the positron and the neutron for the signal events.

%%%%%%%%%%%%%%%%%%%%%%%%%%%%%%%%%%%%%%%%%%%%%%%%%%%%%%%%%%%%%%%%%%
\subsubsection{$^{12}C(\nu_e,e^-)^{12}N_{g.s.}$ BG}
%%%%%%%%%%%%%%%%%%%%%%%%%%%%%%%%%%%%%%%%%%%%%%%%%%%%%%%%%%%%%%%%%%

$\nu_e$'s comes from the $\mu^+$ decay and they can interact with the $^{12}$C of the detector's liquid scintillator, resulting in a $^{12}$N and an electron, i.e.

\begin{equation}
\nu_e + ^{12}\mathrm{C} \rightarrow e^-+^{12}\mathrm{N_{g.s.}} , ^{12}\mathrm{N_{g.s.}}\rightarrow ^{12}C+e^{+}+\nu_{e}
\end{equation}

The nitrogen will be produced in the ground state in about 95\% of the cases, and it will decay with a positron. This decay has an end point energy of $\sim$~16~MeV a half-life of 16~ms. One needs to know the energy of the incident $\nu_e$ to calculate the energy and momentum of the electron, and the description of the nitrogen-12 $\beta^+$ decay, for the positron.

\begin{figure}[htbp]
 \centering
 \includegraphics[width=0.6\textwidth]{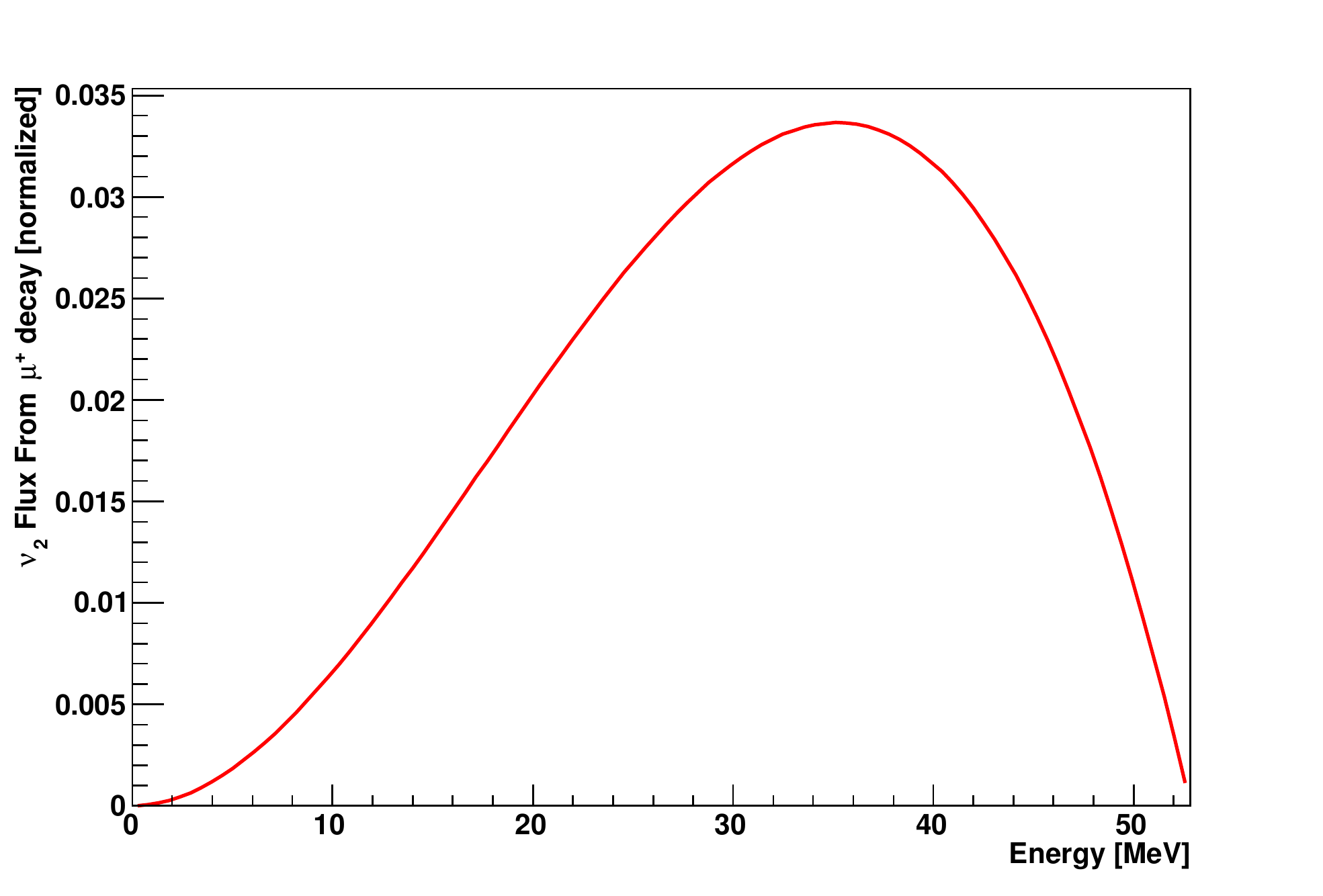}
 \caption{The energy spectrum of  $\nu_e$ from $\mu^{+}$ decay.}
 \label{fig:nue12C_nuEnergy}
\end{figure}

The $\nu_e$ from a $\mu^+$ decay has the well known energy spectrum:

\begin{equation}
\label{eq:nue_from_mu_decay}
\frac{d \Gamma}{dE_{\nu_e}}\sim E^2_{\nu_e} \left( 1 - 2 \frac{E_{\nu_e}}{m_\mu} \right)
\end{equation}

which is represented in Figure~\ref{fig:nue12C_nuEnergy}. 

The positron spectrum is defined as equation~\ref{eq:positron_from_12N}:

\begin{equation}
\label{eq:positron_from_12N}
\frac{d\rm N}{dE_{e}} = P_e E_e (E_{\rm max.} - E_e)^2 \frac{2\pi\eta}{e^{2\pi \eta}-1}
\end{equation}

where 

\begin{equation}
\eta = \frac{Z\alpha}{\beta_e}.
\end{equation}

\begin{figure}[htbp]
 \centering
 \includegraphics[width=15cm]{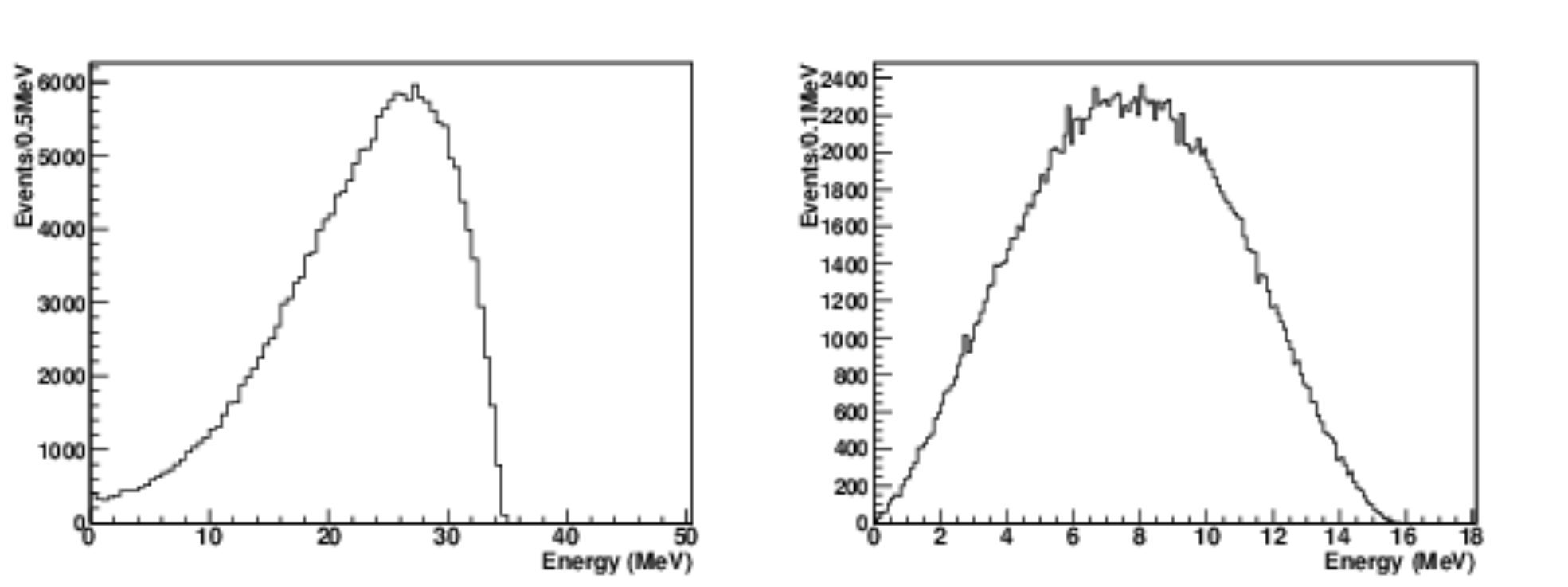}
 \caption{\setlength{\baselineskip}{4mm}
 	Electron and positron energy distribution for the $^{12}C(\nu_e,e^-)^{12}N_{g.s.}$ process. On the left the electron energy is calculated using the kinematics of reaction, and on the right, the positron energy comes from the $^{12}$N decay, as described by equation~\ref{eq:positron_from_12N}.}
  \label{fig:positron_from_12N_decay}
\end{figure}

The positron spectrum generated by the equation is shown in 
Figure~\ref{fig:positron_from_12N_decay}.

\begin{figure}[htbp]
 \centering
 \includegraphics[width=0.45\textwidth]{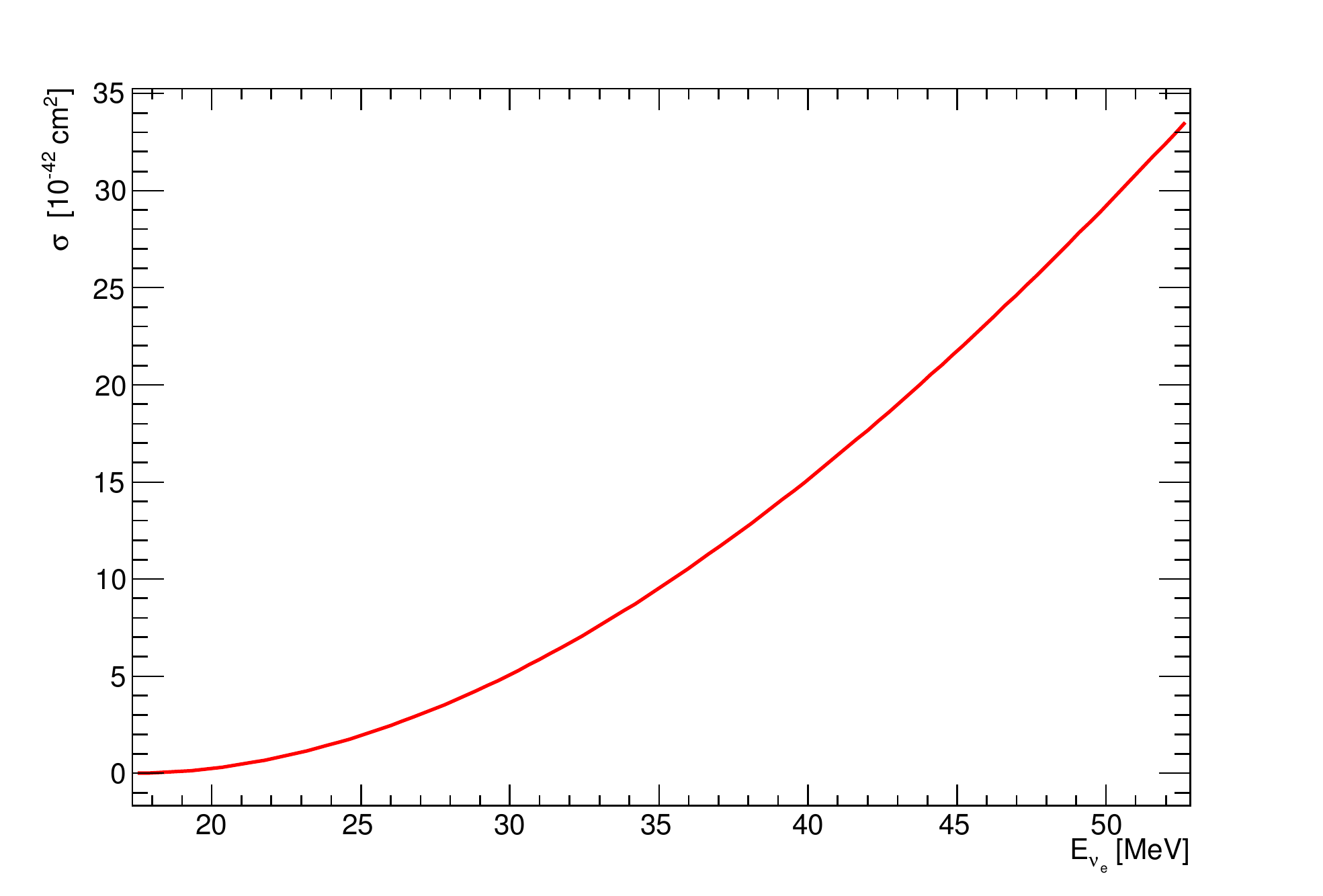}
 \includegraphics[width=0.45\textwidth]{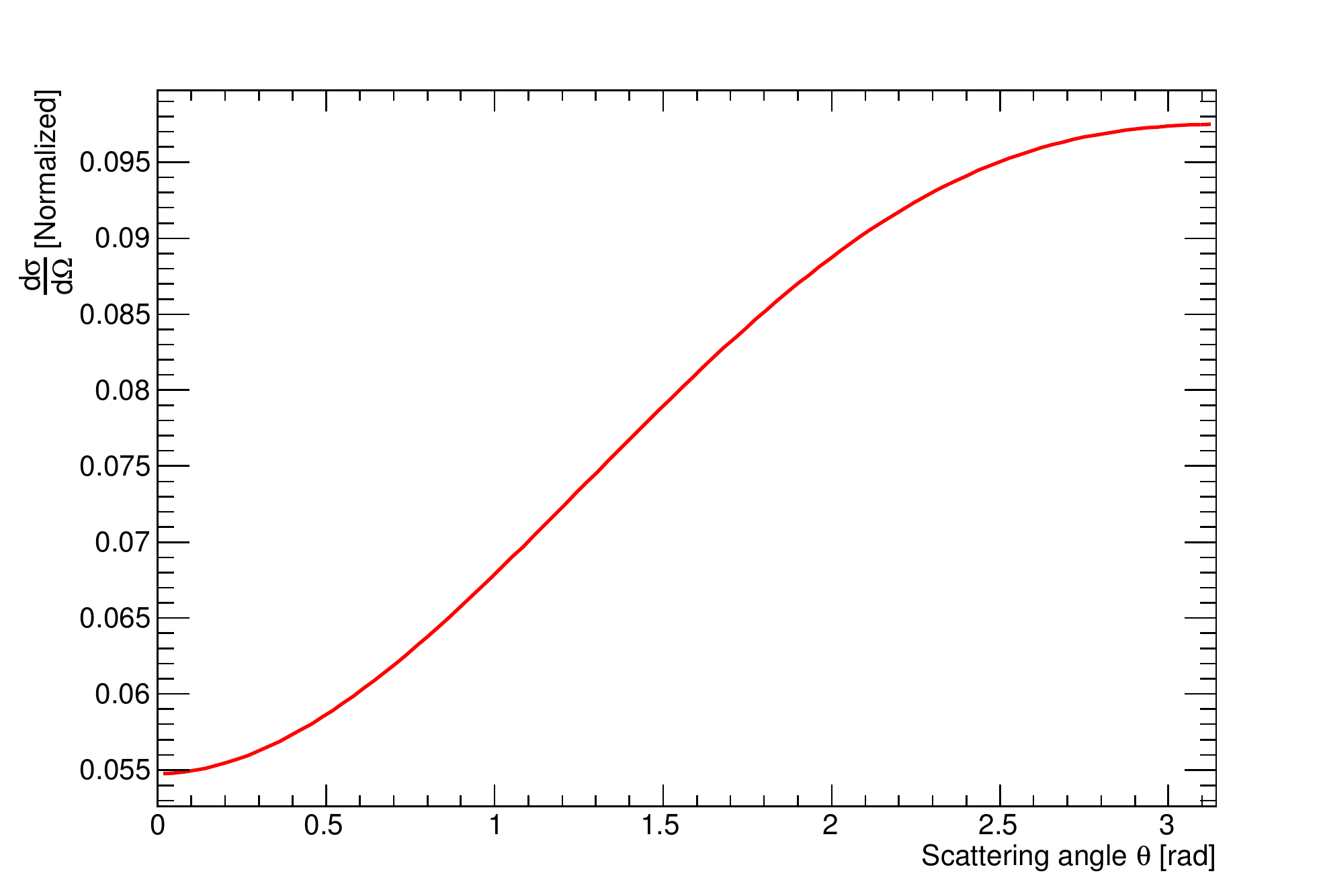}
 \caption{\setlength{\baselineskip}{4mm}
 	Cross section of the $^{12}C(\nu_e,e^-)^{12}N_{g.s.}$ reaction. On the left is the dependence on the neutrino energy and on the right on the angle between the neutrino and the emitted electron~\cite{CCS}.}%Phys. Rev. C 54, 1741 (1996)
  \label{fig:nue12C_crossSec}
\end{figure}

The interaction cross section between the neutrino and the carbon atom has a shape as defined in Figure~\ref{fig:nue12C_crossSec}.

\subsubsection{$\bar{\nu}_e$ from $\mu^-$ decay BG}
Although the $\mu^-$ is absorbed before its decay, a small fraction of them can decay emitting $\bar{\nu}_e$ that will mimic a true $(\bar{\nu}_\mu \rightarrow \bar{\nu}_e)$ oscillation. The energy spectrum is also defined as shown in equation~\ref{eq:nue_from_mu_decay} and the interaction process is the IBD, as for the signal case, having similar distributions as shown in Figures~\ref{fig:nue12C_nuEnergy} for the neutrino and Figure~\ref{fig:IBD_pos_neu_energy} for the IBD products.

\subsubsection{Environmental Gammas}
For the MC simulation of environmental gammas, each of $^{238}$U, $^{232}$Th series, 30 $\gamma$-ray energies having largest branching ratios are considered~\cite{ENVGGEN}, and for $^{40}$K, single $\gamma$-ray of 1.461 MeV are generated in the MC simulation. 

Main environmental gamma sources are gammas from the PMT glass surface and various materials outside of the detector, thus we describe these 
gammas.
 
For the PMT glass surface, this generator was used. Concentrations of $^{238}$U, $^{232}$Th and $^{40}$K in the PMT glass are 13ppb, 61ppb and 3.3ppb, respectively. The gammas are generated isotropically.  
Meanwhile, for the gammas from out side of the detector, it is difficult to generate them precisely from all materials at the experimental site because the geometry is complicated. So assuming that most material generating the gammas is concrete, and the thickness is above 50~cm at least considering with the self shielding of the gammas, the gammas from outside of the detector are generated based on the energy spectrum after passing through 50~cm of the concrete calculated by the MC simulation with this generator. Concentrations of $^{238}$U, $^{232}$Th and $^{40}$K in the PMT glass were calculated by fitting with sum energy spectrum weighted by the concentrations of $^{238}$U, $^{232}$Th and $^{40}$K to measured energy spectrum with the NaI measurement at BL13.

\subsubsection{Cosmic muon}
Cosmic-ray muon generator is based on study by J. Kempa and A. Krawczynaka~\cite{COSMUGEN}. The flux I[cm$^{-2}$s$^{-1}$sr$^{-1}$(GeV/c)$^{-1}$] is expressed as a function of the momentum $p_{\mu}$[GeV/c] and that of zenith angle $\theta$ (deg) by
\begin{equation}
logI=aln^{2}p_{\mu}+blnp_{\mu}+c,
\end{equation}
where $a$ and $b$ are given by
\begin{equation}
Y=p_1/(1/\theta+p_2\theta)+p_3+p_4exp(-p_5\theta),
\end{equation}
where Y means $a$ or $b$ respectively. $c$ is given by 
\begin{equation}
c=p_1\theta^{2}+p_2\theta+p_3+p_4exp(-p_5\theta).
\end{equation}
The coefficients $p_1$,$p_2$,$p_3$,$p_4$ and $p_5$ are listed in Table~\ref{COSMUGENT}. The generated position is
determined by randomizing in a disk with diameter of 60 m and at the height of 6m from ground level.

\begin{table}[htb]
\begin{center}
\begin{tabular}{|c|c|c|c|}\hline
$p_i$&$a$&$b$&$c$\\\hline
$p_1$&-0.8816$\times$10$^{-4}$&0.4169$\times$10$^{-2}$&-0.3516$\times$10$^{-3}$\\\hline
$p_2$&-0.1117$\times$10$^{-3}$&-0.9891$\times$10$^{-4}$&0.8861$\times$10$^{-2}$\\\hline
$p_3$&0.1096&4.0395&-2.5985\\\hline
$p_4$&-0.1966$\times$10$^{-1}$&-4.3118&-0.8745$\times$10$^{-5}$\\\hline
$p_5$&0.2040$\times$10$^{-1}$&-0.9235$\times$10$^{-3}$&-0.1457\\\hline
\end{tabular}
\caption{The values of the coefficients of a,b and c.}
\label{COSMUGENT}
\end{center}
\end{table}

\subsubsection{Cosmic induced fast neutron}
\begin{figure}[htbp]
\begin{center}
\includegraphics[width=15cm]{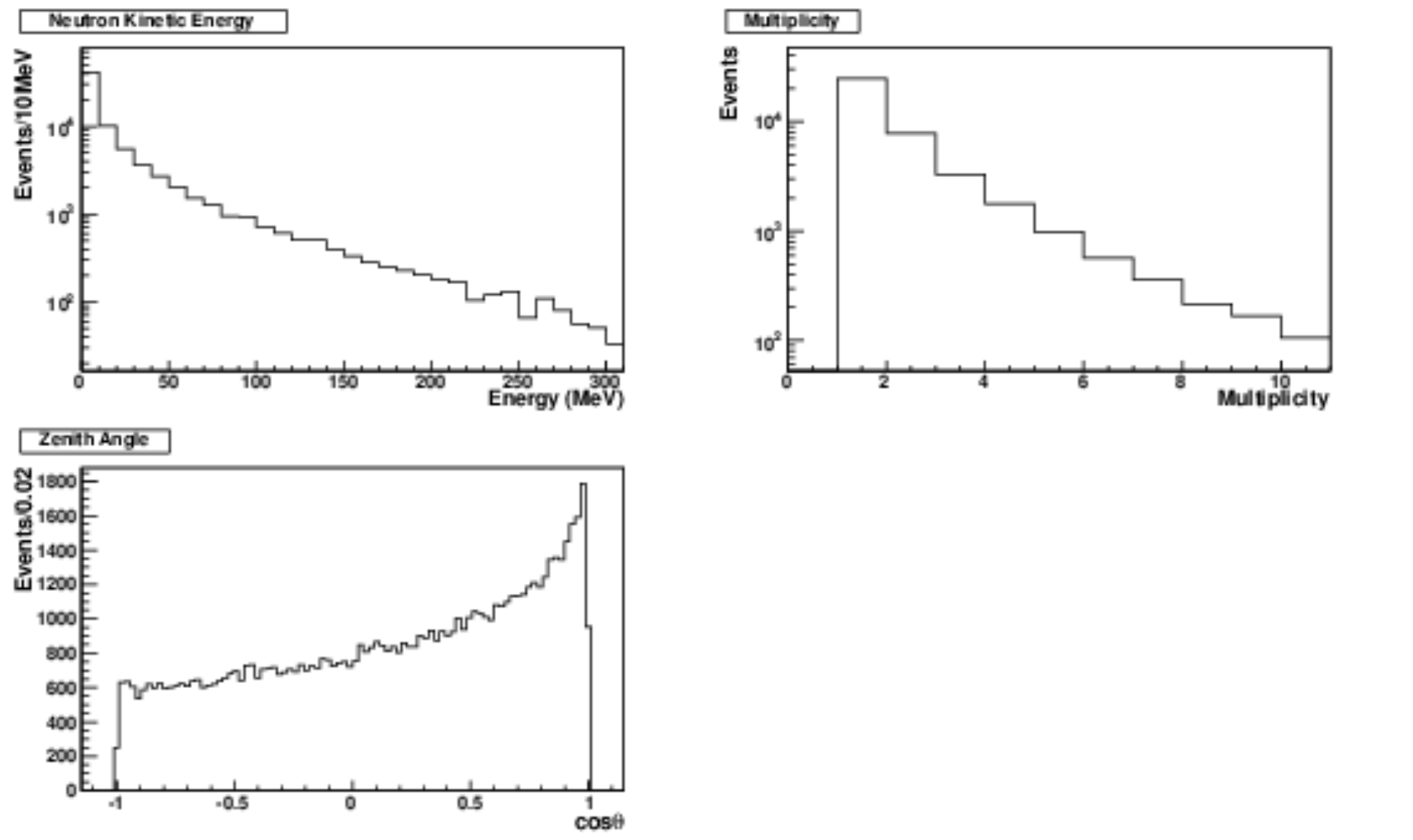}
\end{center}
\caption{\setlength{\baselineskip}{4mm}
	Distributions of the neutron kinetic energy(upper left plot), multiplicity(upper right plot) and zenith angle (lower plot) of the cosmic muon induced fast neutron events generated by the empirical functions in the reference~\cite{COFASTN}.}
\label{COFASTND}
\end{figure}

In order to estimate the cosmic muon induced fast neutron events rate entering in the detector at the candidate site(see Appendix~\ref{COSMUEVE}), empirical functions depending on the muon energy($E_{\mu}$) in reference~\cite{COFASTN} are used for generating kinetic energy spectrum($E_{n}$), multiplicity($M$) and angular distribution($\theta$) of the fast neutrons for the MC simulation. Figure~\ref{COFASTND} shows distributions of the kinetic energy, multiplicity and zenith angle. 
The kinetic energy spectrum,  multiplicity and angular distribution are defined as functions~\ref{COFASTNE}, \ref{COFASTNA} and \ref{COFASTNM}, respectively as follows;
\begin{equation}
\frac{dN}{dE_{n}}=A(\frac{e^{-7E_{n}}}{E_{n}}+B(E_{\mu})e^{-2E_{n}}),
\label{COFASTNE}
\end{equation}
where A is a normalization factor, and $B(E_{\mu})=0.52-0.58e^{-0.0099E_{\mu}}$.

\begin{equation}
\frac{dN}{dM}=A(e^{-A(E_{\mu})M}+B(E_{\mu})e^{-C(E_{\mu})M}),
\label{COFASTNM}
\end{equation}
where $A(E_{\mu})=0.085+0.54e^{-0.075E_{\mu}}$, $B(E_{\mu})=\frac{27.2}{1+7.2e^{-0.076E_{\mu}}}$, $C(E_{\mu})=0.67+1.4e^{-0.12E_{\mu}}$.

\begin{equation}
\frac{dN}{dcos\theta}=\frac{A}{(1-cos\theta)^{0.6}+B(E_{\mu})}
\label{COFASTNA}
\end{equation}
where $B(E_{\mu})=0.699E^{-0.136}_{\mu}$. Then, the fast neutron events were generated uniformly from cylindrical surface with 3.2m of the radius and 6.4m of the height(1m outside from the SUS tank surface).

%%%%%%%%%%%%%%%%
\subsection{Detector simulation}
%%%%%%%%%%%%%%%
\subsubsection{Detector simulation}
In this proposal, Geant4 was used for detector simulation study~\cite{GEANT4}. Geant4 is C++ class library produced by CERN, which provides a calculation of particle tracking in materials. Detector simulator was based on Geant4~(The version is 4.9.0.p1.). The simulator computes interactions between incident particles and the detector materials, and also trajectory of optical photons emitted by the scintillation radiation process. For the hadronic interaction process, QGSP BIC HP model was employed. It comprehends from low energy region under 20MeV such as behavior of thermal and fast neutron to high energy region such as interactions between cosmic-ray muons and materials around the detector. The simulator also follows the trajectory of optical photons emitted in the liquid scintillator due to ionization by charged particles, the optical process includes attenuation and scattering of the photons. The number of these incident photons and the timing for each PMT are stored in the simulator after reduction due to the quantum efficiency($\sim$20\%).

Time distribution of the scintillation lights emission is assumed as sum of 2 exponentials with fast and slow time constants of the liquid scintillator. The fast and slow time constants are 3.6ns and 270ns, respectively, and ratio of the fast and slow constants is 0.57.

\subsubsection{Pulse generation}
\begin{figure}[htbp]
 \begin{center}
 \includegraphics[width=12cm]{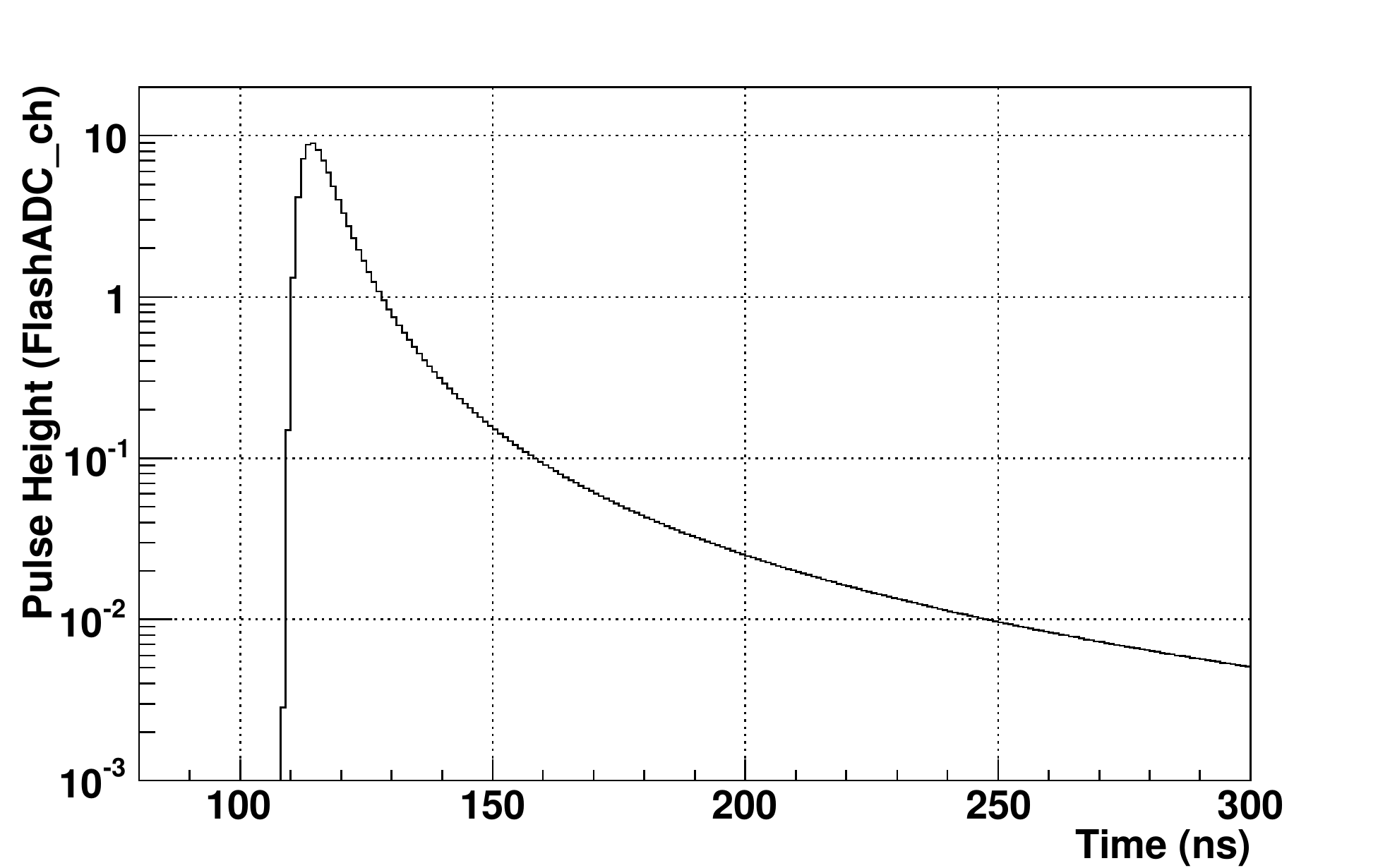}
 \end{center}
 \caption{\setlength{\baselineskip}{4mm}
 	An example of the reconstructed MC pulse of PMT.}
 \label{WFORM}
\end{figure}

The hit times for each PMT from the Geant4 output are shifted with transit time (TT$\sim$60ns) and smeared following Gaussian distribution with transit time spread (TTS$\sim$2.9ns) for  the 10inches PMT(HAMAMATSU R7081) in the data sheet. One photoelectron pulse was reconstructed assuming Landau distribution, then the shifted and smeared time were used for mean time of the Landau function, and the sigma estimated by Double Chooz group was used in our sigma. One pulse for each PMT was reconstructed as sum of the one photoelectron pulses.
Figure~\ref{WFORM} shows an example of the one pulse for each PMT. Finally, the pulse shape depends on sum of time distributions of the scintillation lights emission of fast and slow component in previous paragraph.

\pdfoutput=1
%%%%%%%%%%%%%%%%%%%%%%%%%%%%%%
\section{\setlength{\baselineskip}{4mm}
Estimation of cosmic muon induced backgrounds events}
%%%%%%%%%%%%%%%%%%%%%%%%%%%%%%
\label{COSMUEVE}

Neutron and some unstable isotopes are produced by reactions between cosmic muon and nuclei(especially $^{12}$C) inside the organic liquid scintillator or materials at the experimental site(spallation products). Then especially, IBD mimic signals are induced by fast neutron produced by the reaction between the muon and materials outside of the detector. Basically, neutron produced inside the detector and short-lived unstable isotopes can be vetoed because the detected signals include parent muons passing through the detector, so they do not contribute to IBD mimic signals.

\subsection{Fast neutron from outside of the detector}
\label{FASTN}
\begin{figure}[htbp]
 \begin{center}
 \includegraphics[width=15cm]{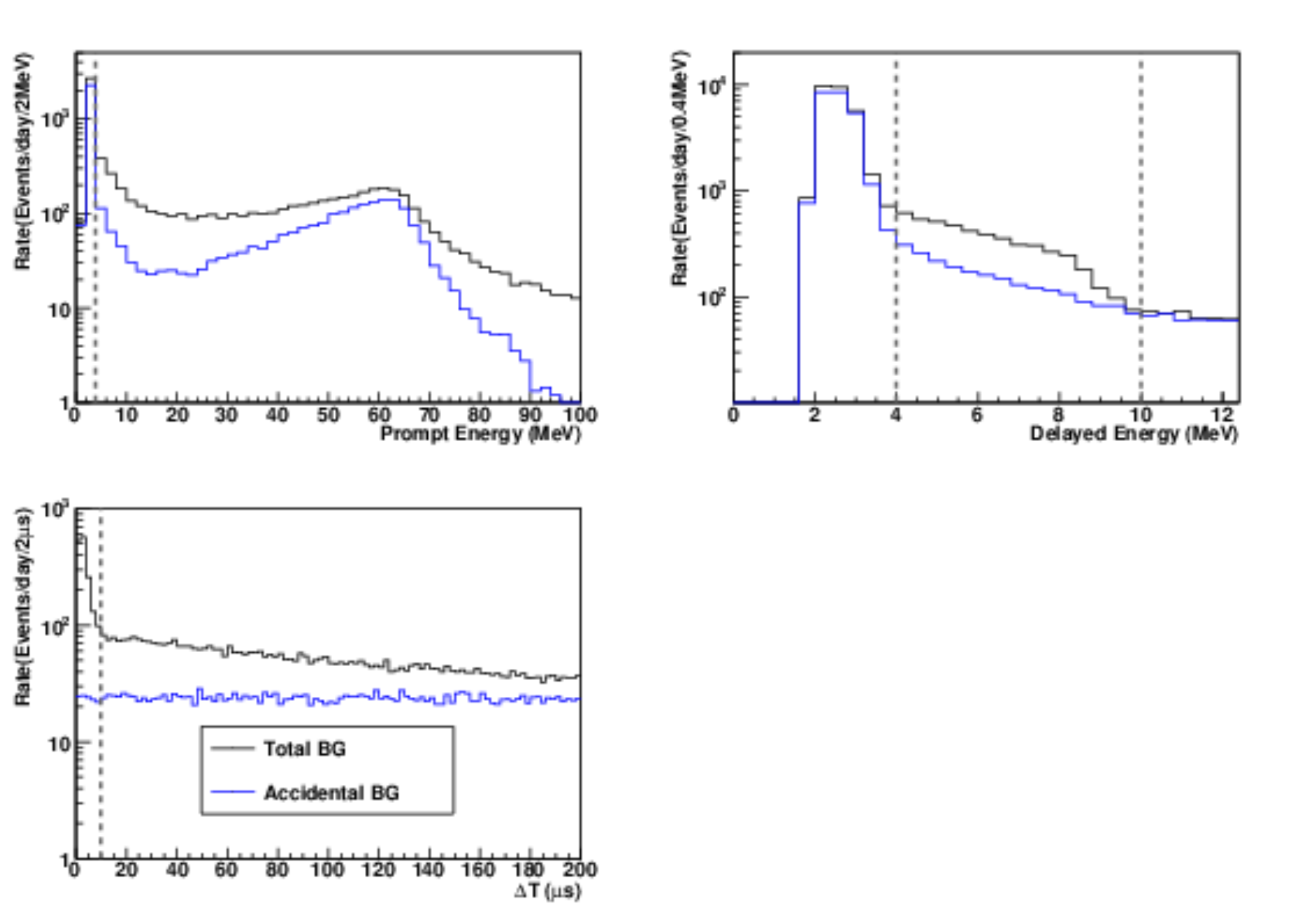}
 \end{center}
 \caption{\setlength{\baselineskip}{4mm}
 	Energy spectra of the prompt(left upper) and delayed signals(right upper), and the $\Delta$t distribution(lower) of muon induced fast neutron candidates with the 200L Gd-loaded liquid scintillator detector. Gray dashed lines show the cut condition. Black lines show the candidates including both correlated and accidental events. Blue lines show only accidental events estimated by off time coincidence~\cite{TOHOKUMONITOR}.}
 \label{FASTN1}
\end{figure}

\begin{figure}[htbp]
 \begin{center}
 \includegraphics[width=15cm]{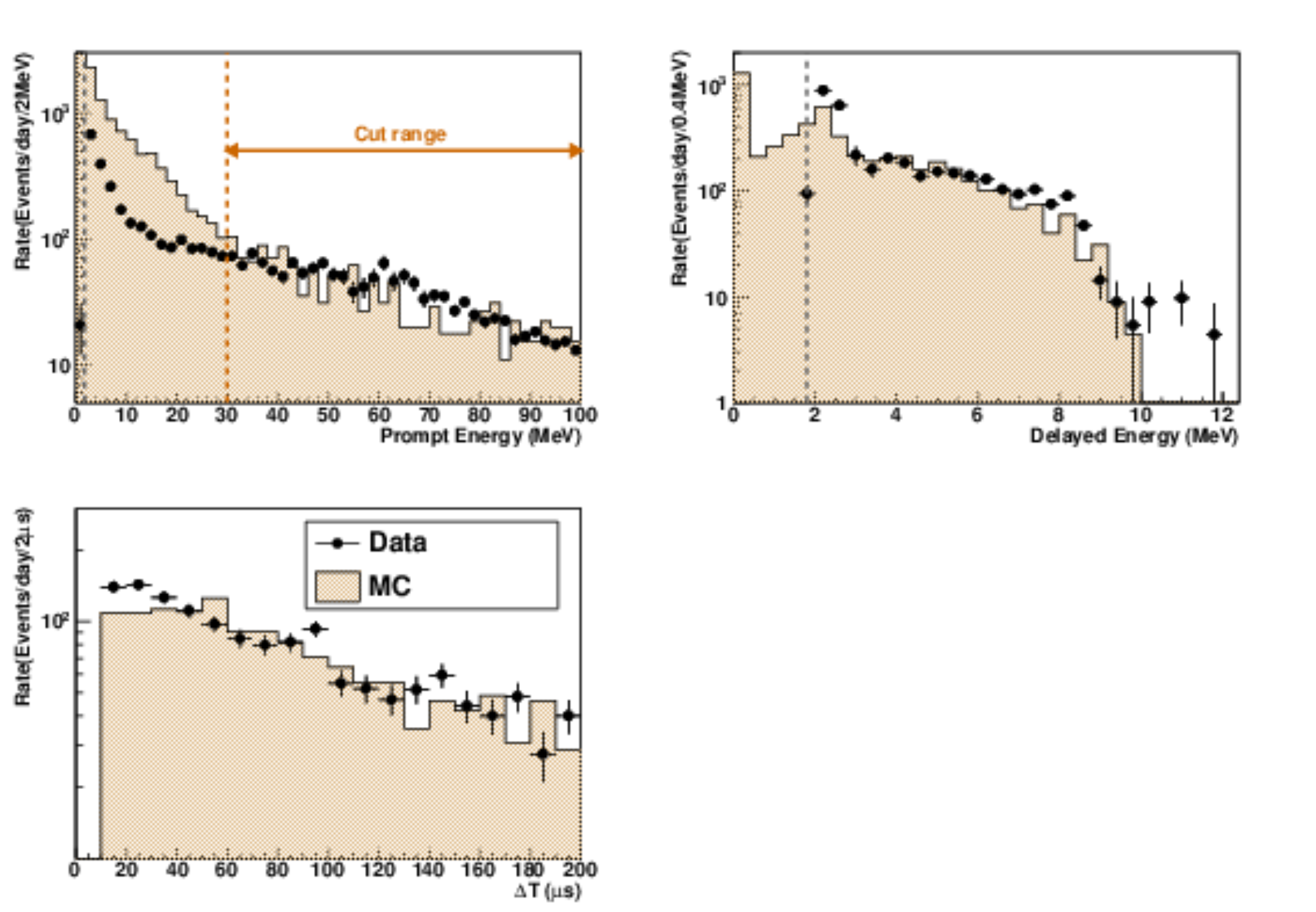}
 \end{center}
 \caption{\setlength{\baselineskip}{4mm}
 	Comparison of distributions of each variables between the measured Tohoku data and the MC samples scaled by the neutron flux calculated by the fitting. Upper left and right figures show the prompt and delayed energy spectra. Lower figure shows the $\Delta$t distribution. Gray lines show the energy threshold level.}
 \label{FASTN_Fit}
\end{figure}

\begin{figure}[htbp]
 \begin{center}
 \includegraphics[width=14cm]{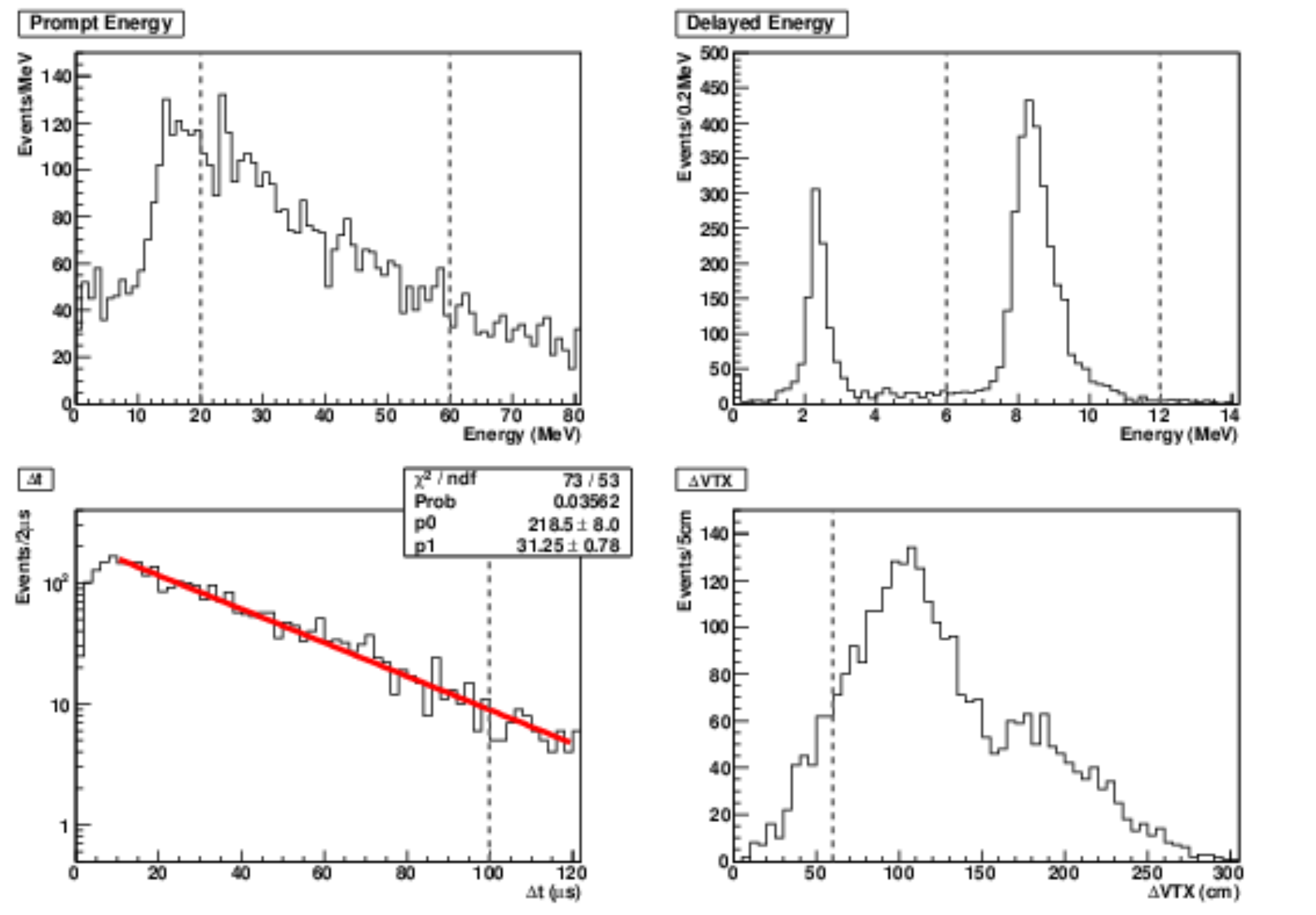}
 \end{center}
 \caption{\setlength{\baselineskip}{4mm}
 	Expected energy spectra of prompt and delayed signals, $\Delta$t and $\Delta$VTX distributions of the muon induced fast neutron events from outside of the detector. (The detector put at the 3rd floor of MLF (to be measured)) }
 \label{FIGFN}
\end{figure}

In this proposal, in order to estimate the fast neutron rate at the candidate site, we used measured data of the fast neutron above ground with the 200L Gd-loaded liquid scintillator detector at an experimental room in Tohoku Univ.~\cite{TOHOKUMONITOR}.
%because we have not carried out some background measurement at the candidate site yet. 
Basically, the fast neutron flux is related to muon energy, flux depending on depth of experimental site and atomic elements composing materials in the experimental site. We can estimate order of magnitude of the fast neutron flux by using the measured data above ground in Tohoku Univ., even if geometry and materials of the experimental room in Tohoku Univ. are differ from ones of the candidate site. We will carry out background measurement with same detector at the candidate site in near future, then we will estimate the fast neutron flux more precisely. The fast neutron candidates from outside of the detector are events after applying selection criteria, which are $4<E_{prompt}<100MeV$, $4<E_{delayed}<10MeV$ and $10<\Delta t<200\mu s$.
Figure~\ref{FASTN1} shows energy spectra of the prompt and delayed signals, and the $\Delta$t distribution after applying the selection criteria. Excesses of black lines in Figure~\ref{FASTN1} show the correlated events. 
(Blue lines show only accidental events estimated by off time coincidence~\cite{TOHOKUMONITOR}.)
The excess in the prompt energy spectra distribute until high energy range above 100MeV. The excess in the delayed energy spectra distribute below around 8 MeV, which is sum of energies of gammas generated via thermal neutron capture on Gd. Most events below 4MeV is environmental gammas. The excess of the $\Delta$t distributions has two components, which distribute based on exponentials with 2.2~$\mu$s of muon life time and about 100~$\mu$s of the thermal neutron capture time on Gd. The selection criteria was defined considering with reducing the environmental gammas and Michel electrons after muon decays in the detector.
The absolute fast neutron flux was calculated by $\chi^{2}$ fitting with the excess of $\Delta$t distribution of the fast neutron candidates above 30 MeV of the prompt energy in measured data to one in the MC samples with the fast neutron generator based on a reference ~\cite{COFASTN}(see~\ref{EVENTGEN}). Figure~\ref{FASTN_Fit} shows comparison of distributions of each variables between the measured data and the MC samples scaled by the neutron flux calculated by the fitting. The prompt energy spectra of measured data and MC samples are not consistent in whole range below 100MeV. Events around several tens MeV include not only fast neutrons entering from outside of the detector but also muon events following neutrons produced inside the detector, because the detector does not implement veto counter for cosmic muons, the muon events following neutrons are remaining in the selection criteria. 
%And the prompt energy spectra of measured data in KARMEN experiment and the MC samples with this generator do not correspond in the lower energy range~\cite{COFASTN}. 
Therefore the fast neutron candidates above 30 MeV of the prompt energy was considered conservatively, and the $\Delta$t distributions were used for the fitting because the distribution is known as a exponential depending on the thermal neutron capture time on Gd. The measured fast neutron flux is 17~Hz/m$^{2}$. The uncertainty of the fitting is $\sim$3~\%.
Figure~\ref{FIGFN} shows the energy spectra of prompt and delayed signals, $\Delta$t and $\Delta$VTX distributions of MC samples for the detector of MLF experiment. The neutron events were generated from cylindrical surface of the SUS tank with the fast neutron generator. 
Finally, the fast neutron rate entering in surface of the SUS tank is expected to be 747~Hz. The remaining events rate after applying the IBD selection criteria, is 0.68~Hz. Furthermore, it is effective for reducing the fast neutron events to apply that there is no veto signal, because sometime there are signals due to recoiled protons in the veto volume before entering the buffer volume. The rate after applying also no veto signal is reduced by $\sim$1/4, it is expected to be 0.16~Hz.
But all of the rate do not contribute to the background rate, only coincidence rate with the beam timing contributes to the background rate. Then the coincidence rate(R$_{coin}$) considering with the prompt time window is calculated as follows.
\begin{equation}
R_{coin}=0.16Hz \times 9.0\mu s = 1.4\times10^{-6}/spill
\end{equation}
Then, the number of the fast neutron events considering two detectors is $R_{con}\times$ 3.6$\times$ 10$^{8}$ $\times$ 2 detectors $\times$ 4~years = 4150~events. The real number of the fast neutron events will be measured in near future at the candidate site. If there is several hundred events per year such as above estimation, it is necessary for us to reduce the events. The number can be reduced by at least 1/100 by implementing some paraffin shielding with several tens cm of the thickness(can be reduced by $\sim$1/10) and liquid scintillator with capability of particle identify like LSND experiments(can be reduced by $\sim$1/100).

\subsection{Spallation products}
\label{CSPP}
Rates of neutron and the long-lived unstable isotopes produced by the spallation reactions depend on muon energy, flux and carbon concentration in the liquid scintillator. KamLAND group measured rates of the various isotopes~\cite{KamSP}, so the rates at the candidate site are extrapolated by the formula below using ratios of target masses(M$_{i}$), mean muon energies(E$_{i}$), fluxes($\phi_{i}$) and carbon concentrations in the scintillator(N$_i$) between our experiment($i=JP$) and KamLAND($i=KL$). Values for the calculation are shown in Table~\ref{VSP}. Sum of the mass of the fiducial and the buffer scintillation region was considered for the calculation.

\begin{equation}
R^{JP}_{isotope}=\frac{M_{JP}}{M_{KL}}\times \frac{N_{JP}}{N_{KL}}\times  \left( \frac{E_{JP}}{E_{KL}}\right)^{\alpha} \times \frac{\phi_{JP}}{\phi_{LK}} \times R^{KL}_{isotope}
\end{equation}
Where, $\alpha$ is a constant of power law for correlation between the mean muon energy and production rates of the spallation isotopes($\alpha$=0.74). The $R^{KL}_{isotope}$ of each isotope was calculated using values in \cite{KamSP}. Considering the prompt energy cut range in the IBD selection criteria, the spallation products is expected to contribute the delayed like signals for the accidental backgrounds. Then the rate per spill for each isotope in 100~$\mu$s time window of the IBD selection is calculated as $R^{JP}_{isotope} \times 100\mu s$.
Table~\ref{SP} shows a summary of the rates per spill per one detector. Basically, the spallation products are expected to be negligible as IBD mimic events.

\begin{table}[htb]
\begin{center}
\begin{tabular}{|c|c|c|c|c|}\hline
&M$_{i}$&N$_{i}$&E$_{i}$&$\phi_{i}$ \\\hline
KamLAND&1000&4.30$\times$10$^{22}$&260&1.49$\times$10$^{-3}$ \\
(i=KL)&(ton)&(carbon/g)&(GeV)&(/m$^{2}$/sec)\\\hline
J-PARC&37&4.30$\times$10$^{22}$&4&100 \\
(i=JP)&(ton)&(carbon/g)&(GeV)&(/m$^{2}$/sec)\\\hline
\end{tabular}
\caption{Values for calculation of rates of spallation products.}
\label{VSP}
\end{center}
\end{table}

\begin{table}[htb]
\begin{center}
\begin{tabular}{|c|c|c|c|}\hline
Isotope&Life time&Radiation energy&R$^{JP}_{isotope}$\\
&&(MeV)&(events/spill)\\\hline \hline
Neutron&-&8(Gd capture)&2.6$\times$10$^{-4}$\\\hline
$^{12}$B&29.1ms&13.4($\beta^{-}$)&5.0$\times$10$^{-5}$\\\hline
$^{12}$N&15.9ms&17.3($\beta^{+}$)&1.5$\times$10$^{-6}$\\\hline
$^{8}$Li&1.21s&16.0($\beta^{-}\alpha$)&3.9$\times$10$^{-5}$\\\hline
$^{8}$B&1.11s&18.0($\beta^{-}\alpha$)&1.1$\times$10$^{-5}$\\\hline
$^{9}$C&182.5ms&16.5($\beta^{+}$)&1.9$\times$10$^{-6}$\\\hline
$^{8}$He&171.7ms&10.7($\beta^{-}\gamma n$)&5.6$\times$10$^{-7}$\\\hline
$^{9}$Li&257.2ms&13.6($\beta^{-}\gamma n$)&5.6$\times$10$^{-6}$\\\hline
%$^{11}$C&29.4min&1.98($\beta_^{+}$)&-\\\hline
%$^{10}$C&27.8s&3.65($\beta_^{+}\gamma$)&-\\\hline
$^{11}$Be&19.9s&11.5($\beta^{-}$)&1.5$\times$10$^{-6}$\\\hline
%$^{6}$He&1.16s&3.51($\beta_^{-}$)&0.12\\\hline
%$^{7}$Be&76.9day&0.478(EC$\gamma$)&-\\\hline
\end{tabular}
\caption{Spallation products}
\label{SP}
\end{center}
\end{table}

\pdfoutput=1

%=============================
\section{Energy and Vertex Position Calibrations}
%----------------
\indent

     The energy range of the MLF experiment to observe is from 8 to 53~MeV. 
     Intrinsic non-linearity of the light output caused by the quenching effect and Cherenkov threshold effect is supposed to be small but on the other hand, non-linearity due to PMT or electronics saturation may be large. 
 Radioactive sources can be put in the target region from the calibration port at the top.  
 A Cf fission source will be used to calibrate the 8~MeV Gd energy and efficiency. 
 For higher energy prompt signals, there is no radioactive sources to cover the necessary energy range.
 However, the Michel electron has well known and similar energy spectrum as expected 
 $\bar{\nu}_e$ energy spectrum and can be used to calibrate the energy.
 High energy cosmic spallation signals will also be used for the energy calibration. 
 
 The fiducial volume is defined by the Gd signal which occurs only in the neutrino target region and no fiducial cut will be applied at the analysis. 
  Thus the vertex position uncertainty does not affect the accuracy of the detection efficiency for the first order. 
 However, reconstructed vertex position will be used to measure baseline dependence of the oscillation, for event selection cut based on the distance between prompt and delayed signals, and to correct position dependence of energy response.  
 Position along the $z$-axis can be calibrated by the deployed sources from the calibration port. 
 For $r$-direction, the radius of the acrylic vessel wall can be identified by looking at the edge effect of signals.
Non uniformity of signal distribution can be checked by looking at the uniformly distributed Michel electron or spallation signals.

\pdfoutput=1
\section{MLF radiation survey}
\indent

Radiation survey has been performed by MLF facility people for the
safety issue. Especially, the survey held on 22-Oct-2012 is important
measurement since the beam power is 284 kW, close to the beam power
at the 1 ton scintillation measurement at BL13.

The survey was performed at many places at the MLF facility, and it 
showed that radiation level of BL13, where the 1 ton scintillator detector is
located,
is higher than that at any other points although it is much safer compared to 
the limit of radiation. 
Table~\ref{Tab:radiation} summarizes 
the survey result for BL13 and the candidate site.

\begin{table}[h]
\begin{center}
  \begin{tabular}{|c|c|c|c|}
  \hline
  Point & neutron & gamma & comments \\ \hline \hline
  BL13 & 1.2$\mu$Sv/h & 0.5$\mu$Sv/h & digit of gamma monitor is 0.1$\mu$Sv/h    \\ \hline
  Candidate site & 0.025$\mu$Sv/h & 0.1$\mu$Sv/h & digit of neutron monitor is 0.01$\mu$Sv/h.   \\ 
  \hline
  \end{tabular}
  \caption{Radiation Survey results}
	\label{Tab:radiation}
\end{center}
\end{table}

This survey monitor counts any energy of the gamma and neutrons, 
therefore the numbers are not guranteed to be directly related to 
amount of the fast and slow neutrons, and gammas. However, the 
radiation from the neutrons, which are source of the many backgrounds, 
are reduced by almost two order of 
magnitudes compared between BL13 and the candidate site.   
Note that measured digit of gamma monitor is 0.1$\mu$Sv/h, while
a digit of neutron monitor is 0.01$\mu$Sv/h. 

This radiation survey reports support the PHITS conclusion.    

\pdfoutput=1
%%%%%%%%%%%%%%%%%%%%%%%%%%%%%%%%%%%%%%%
\section{Consideration on the detector}
%%%%%%%%%%%%%%%%%%%%%%%%%%%%%%%%%%%%%%%

%%%%%%%%%%%%%%%%%%%%%%%%%%%%%%%%%%%%%%%%%%%%%%%%%%%%%%%%%%%%%
\subsection{An alternative detector concepts- pros and cons}
%%%%%%%%%%%%%%%%%%%%%%%%%%%%%%%%%%%%%%%%%%%%%%%%%%%%%%%%%%%%%
\indent

An alternative option of the detector is segemnted detector like
KARMEN~\cite{cite:KARMEN}. 
Compared to the LSND type detector, there are following pros and
cons. 

Pros; 
\begin{itemize}
\item A vertex resolution is determined by the size of one module. 
     The detector also recognizes the multi-vertices easily. Michel 
     electrons originated by fast neutrons' interactions can be rejected
     with the features.    
\item The self-shielding region of the detector for gammas from PMTs 
    and enviroment can be reduced since PMTs are attched outside 
    of the detector.     
\item The pilot runs for physics with small number of modules can be 
    performed. The prototype module can be tested easily.
\end{itemize}

Cons;
\begin{itemize}
\item Number of PMTs are larger than that in LSND type.
\item Particle Identification with Cherenkov is difficult. 
\end{itemize}

%%%%%%%%%%%%%%%%%%%%%%%%%%%%%%%%%%%%%%%%%%%%%
\subsection{Bases of detector type choice}
%%%%%%%%%%%%%%%%%%%%%%%%%%%%%%%%%%%%%%%%%%%%%
\indent

As described in the main text, fast neutrons can create Michel 
electrons via charged pion production, thus we have to remove the 
events carefully. 

LSND type detectors cannot reconstruct multi-vertices 
correctly, therefore the most of events which have beam on-bunch 
activities have to be cut. 
This strategy may be damaged if the number of beam spills, which 
have on-bunch activities, is larger than those has no on beam bunch 
activity.

On the other hand, the segmented detector can reconstruct the on-
bunch multi-vertices events correctly. The reconstructed vertices 
are used for the spatial veto for the Michel electrons without 
thrown away the event itself since the distance between
the signal from charged pions (on-bunch) and Michel electrons are 
correlated as shown in Section 4.       

In conclusion, the next background measurement determines the 
detector choice. If the on-bunch activities are manageable with a 
LSND type detector, the strategy in this proposal of the detector is 
kept, however it could be necessary to change it if the on-bunch 
activities requires it.

%%%%%%%%%%%%%%%%%%%%%%%%%%%%%%%%%%%%%%%%%%%
\subsection{\setlength{\baselineskip}{4mm}
Vertex and Energy resolution for the alternative detector}
%%%%%%%%%%%%%%%%%%%%%%%%%%%%%%%%%%%%%%%%%%%
 
KARMEN detector consists of liquid scintillator modules, which have 
size of $\sim$18cm $\times$ 18cm $\times$ 350cm. 
Four 3 inch PMTs are attached to both sides of the 350 cm length. 

Vertex resolution for two directions are determined by module size.
As known well, typical resolution of tracking is calculated that 
the size detector devided by $\sqrt{12}$, that is, 5.2 cm.
The resolution for the long detector side is determined by the timing
resolution of the optics (a path length of scintillation light and 
PMT response). The timing resolution of the KARMEN detector is 350 ps
due to careful scintillation ray-trace simulation, and it provides
5 cm vertex resolution.  

To identify the multiple vertices in an events, 5 cm vertex resolution is
good enough since there are a few events in 50 tons fiducial volume, 
the overlap of the vertices are negliegibly small.

\hspace*{5mm}
 
For the energy resolution, KARMEN achieved the 11.6$\% \sqrt{E}$. It is 
similar or better than that of LSND type detector in the proposal.

\pdfoutput=1
\section{\setlength{\baselineskip}{4mm}
Possible future extension using a large detector at longer distance}

If neutrino oscillation is not found in the high $\Delta m^2$ region by 
the detectors at 17~m, an extension of the experiment should be considered 
to search for the sub-eV$^2$ region. 
One possibility is to use $\sim$1~kton detector at  $\sim$60~m from the target. 
The longer baseline extends the sensitivity to the lower $\Delta m^2$ region.
By using the 17~m detectors as near detector, systematic uncertainties are 
widely canceled out. 
Moreover, it becomes possible to separate physical $\bar{\nu}_e$ background 
from signal $\bar{\nu}_e$ 
because the background $\bar{\nu}_e$ flux is proportional to $1/L^2$ 
while the signal $\bar{\nu}_e$ flux is proportional to 
$\sin^2(\Delta m^2L/4E)/L^2$.
 
\begin{figure}[htbp]
   \begin{center}
    \includegraphics[keepaspectratio=true,height=120mm]{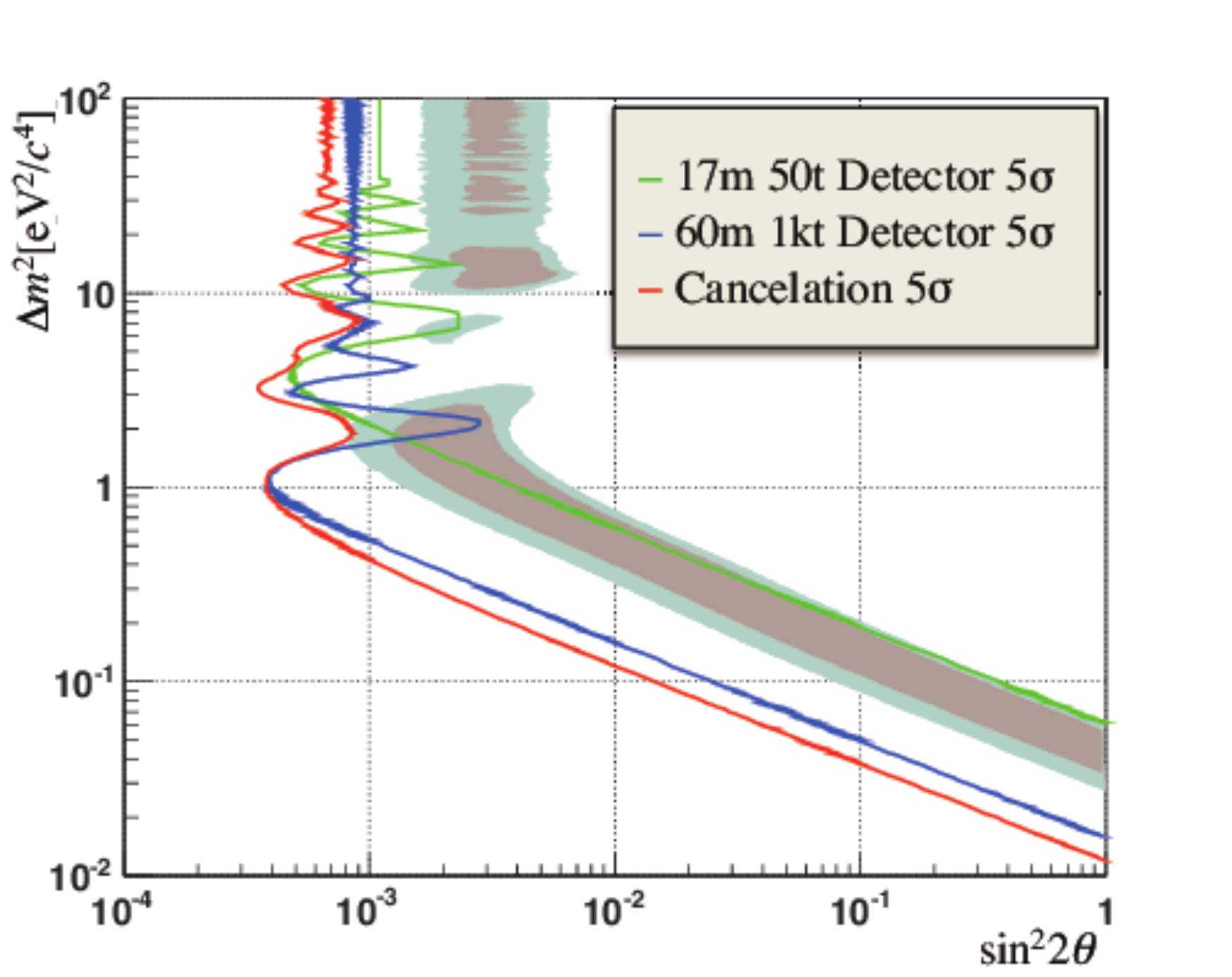}
    \caption{ \setlength{\baselineskip}{4mm}
Typical sensitivity using 1.0~kton detector with 60~m baseline. 
The green line is the sensitivity of the experiment in this proposal at 17~m. 
The blue line is the far-detector-only sensitivity at 60~m and 
the red line is the far+near detector sensitivity. 
For 17~m and the far detector only sensitivity calculations, 10~\% of flux 
systematics and 50~\% of the $\bar{\nu}_e$ background uncertainty are assumed. 
For far+near sensitivity calculation, it is assumed that 2\% of the systematic 
uncertainty remains after the cancellation and the 
$\bar{\nu}_e$ background is introduced as a free parameter. 
}
    \label{fig:FarSensitivity}
    \end{center}
\end{figure}

Figure~\ref{fig:FarSensitivity} shows 5~$\sigma$ sensitivities of these cases.
The green line is the sensitivity of the experiment in this proposal at 17~m. 
The blue line is the far-detector-only sensitivity at 60~m and 
the red line is the far+near detector sensitivity. 
For 17~m and the far detector only sensitivity calculations, 10~\% of flux 
systematics and 50~\% of the $\bar{\nu}_e$ background uncertainty are assumed. 
For far+near sensitivity calculation, it is assumed that 2\% of the systematic 
uncertainty remains after the cancellation while the amount of the 
$\bar{\nu}_e$ background is introduced as a free parameter.  
Here we assumed a detector with 1.0~k~ton is used, the detection efficiency 
is 50\%, MLF beam power is 1~MW and exposure is 2~years (4,000~hours/year). 
Only $\bar{\nu}_e$ from $\mu^{-}$ background, which is supposed to be most 
severe one, is considered. 

\clearpage

%%%%%%%%%%%%%%%%%%%%%%%%%%%%%%%%%%%%%%%%%%%%%%%%%%
% bibliography
%%%%%%%%%%%%%%%%%%%%%%%%%%%%%%%%%%%%%%%%%%%%%%%%%%


\begin{thebibliography}{99.}
\bibitem{cite:SK}
Y. Ashie et al. Super-Kamiokande Collaboration, 
Phys. Rev. D 71, 112005 (2005);
Y. Fukuda et al. (Super-Kamiokande Collaboration), 
Phys. Rev. Lett. 81, 1562窶・567 (1998).

\bibitem{cite:atm}
K. S. Hirata et al. [Kamiokande-II Collaboration], Phys. Rev. Lett. 63, 16 (1989);
J. Abdurashitov et al. [SAGE Collaboration], Phys. Lett. B 328, 234 (1994);
P. Anselmann et al. [GALLEX Collaboration], Phys. Lett. B 327, 377 (1994);
S. Fukuda et al. [Super-Kamiokande Collaboration], Phys. Rev. Lett. 86, 5651 (2001);
Q. R. Ahmad et al. [SNO Collaboration], Phys. Rev. Lett. 89, 011301 (2002);
C. Arpesella et al. [Borexino Collaboration], Phys. Rev. Lett. 101, 091302 (2008).

\bibitem{cite:acc}
H. Ahn et al. [K2K Collaboration], Phys. Rev. D 74, 072003 (2006);
P. Adamson et al. [MINOS Collaboration], Phys. Rev. Lett. 108, 191801 (2012);
K. Abe et al. [T2K Collaboration], Phys. Rev. D 85, 031103 (2012);
N. Agafonova et al. [OPERA Collaboration], Phys. Lett. B 691, 138 (2010).

\bibitem{cite:solar}
K. S. Hirata et al. [Kamiokande-II Collaboration], Phys. Rev. Lett. 63, 16 (1989);
J. Abdurashitov et al. [SAGE Collaboration], Phys. Lett. B 328, 234 (1994);
P. Anselmann et al. [GALLEX Collaboration], Phys. Lett. B 327, 377 (1994);
S. Fukuda et al. [Super-Kamiokande Collaboration], Phys. Rev. Lett. 86, 5651 (2001);
Q. R. Ahmad et al. [SNO Collaboration], Phys. Rev. Lett. 89, 011301 (2002);
C. Arpesella et al. [Borexino Collaboration], Phys. Rev. Lett. 101, 091302 (2008).

\bibitem{cite:reactor}
F. P. An et al. [Daya Bay Collaboration], Phys. Rev. Lett. 108, 171803 (2012);
J. K. Ahn et al. [RENO Collaboration], Phys. Rev. Lett. 108, 191802 (2012);
Y. Abe et al. [Double Chooz Collaboration], Phys. Rev. D 86, 052008 (2012);
B. Achkar et al, Nucl. Phys. B 434, 503 (1995).

\bibitem{cite:PMNS}
B. Pontecorvo, JETP34, 172 (1958);
V. N. Gribov and B. Pontecorvo, Phys. Lett. B 28, 493 (1969);
Z. Maki, M. Nakagawa and S. Sakata, Prog. Theor. Phys. 28, 870 (1962).

\bibitem{cite:LEP}
ALEPH, DELPHI, L3, OPAL, and SLD Collaborations, and LEP Electroweak Working Group, and SLD Electroweak Group, and SLD Heavy Flavour Group, Phys. Reports 427, 257 (2006).

\bibitem{cite:theory}
T. Asaka, S. Blanchet, and M. Shaposhnikov, Phys. Lett. B631, 151 (2005);
A. Boyarsky, O. Ruchayskiy, and M. Shaposhnikov, Ann.Rev.Nucl.Part.Sci. 59, 191 (2009).

\bibitem{LSND}
A. Aguilar et al., Phys. Rev. D64, 112007 (2001).

\bibitem{cite:MiniBooNE}
A.\,A.\,Aguilar-Arevalo\,et\,al.\,[MiniBooNE\,Collaboration], Phys.Rev.Lett.110.161801,2013.

\bibitem{GaAnomaly}
C. Giunti and M. Laveder, Phys.Rev.C 83 (2011) 065504
\bibitem{cite:ReactorAnomaly}

Th. A. Mueller et al., Phys.Rev.C 83 (2011) 054615.
P. Huber, Phys.Rev.C 84 (2011) 024617.

\bibitem{cite:tension}
J. Kopp et al., arXiv:1303.3011.

\bibitem{cite:contmue}
M. Antonello et al., arXiv:1307.4699 [hep-ex].

\bibitem{cite:contee}
C. Giunti et al., Phys.Rev. D86 (2012) 113014.

\bibitem{cite:IBD}
P. Vogel and J. F. Beacom, 
Physical Review D60 (053003).

\bibitem{cite:XSECgraph}
M. Shaevitz, 
2011 Workshop on Baryon \& Lepton Number Violation".

\bibitem{FLUKA}
G. Battistoni, S. Muraro, P.R. Sala, F. Cerutti, A. Ferrari,
S. Roesler, A. Fasso`, J. Ranft,
Proceedings of the Hadronic Shower Simulation Workshop 2006,
Fermilab 6--8 September 2006, M. Albrow, R. Raja eds.,
AIP Conference Proceeding 896, 31-49, (2007);
A. Ferrari, P.R. Sala, A. Fasso`, and J. Ranft,
CERN-2005-10 (2005), INFN/TC\_05/11, SLAC-R-773.

\bibitem{GEANT4}
J. Alison et al., 
 IEEE Transactions on Nuclear Science 53 No. 1 (2006) 270-278;
S. Agostinelli et al., 
Nuclear Instruments and Methods A 506 (2003) 250-303.

\bibitem{mudecay}
T.Suzuki et.al.,  Phys. Rev. C 35 (1986) 2212.

\bibitem{PHITS}
H. Iwase, K. Niita,T. Nakamura, 
JOURNAL OF NUCLEAR SCIENCE AND TECHNOLOGY 39 (11): 1142-1151
NOV 2002.

\bibitem{GdLS} M.~Yeh, A.~Garnov, and R.L.~Hahn, 
Nucl. Instrum. and Meth. A578 (2007) 329-339.

\bibitem{PMT} 
T.Matsubara, et al.,
Nucl. Instrum. and Meth. A661 (2012) 16-25.
%
\bibitem{cite:noumachi}
M. Nomachi, S. Ajimura, IEEE Transactions on Nuclear science, vol. 53, 2849-2852, (2006).

\bibitem{cite:umehara}
S. Umehara, et. al., IEEE Nuclear Science Symposium Conference Record, 2091-2094, (2011).

\bibitem{cite:SpW2013}
Masaharu Nomachi, Shuhei Ajimura, Takayuki Yuasa, Tadayuki Takahashi, Iwao Fujishiro, Fumio Hodoshima; SPACEWIRE BACKPLANE FOR GROUND EQUIPMENT, International Spacewire Conference 2013

\bibitem{cite:XSEC12C}
E. Kolbe, K. Langanke, G. Martﾃｭnez-Pinedo, P. Vogel,
J.Phys.G29:2569-2596,2003;

\bibitem{cite:PDG}
J. Beringer et al. (Particle Data Group), Phys. Rev. D86, 010001 (2012).

\bibitem{TOHOKUMONITOR}H. Furuta et al.,
"Pulse Shape Discrimination study with Gd loaded liquid scintillator for reactor neutrino monitoring",
IEEE proceedings of the 2nd int. conf. ANIMMA 2011, Ghent (2011).

\bibitem{COFASTN}Y-F. Wang et al., 
"Predicting Neutron Production from Cosmic-ray Muons",
 Phys. Rev. D 64, 013012 (2001). 

\bibitem{KamSP}S. Abe et al. [KamLAND Collaboration],
 "Production of radioactive isotopes through cosmic muon spallation in KamLAND",
 Phys.Rev. C 81, 025807 (2010).

\bibitem{CCS}E. Kolbe, 
"Differential cross sections for neutrino scattering on $^{12}$C",
Phys. Rev. C 54, 1741 (1996).

\bibitem{ENVGGEN}H. Furuta et al.,
"On-site underground background measurements for the KASKA reactor-neutrino experiment",
Nuclear Instruments and Methods, A,568,(2006),710-715.

\bibitem{COSMUGEN}J. Kempa and A. Krawczynska,
"Low energy muons in the cosmic radiation",
Nuclear Physics B (Proc. Suppl.) 151(2006) 299-302.

\bibitem{cite:KARMEN}
B.~Armbruster {\it et al.}  [KARMEN Collaboration],
  Phys.\ Rev.\ D {\bf 65}, 112001 (2002)
  [hep-ex/0203021].
  
\bibitem{cite:QgspBicG4}
http://geant4.web.cern.ch/geant4/G4UsersDocuments/UsersGuides\\
/PhysicsReferenceManual/html/node130.html.


\end{thebibliography}
\end{document}